\documentclass[11pt]{article}
\usepackage{graphicx}
\usepackage{booktabs}
\usepackage[letterpaper,margin=20mm]{geometry}
\usepackage{xcolor}
\usepackage{float}
\usepackage{pdflscape}
\usepackage{hyperref}
\usepackage[font=footnotesize]{caption}
\usepackage{subcaption}
\usepackage{comment}
\usepackage{amsmath}
\usepackage{listings}
\usepackage{makecell}
\usepackage{mathtools}
\usepackage[T1]{fontenc}
\usepackage[utf8]{inputenc}
\usepackage{authblk}

\date{}
\usepackage[
  style=ieee,
  citestyle=numeric-comp,
  backend=biber,
  defernumbers=true
]{biblatex}
\begin{document}
\title{Empirical Trajectory Sparsity Biases \\ Mobility-Informed Epidemic Modeling}

\author[1]{Federico Delussu}
\author[2]{Francisco Barreras}
\author[1,5,6]{Yuan Liao}
\author[2,3,4]{Duncan J. Watts}
\author[1]{Laura Alessandretti\thanks{Corresponding author: \href{mailto:lauale@dtu.dk}{lauale@dtu.dk}}}

\affil[1]{Department of Applied Mathematics and Computer Science, Technical University of Denmark, Lyngby, Denmark}
\affil[2]{Department of Computer and Information Science, University of Pennsylvania, Philadelphia, PA, USA}
\affil[3]{Annenberg School of Communication, University of Pennsylvania, Philadelphia, PA, USA}
\affil[4]{Operations, Information and Decisions Department, Wharton School, University of Pennsylvania, Philadelphia, PA, USA}
\affil[5]{Department of Space, Earth and Environment, Chalmers University of Technology, Gothenburg, Sweden}
\affil[6]{Department of Human Geography, Lund University, Lund, Sweden}

\maketitle

\begin{abstract}
\noindent
GPS mobility data are increasingly used in epidemic modeling, allowing the construction of co-location networks or population flows.
These trajectories typically exhibit high temporal sparsity because data collection is opportunistic and tied to phone use.
Despite growing awareness of this limitation, the analysis and treatment of biases derived from it have been largely overlooked in existing epidemic modeling studies, raising concerns about the robustness of downstream inferences.
We introduce a principled framework to quantify the impact of trajectory sparsity on key epidemic modeling outcomes across different levels of data missingness.
Our approach leverages a highly complete dataset that exhibits both near-complete and sparse GPS trajectories.
Near-complete trajectories provide baseline epidemic outcomes, while sparse trajectories provide realistic missingness patterns that we impose on the baseline to measure bias.
In this way, we show how missing records can result in substantial underestimation of key measures of epidemic intensity, explained not only by the amount of missing data, but by more complex features of data missingness that should be taken into account when designing correction methods.
Finally, we propose and evaluate a correction based on inverse probability weighting of the contact network before epidemic model calibration, which is shown to reduce bias and parameter misspecification.
We also demonstrate this correction on a separate anonymized sample from a commercial GPS mobility dataset and report on its effect.
Together, our findings provide a first rigorous quantification of trajectory-sparsity bias in epidemic modeling, offering initial guidance on the treatment of this issue.
\end{abstract}

\begin{refsection} % MAIN

\section{Introduction}

The COVID-19 pandemic demonstrated that human mobility data can be successfully integrated into a rich variety of experiments and decision-support tools, leading to their widespread adoption in epidemiology and other domains.
Behind this trend was the sharing of commercial GPS data from mobile phones with researchers and public institutions, as part of ``Data for Good'' initiatives by data providers~\cite{chafetz2022data4covid19,aktay2020google,apple2020covid,descartes2020aggregated}.
Compared with other sources of mobility data, GPS traces provide richer insights into the relationship between mobility and epidemic spreading~\cite{chang2021mobility, vanni2021use, buckee2020aggregated}, due to their high spatiotemporal resolution and coverage, often encompassing a few percent of national populations \cite{barreras2024exciting, yabe2020understanding, Crawford2022ImpactData, kang2020multiscale}.
Such high precision enabled the use of co-location networks for epidemic simulations~\cite{lu2026human}, the identification of infection hotspots \cite{aleta2022quantifying, chang2021mobility}, the evaluation of the effectiveness of vaccinations~\cite{chen2022strategic,gozzi2023estimating} and non-pharmaceutical interventions~\cite{kraemer2020effect, wellenius2021impacts}, and the study of behaviors associated with stay-at-home mandates \cite{Painter2021PoliticalMandates, weill2020social}.
These data have also informed human behavior in other domains, like disaster response and displacement~\cite{makinoshima2025large,giardini2023using,kieu2025modelling}, transportation systems and travel demand~\cite{abbiasov202415, magyar2025utilizing}, sustainable urban planning~\cite{gauvin2020gender,
xu2018planning, liao2026urban}, experienced segregation~\cite{liao2025effect}, and the broader determinants of human mobility and behavior~\cite{zheng2008understanding,eagle_inferring_2009,alessandretti2020scales, simini2012universal}.

Despite their potential, the robustness of findings derived from commercial mobility datasets can be limited by two types of bias: demographic biases, because individuals in the dataset are users of specific smartphone apps that do not fully represent the general population~\cite{barreras2024exciting, coston2021leveraging, grantz2020use, yabe2024enhancing}; and data-missingness biases, because location records are mainly sampled opportunistically when those apps are being used, and subject to throttling by battery-saving software settings linked to geofencing, detected movement, and battery level~\cite{zhou2020demystifying, graser2021exploratory, chen_lbs_biases_2026, bahr_missing_2022, mccool_maximum_2024, sanchez2026correcting}.
Such gaps in the data, or temporal sparsity, can range from minutes to several days, substantially affecting the coverage of commonly used panels~\cite{ugurel_correcting_2024, barreras2024exciting, chen_lbs_biases_2026, wang_exploring_2025, wu_location-based_2024}.
As an example, our analysis of a sample from Cuebiq---a widely used commercial GPS provider---with approximately 1,000 trajectories from Ithaca, NY, spanning September--November 2024, shows that over 80\% of users have no data in more than half of the hours (see Supplementary Section \ref{supp:Spectus}).
These numbers are in agreement with other samples from the same provider \cite{ugurel_correcting_2024}.
More importantly, these gaps can be correlated with contextual and behavioral variables, distorting estimated mobility metrics.
Sparsity appears to be higher at night, on weekdays, and when devices are at home~\cite{bahr_missing_2022, yoo2020quality, couture2022jue, ozaki2022direct}, while other work links it to the type of venue being visited~\cite{sanchez2026correcting}, social demonstrations~\cite{Budak2015DissectingMovement}, affluence~\cite{schlosser2021biases}, and inferred radius of gyration~\cite{Wesolowski2013TheMobility}.

\begin{figure}[h!]
\centering
\includegraphics[width=\textwidth]
{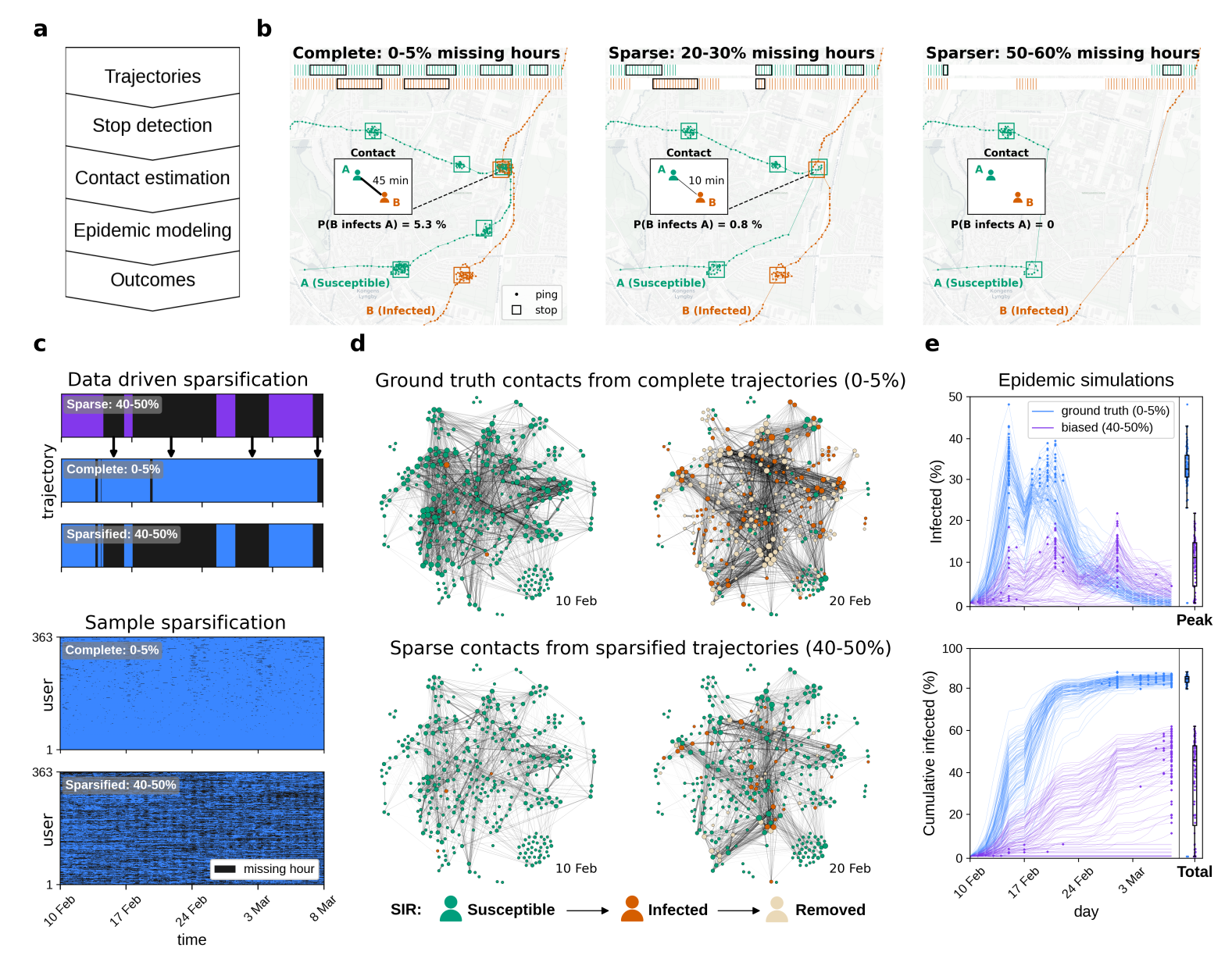}
\caption{\textbf{Quantifying the effect of trajectory sparsity on epidemic modeling outcomes.}
\textbf{(a)} Overview of the processing pipeline transforming raw individual GPS trajectories into epidemic modeling predictions.
\textbf{(b)} Synthetic example illustrating how trajectory sparsity affects the estimated probability that infected individual B transmits the infection to susceptible individual A.
The panels show three sparsity regimes: 0--5\% (left), 20--30\% (center), and 50--60\% (right) missing hours.
Stop locations (squares) are inferred from GPS pings (dots), with sparse timestamps (top insets).
Sparsity reduces contact duration and infection probability (central insets).
\textbf{(c)} Data-driven trajectory sparsification procedure, in which temporal gaps from a sparse trajectory are overlaid onto a complete one (top panel).
Sampling random sparsity patterns and applying them to the baseline complete trajectories results in a set of trajectories with realistic sparsity (bottom panel).
\textbf{(d)} Example SIR simulation on daily contact networks derived from complete trajectories (top) and sparsified trajectories (bottom), shown at initialization (seeded with 3 random nodes) and 10 days in.
Node colors indicate the individual's compartment (S, I, or R) (see Methods Section \ref{sec:methods}).
\textbf{(e)} Epidemic outcomes in the ground-truth scenario (blue) vs. the 40--50\% sparsity regime (purple), based on 100 simulation runs.
Top: daily infections over time, with dots indicating each simulation's peak infection count $I_{\texttt{peak}}$.
Bottom: cumulative infections, with dots indicating total infections $I_{\texttt{tot}}$.
Insets show boxplots for $I_{\texttt{peak}}$ and $I_{\texttt{tot}}$, with boxes representing IQR and whiskers denoting the 95\% confidence interval.}
\label{Fig:1}
\end{figure}

In response to these two sources of bias, researchers have developed much clearer guidance for handling demographic bias than for data-missingness bias.
Mobility panels can be linked to high-quality administrative data, like Census data, through inferred home areas.
Thus, validating panel composition against ground-truth sociodemographic data and the use of post-stratification to mitigate biases have become standard practice \cite{coston2021leveraging,yabe2024enhancing,li_understanding_2024,wang2018urban,Moro2021MobilityCities,athey_estimating_2021,yabe_behavioral_2023, safegraph_what_2019}, including in epidemiology \cite{chang2021mobility, couture2022jue, aleta2020modelling, aleta2022quantifying,  ozaki2022direct}.
In contrast, there is no comparable standard for the treatment of biases caused by trajectory data missingness.
Some studies propose methods to debias specific mobility metrics, mostly through imputation \cite{chen2019complete, couture2022jue, mccool_maximum_2024, Pepe2020COVID-19Lockdown, hwang2022comparison, schlosser2021biases} but also with weights \cite{liao2025effect}.
These approaches, while innovative and possibly effective, are usually ad hoc and typically not validated against high-quality ground truth.
Other studies do validate debiasing methods with external auxiliary data, including official hourly time-use surveys and direct follow-up surveys of frequently visited locations \cite{sanchez2026correcting,yoo2020quality}, but such benchmarks are not as comprehensive or granular as administrative data used for post-stratification.
Absent such benchmarks, another principled approach is to start from a baseline of relatively complete trajectories, ``sparsify'' them by removing records at random, and measure the induced error in metrics such as entropy, distance traveled, or radius of gyration, with and without correction \cite{barnett2020inferring, song2010limits}.
This approach enables validation against a known target statistic, echoing established practice in survey nonresponse analysis~\cite{Sarndal2005EstimationNonresponse, santos2019generating, schouten2018generating} and in graph theory \cite{mercier2022effective, spielman2008graph}.

In epidemic modeling, GPS data missingness propagates into epidemic outcomes by making network inputs---such as co-location estimates, flow networks, or point-of-interest visitation tables---less dense and connected.
The typical processing pipeline through which these errors propagate is summarized in Figure~\ref{Fig:1}a, while Figure~\ref{Fig:1}b gives an example of reduced connectivity due to trajectory sparsity.
These changes slow the spreading dynamics in ways well-understood by analytical and simulation studies of sparse, random, and synthetic temporal networks~\cite{karrer2014percolation,pastor2015epidemic,salathe2010dynamics,valdano2015analytical,nadini2018epidemic, van2008virus, Nowzari2016AnalysisNetworks}.
However, a rigorous treatment of trajectory data-missingness bias faces two limitations.
First, many epidemiological studies integrating human mobility data begin from already-aggregated inputs, in which the sparsity in the population contact patterns conflates with the sparsity induced by missing trajectory data~\cite{pei2020differential,chang2021mobility,kang2020multiscale,savi2023standardised,tizzoni2014use,schlosser2020covid,vanni2021use}.
This conflation makes sparsity bias correction infeasible, since inputs requiring correction may be indistinguishable from inputs with no missingness-induced bias.
It also limits studies quantifying bias by pruning or adding edges to already aggregated flow networks, and observing changes in epidemic outcomes \cite{gallotti2024distorted, schlosser2021biases, mercier2022effective}.
Second, direct benchmarks for bias correction---ground-truth contact patterns, epidemic outcomes, or nearly complete trajectories for data amputation---are rarely available.
Parameter calibration may partly correct this data-missingness bias, especially when epidemic models are fitted to historical incidence curves and mobility data~\cite{davis2020estimating,chang2021mobility,bhouri2021covid}; yet the extent to which model calibration debiases outcomes, and the degree of parameter misspecification it would introduce, remains an open question~\cite{bhouri2021covid,kheifetz2022parametrization}.

These limitations point to a gap between general awareness that GPS mobility data can exhibit sparsity, and the limited treatment of data missingness bias in epidemic modeling research.
In contrast to demographic bias, for which validation experiments and post-stratification are common practice~\cite{liu2024nonrepresentativeness, chin2025bias, pullano2024mobility, chang2021mobility}, biases from trajectory data-missingness are usually left unaddressed \cite{Crawford2022ImpactData,lucchini2021living, ross2021household, aleta2022quantifying, schlosser2020covid, kang2020multiscale, savi2023standardised, wesolowski2016connecting, pei2020differential, yabe_behavioral_2023, klein_characterizing_2024}, or treated with ad-hoc imputation procedures \cite{Pepe2020COVID-19Lockdown, couture2022jue, schlosser2021biases}.
To our knowledge, there is still no rigorous analysis of the effect of trajectory sparsity on key epidemic modeling outcomes, nor is there any basic guidance on how to correct biases arising from it.

In this paper, we address this gap by introducing a rigorous framework to quantify bias from temporal sparsity in GPS trajectory data and evaluate correction methods against a near-complete baseline, trajectories with data missing in less than 5\% of study-period hours.
We implement it by adding empirical temporal sparsity to near-complete anonymized GPS trajectories from approximately 1,000 students at the Technical University of Denmark (DTU) campus~\cite{stopczynski2014measuring}.
This dataset is well suited to our framework thanks to two key features.
First, it captures a tightly connected population with interactions concentrated in overlapping environments, facilitating simulations of epidemic spread.
Second, it combines fixed-interval measurements, which provide long highly complete sub-trajectories for some users, while other users exhibit opportunistic sampling triggered by app use and software settings, resembling commercial datasets.
This combination lets us compare epidemic simulations on near-complete trajectories against counterfactuals in which observed sparsity patterns from the opportunistic sampling have been added, with and without corrections.

Our contributions are threefold.
First, our framework enables analyses of trajectory-sparsity bias in epidemic modeling, where auxiliary benchmark datasets for direct bias measurement are rarely available.
We show that data missingness can result in underestimation of final epidemic size, peak incidence, and probability of outbreak, with the errors worsening as sparsity increases.
Second, we show that analyzing these biases using empirical missingness results in larger quantified biases compared to analyses removing records completely at random.
Thus, our data-driven sparsification approach improves upon previous studies of robustness assuming missingness completely at random (MCAR)~\cite{barnett2020inferring,song2010limits}, which can lead to erroneous assessments and treatment of trajectory-sparsity bias.
Third, we evaluate inverse probability weighting (IPW) of contact durations followed by epidemic model calibration as a correction grounded in survey nonresponse theory.
In experiments with DTU data, it successfully mitigates bias in key epidemic outcomes while reducing parameter misspecification relative to calibration using uncorrected mobility inputs.
Our experiments with sparse commercial GPS data from Cuebiq further show that the method remains practical even when high sparsity prevents a baseline of near-complete trajectories.
Together, these results raise concerns about the magnitude of these biases and the serious inferential errors that can occur when this problem is not addressed, while providing initial guidance for correcting data-missingness bias in the field, a step largely omitted in current epidemic modeling practice.

\section{Results}

To evaluate the effect of varying levels of trajectory sparsity on epidemic modeling outcomes, we use a sample of 363 near-complete trajectories from the DTU Campus data as a baseline for epidemic simulations.
Each trajectory is a sequence of device-level spatiotemporal records, or ``pings''.
Ping-table formats vary but a typical record has the form ``(device\_id, lon, lat, t)'', where ``device\_id'' are pseudo-anonymized device-level identifiers, ``lat'' and ``lon'' are WGS84 coordinates, and ``t'' is Unix time in seconds (see Methods Section \ref{sec:methods}).
Following~\cite{song2010limits}, we quantify sparsity as the percentage of hours without data, and define a trajectory as complete when its sparsity over the study period is below 5\%.

We then construct five sparsity regimes with target average sparsity levels between 10\% and 60\% (10--20\%, 20--30\%, 30--40\%, 40--50\%, and 50--60\%), consistent with typical patterns observed in commercial location datasets (Supplementary Section~\ref{supp:Spectus} reports the sparsity levels in data gathered from Cuebiq, a leading location intelligence company).
For each sparsity regime, we first select from the original dataset the trajectories whose empirical sparsity falls within the corresponding range; these trajectories provide data-driven templates of missingness.
We then replicate sparsity patterns by adding temporal gaps to the initially complete trajectories (Fig.~\ref{Fig:1}c).
This approach assumes that empirical missingness is independent of users and their mobility since we sample the gaps from individuals who are not part of the complete sample.

These near-complete trajectories therefore provide the baseline epidemic outcomes for the missingness process studied here.
Although this sample of users likely exhibits selection bias, we can be sure that statistics computed from such trajectories will not exhibit any data-missingness bias, hence epidemic simulations on these trajectories can be treated as ``ground-truth'' or target outcomes to compare against.
The sparsified versions then act as counterfactual trajectories, in which the only differences are the added sparsity patterns.
Thus, comparing epidemic simulations before and after adding these gaps allows us to isolate the bias.
This type of analysis is often referred to as data amputation or synthetic missingness in missing-data simulation studies~\cite{santos2019generating,schouten2018generating}.

Our simulations integrate a simple SIR compartmental model over a temporal network of contacts derived from the location trajectories of each pair of users.
As an example, Fig.~\ref{Fig:1}d shows two daily snapshots of a single SIR simulation run on contact networks derived from complete trajectories and from sparsified trajectories in the 40--50\% sparsity regime.

Repeating this simulation pipeline allows us to estimate variation in the epidemic outcomes while accumulating two sources of uncertainty: randomness from the epidemic dynamics and randomness from the sparsification process.
For each sparsity regime, we therefore use an ensemble of $5{,}000$ simulations, consisting of 100 epidemic simulations for each of 50 missingness realizations.
In the following sections, we report the mean and its standard deviation over the realizations of missingness for all epidemic outcomes.
Additional indicators, including confidence intervals, are reported in the Supplementary Information.

\subsection{Trajectory sparsity leads to underestimated intensity and frequency of epidemic outbreaks}\label{sec:bias-on-emos}

The adverse consequences of an epidemic outbreak are directly linked to how easily a pathogen can spread over the population.
We therefore focus on quantifying bias in model outcomes that describe the likelihood of an outbreak occurring as well as its intensity if it occurs.
First, the probability that an infection seed grows into an outbreak, $p_{\texttt{outbreak}}$, is central to early risk assessment and to deciding whether containment efforts are justified before wider spread occurs.
Given the ensemble of stochastic simulations, we define $p_{\texttt{outbreak}}$ as the fraction of simulations where $I_{\texttt{tot}}$ exceeds 5\% of the total population. Over the resulting set of outbreaks, we then define the total fraction of individuals infected by the end of the epidemic, $I_{\texttt{tot}}$, which reflects its overall burden and the human cost to society, and the day of last infection, $t_{\texttt{tot}}$, which captures how long the crisis is expected to last (Fig.~\ref{Fig:1}e).
Finally, the peak incidence, $I_{\texttt{peak}}$, together with its timing, $t_{\texttt{peak}}$, capture the moment of greatest pressure on health-system capacity, indicating how long authorities have to prepare to handle the crisis (see Fig.~\ref{Fig:1}e, bottom).
For our results, we compute point estimates, quartiles and 95\% CIs of the epidemic descriptive statistics, emphasizing again that we use only the ensemble of realized outbreaks (excluding the extinguished outbreaks allows us to consider only epidemic curves which are meaningful). 
In the following, we describe in detail the results obtained using the data-driven sparsification approach. 
Fig.~\ref{Fig:2}a-c (purple) and Fig.~\ref{Fig:2}d (left) illustrate and characterize the data-driven sparsification, aspects which are described in more detail against random baselines later in section ~\ref{main:random_sparsity}. 

For our parametrization of the SIR model, increasing trajectory sparsity leads to noticeable shifts in epidemic outcomes.
From Fig.~\ref{Fig:2}e (purple boxplot), we observe a decrease in epidemic size as empirical sparsity increases.
The ground-truth values of $ I_{\texttt{tot}} = 85.29 \pm 0.03\% $ and $I_{\texttt{peak}} = 33.86 \pm 0.06\% $ reach a minimum of $24.92 \pm 0.19\%$ and $6.87 \pm 0.05\%$ under 50--60\% sparsity, respectively.
This corresponds to a relative reduction of $70.78 \pm 0.23 \%$ for $I_{\texttt{tot}}$ and $79.72 \pm 0.16 \%$ for $ I_{\texttt{peak}} $.
Trajectory sparsity also reduces the probability of an outbreak.
Ground-truth simulations show a $p_{\texttt{outbreak}}$ of $98.32 \pm 0.15\%$ which decreases to $65.06 \pm 0.77\%$ for the 50--60\% range.
This corresponds to a relative reduction of $33.26 \pm 0.77 \%$.
Moreover, the variance in the probability of outbreak $p_{\texttt{outbreak}}$ increases with the sparsity levels (as detailed in Supplementary Table \ref{tab:mm_realized_outbreaks}).

From Fig.~\ref{Fig:2}f (top panel, purple plot) we observe that the day on which the number of cases peaks, $t_{\texttt{peak}}$, is delayed by over a week when sparsity reaches levels of 40\% or greater (as detailed in Supplementary Table \ref{tab:mm_dynamic_peak_day}).
Additionally, under high sparsity regimes, the epidemic simulations do not reach the extinction phase within the considered time window.% of 26 days.
Hence, the estimates of $I_{\texttt{tot}}$ and the day of last infection $t_{\texttt{tot}}$ are biased, and the variance in $I_{\texttt{tot}}$ is larger (see bottom panel of Fig.~\ref{Fig:2}f, and also bottom panel of Fig.~\ref{Fig:1}e, purple plots).

These results can be interpreted as follows: Holding all other factors constant, predictions using a sparser dataset decisively flatten the curve of cases, underestimating the intensity of the epidemic and probability of outbreaks.
As expected, trajectory sparsity reduces the number and duration of contacts, corresponding to the edges of the contact network, which in turn results in a decreased epidemic intensity.
The ground-truth scenario accounts for 65,766 daily detected contacts over the study period and an average time in contact (across non-zero edges) of 114 $\pm$ 0.5 minutes (detailed in Supplementary Section \ref{supp:mm_stats_contacts_infectivity}).
Both measures decline steadily as sparsity increases, reaching about 23,309 $\pm$ 90 daily contacts ($-64.56 \pm 0.14\%$ from the ground truth) and 87 $\pm$ 0.19 minutes of average contact ($-23 \pm 0.4\%$ from the ground truth) at the 50--60\% range.
A more direct proxy of the network's sparsity that takes into consideration the epidemic parameters is the average reproduction number $\overline{R}_0$ (detailed in Supplementary Section \ref{supp:epid_modeling}), which also exhibits a commensurate decrease (Supplementary Table \ref{tab:mm_avg_R0}).

\subsection{Assuming random sparsity distorts bias estimates}\label{main:random_sparsity}

\begin{figure}[H]
\centering
\includegraphics[width=.95\textwidth]
{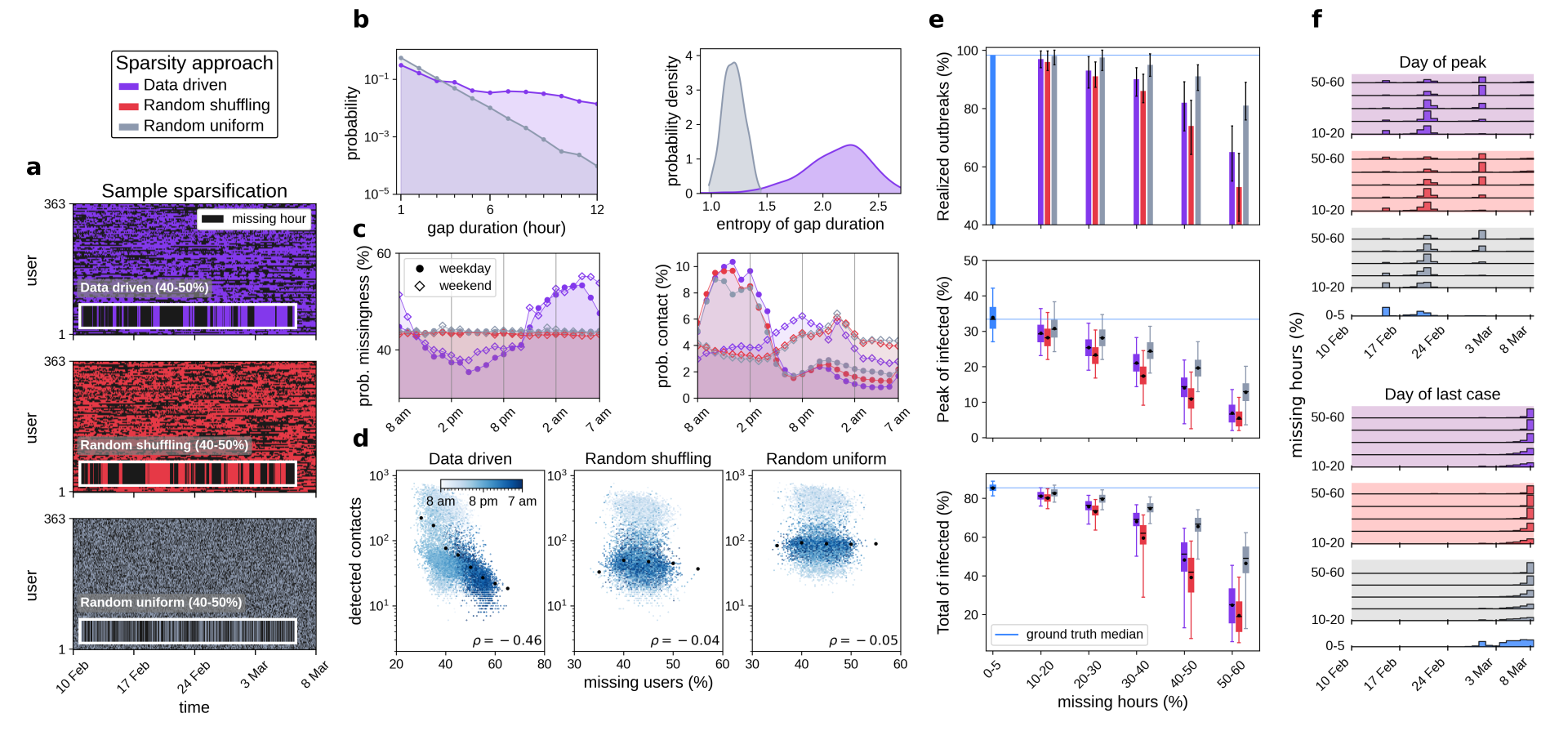}
\caption{\textbf{Data-driven sparsity differs from random baselines.}
\textbf{(a)} Trajectory sparsity under different sparsification approaches for the 40--50\% level of missing hours.
The data-driven approach (top, purple) is compared to two random baselines: (center, red) random-shuffling and (bottom, gray) random-uniform removal of records.
Rows indicate the users while columns indicate the time at the hourly resolution.
Missing hours are shown in black and hours with data are shown in color.
The inset in each panel represents a single trajectory.
\textbf{(b)} Distribution of gap duration (left) and gap duration entropy (right) for the three sparsification approaches.
The random-shuffling is equivalent to the data-driven approach since gap durations are preserved (gap statistic computation is detailed in Supplementary Section \ref{supp:mm_stats_sparsity}).
\textbf{(c)} Probability of missing data (top) and probability of contact between users (bottom) over hours of the day during weekdays (dots) and weekends (diamonds).
\textbf{(d)} Number of detected contacts vs. the percentage of missing users within each hour of data (dots).
Colors indicate the hour of the day (see color bar).
Black dots show median values within bins along the x-axis.
Results in panels a-d are shown for the 40--50\% sparsity level.
\textbf{(e)} Outbreak probability (top), peak daily infections (center), and total infections (bottom) for the three sparsification strategies.
The sample size of the realized outbreaks is reported in Supplementary Table \ref{tab:mm_realized_outbreaks_sample_size}.
Boxplots display the IQR with whiskers marking the 95\% CI.
Vertical lines denote the median and dots indicate the mean.
\textbf{(f)} Occurrence of peak day and last infection day over the ensemble of realized outbreaks.}
\label{Fig:2}
\end{figure}

As mentioned above, prior work on analyzing bias in mobility metrics due to sparsification or data amputation~\cite{song2010limits, barnett2020inferring} assumes missingness completely at random (MCAR), in which each record is deleted with a constant probability.
Superficially, the MCAR assumption seems implausible given extensive evidence of intertemporal correlations in behavioral data \cite{bahr_missing_2022, yoo2020quality, couture2022jue, sanchez2026correcting, Budak2015DissectingMovement}.
Nonetheless, it is of interest to study how different our analysis of bias is by performing data amputation according to the empirical data missingness patterns which, after all, are the type of missingness that would need to be addressed in practice.
To model missingness assuming MCAR, it is sufficient to know the volume of observed data.
In contrast, we expect empirical sparsity to be more heterogeneous, likely dependent on time, with longer gap durations, and even correlations in the missingness across users if they share the same context or use the same apps.

We evaluate the effect of sparsity heterogeneity in empirical missingness on the epidemic modeling outcome biases by comparing the data-driven outcomes with those obtained from two baselines:
(i) a random-shuffling approach, in which we sample from the sparse trajectories while preserving the duration of gaps and intervals with records, but randomize their order by alternating between them; and (ii) a random-uniform approach, in which hourly gaps are added homogeneously to match the prescribed sparsity level (Fig.~\ref{Fig:2}a).
Compared to the data-driven sparsity, the random-shuffling has the same distributions and entropies of gap duration by design (Fig.~\ref{Fig:2}b) but it breaks the temporal patterns of sparsity.
In fact, the probability that a record is missing is constant through time in contrast to the data-driven case (Fig.~\ref{Fig:2}c, top panel).
Instead, the random-uniform approach removes all dependencies and assumes that each hour's missingness is independent and identically distributed, preserving only the amount of missing hours but distorting both the gap distribution and the temporal patterns.
Supplementary Fig.~\ref{fig:mm_gap_statistics} compares the distribution and entropy of gap duration between the data-driven and random-uniform approaches.

On one hand, when the random-shuffling approach preserves real gap durations while breaking the time dependence, we observe a larger bias, with a steeper decline in epidemic intensity as sparsity increases compared to our chosen approach (gray versus purple in Fig.~\ref{Fig:2}e).
We observe a relative reduction of $I_{\texttt{tot}}$ for the 50--60\% level of $77.06 \pm 0.22$\% for $I_{\texttt{tot}}$ compared to the data-driven reduction of $70.78 \pm 0.23$\%.
On the other hand, the random-uniform approach underestimates the biases across all metrics of epidemic intensity (red versus purple in Fig.~\ref{Fig:2}e) with a relative reduction of $45.53 \pm 0.22\%$ (Supplementary Table \ref{tab:mm_size_total}).
For both baselines, we do not observe noticeable differences in the occurrence of the peak and of last infection (Fig.~\ref{Fig:2}f and Supplementary Tables \ref{tab:mm_dynamic_peak_day},\ref{tab:mm_dynamic_lastcase_day}).
Supplementary Section \ref{supp:mm_emobias} reports the variation of the epidemic outcome metrics for all sparsity ranges.

Empirical missingness deviates from the random-uniform approach in both gap duration and the temporal distribution of gaps, whereas the deviation from the random-shuffling approach lies solely in the time dependence.
However, these deviations distort the EMOs bias estimation in opposite directions: the random-uniform approach underestimates, while the random-shuffling approach overestimates the magnitude of the downward bias in epidemic intensity and probability of outbreak.
There are two main reasons underlying the result above.
First, the pre-processing of the data, namely, the stop-detection algorithm is robust to small temporal gaps, of which there are more in the random-uniform approach (Fig.~\ref{Fig:2}b and Supplementary Fig.~\ref{fig:mm_duration_contact_stops}).
Second, the random-shuffling approach breaks the dependence on the hour of the day, resulting in a higher share of gaps added during the day compared to what is observed under empirical missingness.
In turn, this results in relatively fewer estimated contacts since most of these happen during daytime hours (see Fig.~\ref{Fig:2}c, bottom panel).
Additionally, under data-driven sparsity, we measure a negative correlation ($\rho = -0.46$ for 40--50\% sparsity) between the fraction of users without records and the number of detected contacts in a given hourly bin (Fig.~\ref{Fig:2}d, left).
Instead, this correlation is much less pronounced for the two random baselines, $\rho \sim -0.05$ for 40--50\% sparsity (see Fig.~\ref{Fig:2}d, center and right panels and Supplementary Table \ref{tab:mm_correlation_contacts_missingusers} for the correlations over all sparsity ranges).

The bias discrepancies between the empirical and the two baselines reveal that independence assumptions about the data-missingness patterns, neglecting the heterogeneity present in real datasets, can result in incorrect estimations of sparsity-induced biases, and misinform the design of correction mechanisms.
These results are consistent across different levels of sparsity (see Supplementary Section \ref{supp:mm_stats_sparsity}).

\subsection{Debiasing EMOs with contact rescaling and calibration}\label{sec:debiasing-methods}

\begin{figure}[H]
\centering
\includegraphics[width=\textwidth]
{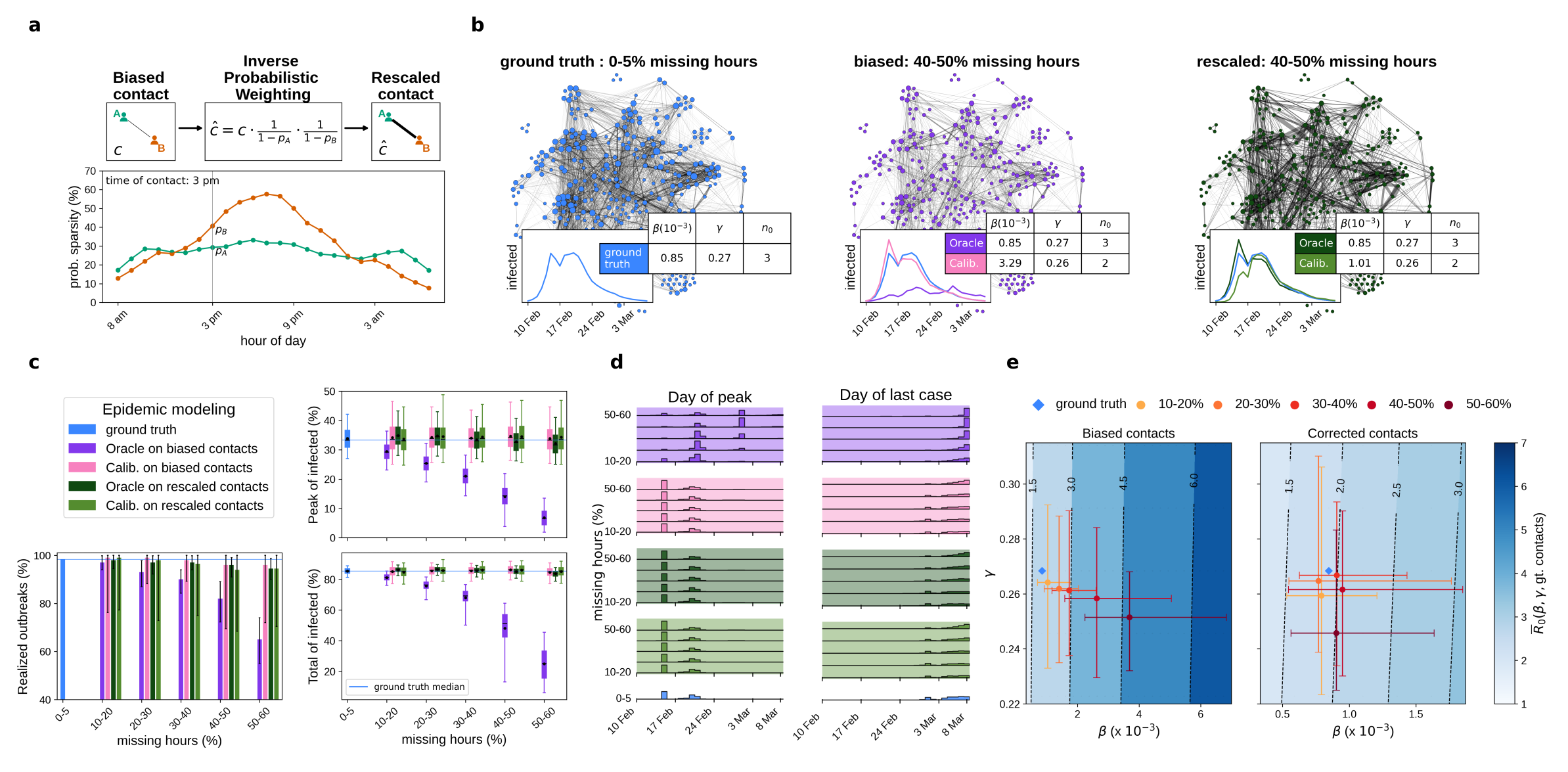}
\caption{\textbf{Debiasing approaches mitigate errors in epidemic simulations.}
\textbf{(a)} Contact rescaling methodology (top row): a biased contact duration is rescaled through inverse probability weighting (IPW) based on the probabilities that data are available at that hour for each of the users.
The probabilities used for weighting are computed at the hourly resolution for both weekend and weekday periods (Supplementary Section \ref{supp:methods_debiasing} details the rescaling procedure).
\textbf{(b)} Instance of oracle and calibration modeling on biased contacts (center) and rescaled contacts (right).
Oracle modeling employs the ground-truth simulation parameters (left table) while calibration searches for the optimal parameters which provide the best fit to a ground-truth curve of cases (left inset) derived from the ground-truth network and ground-truth simulation parameters (left).
Calibration on biased contacts recovers the ground truth (center inset) at the expense of overestimating the infection probability $\beta$ (center table).
The bias in this specific example is mitigated when calibrating on rescaled contacts (right table).
\textbf{(c)} Realized outbreak probability (bottom left), peak daily infections (top right), and total infections (bottom right).
The sample size of the realized outbreaks is reported in Supplementary Table \ref{tab:debiasing_realized_outbreaks_sample_size}.
Boxplots display the IQR with whiskers marking the 95\% CI.
Vertical lines denote the median and dots indicate the mean.
\textbf{(d)} Occurrence of peak day and last infection day over the ensemble of realized outbreaks.
\textbf{(e)} The ground-truth simulation parameters (blue diamond) are compared to the estimated parameters derived from calibration on biased contacts (left) and rescaled contacts (right).
Dots indicate the median and error bars denote the 95\% CI over the missingness realizations.
The underlying heatmap displays the average $\overline{R}_0$ computed from the ground-truth contacts over the $(\beta, \gamma)$ parameter space (Supplementary Section \ref{supp:debiasing} details the outcomes from calibration).}
\label{Fig:3}
\end{figure}

The simulations in our previous results use the ground-truth parameters, which are not available to practitioners wanting to apply epidemic modeling.
In some cases, parameter estimates are borrowed from clinical experiments or other prior work and, thus, treated as ground truth \cite{li_substantial_2020, chang2021mobility, birge_controlling_2022, aleta2020modelling}, a scenario in which our observation of growing bias with increased sparsity could apply and be mitigated by ``rescaling'' the network of contacts, compensating the loss of connectivity.
We will refer to simulations using ground-truth parameters as oracles.
However, the typical experimental design in mobility-informed epidemiology involves calibrating some, if not all, of the model parameters to target caseload data.
This calibration can reduce the bias in key outcomes, particularly if these are part of the target for calibration, without directly addressing trajectory sparsity. 
We therefore ask whether calibration alone is sufficient to address trajectory-sparsity bias, or whether direct treatment of sparsity in the mobility inputs is necessary.
To answer this question, we compare model calibration, rescaling of the contact network, and their combination.

To rescale the contact network, we re-weight each observed contact duration using hourly weights that reflect heterogeneity in trajectory sparsity.
Specifically, we apply inverse probability weighting (IPW), rescaling each observed contact by the inverse of the probability that both users have location records at that hour (Fig.~\ref{Fig:3}a).
We propose the use of IPW as it is a standard method from survey nonresponse theory and can be implemented using trajectory-completeness statistics computable from the raw data.
For model calibration, we fit the epidemic model parameters to reduce discrepancies between predicted and ``observed'' caseloads without further corrections to the biased mobility data (Fig.~\ref{Fig:3}b).
We implement this calibration using Bayesian kernel fitting~\cite{akiba2019optuna}.
The Methods Section \ref{sec:methods} describes both approaches in detail.
We first evaluate contact rescaling against no rescaling under the oracle setting to isolate its effect on EMOs, and then calibrate the model on both the uncorrected and rescaled contact networks using the same ground-truth target caseload curve.

Calibration accurately recovers $I_{\texttt{peak}}$, $I_{\texttt{tot}}$, $t_{\texttt{peak}}$, and $t_{\texttt{tot}}$, consistent with the fact that these quantities are descriptive statistics of the target caseload curve used for fitting (Fig.~\ref{Fig:3}c and Fig.~\ref{Fig:3}d).
More notably, contact rescaling alone also mitigates bias in these four outcomes under the oracle setting, even though it only re-weights non-zero contact durations while preserving the sparser structure due to missed contacts (Fig.~\ref{Fig:3}a).
The point estimates of $p_{\texttt{outbreak}}$ are closer to the ground truth for all debiasing methods, over 90\% versus 98\% for ground truth, albeit the variance is quite high for the calibration methods (Fig.~\ref{Fig:3}c).
Nonetheless, in the regimes with the highest sparsity, this statistic remains underestimated even after debiasing.
For instance, the complementary probability that a seeded outbreak extinguishes is overestimated threefold, over 9\% versus 2\% in the ground-truth case, in both the oracle and calibrated models using corrected contacts (Supplementary Table \ref{tab:da_realized_outbreaks}).

\begin{figure}[h!]
\centering
\includegraphics[width=0.65\textwidth]
{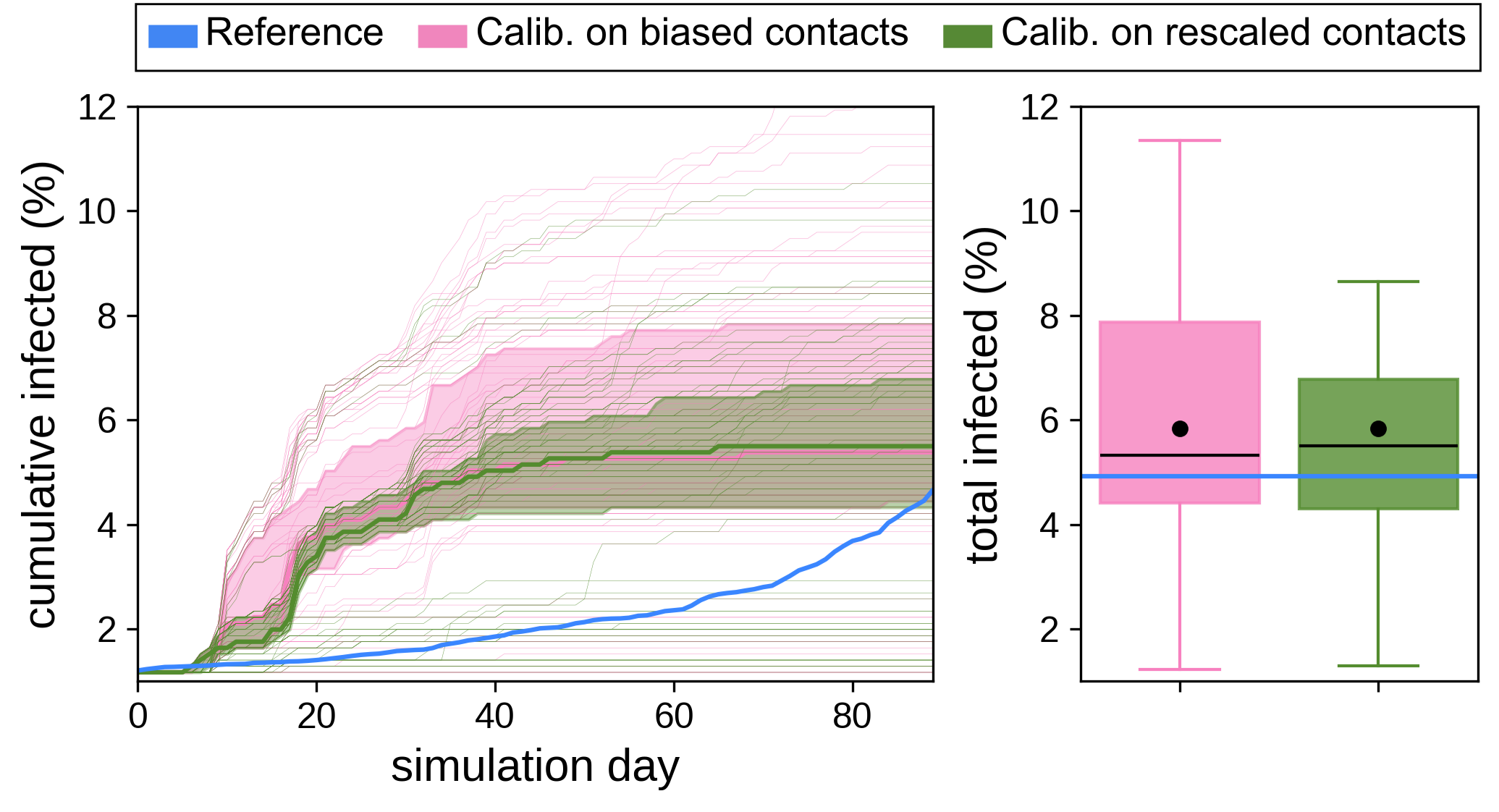}
\caption{\textbf{Epidemic modeling using Cuebiq data.
Contact rescaling reduces the uncertainty in total infected predictions.}
(left) A reference of cumulative infected cases (blue curve) is compared to an ensemble of 100 SIR simulations using parameters calibrated on biased contacts (purple curves, $(\beta, \gamma) = (0.704, 0.09)$) and on rescaled contacts (green curves, $(\beta, \gamma) = (0.005, 0.08)$).
Thin lines show the ensemble curves, the shaded region denotes the IQR and the thicker line represents the median across the ensemble.
(right) Total infected predictions from calibration on biased contacts (purple) and rescaled contacts (green).
Each boxplot summarizes the distribution of total infected across the ensemble: boxes denote the IQR, whiskers indicate the 95\% CI, black dashes mark the median, and black points denote the mean.
The blue horizontal line indicates the total infected for the reference curve.}
\label{Fig:Spectus_right_only}
\end{figure}

In order to demonstrate the practical utility of our proposed debiasing approach, as well as its effect on EMOs, we applied it to an equivalent model using mobility data from commercial provider Cuebiq (Supplementary Section \ref{supp:Spectus}).
Due to the high sparsity in this dataset, a reliable ground-truth baseline of nearly complete users was not available, preventing a full replication of our validation results.
Nevertheless, comparing debiased against biased outcomes, we find a notable reduction in uncertainty in the predictions of the total number of infected (Fig.~\ref{Fig:Spectus_right_only}) when using contact rescaling. Furthermore, the inferred infection rate $\beta$ is orders of magnitude higher without debiasing the network. However, unlike with the DTU mobility data and the synthetic epidemic, the best fitting models for each approach reach peak incidence earlier than the observed COVID-19 caseload---possibly due to limitations in the time window and the epidemic model itself.

We highlight two points from these results.
First, contact rescaling via IPW can mitigate bias in EMOs while requiring only hourly and trajectory-level completeness metrics.
While more sophisticated methods exist and richer auxiliary data could be used to inform the debiasing, our applications using DTU data and sparse commercial mobility data from Cuebiq suggest that IPW can be a practical, effective correction to the underlying epidemic networks provided granular data are available.
Second, calibration of the epidemic model's parameters using the biased networks can also recover several EMOs, which is relevant for setups in which the input epidemic networks are already aggregated.
When combining contact correction and calibration, however, caution is warranted because both the rescaling weights and the fitted parameters are random variables estimated from the data.
As such, errors can compound and increase the variance of the estimated EMOs (Supplementary Table \ref{tab:da_size_total}).
Furthermore, calibrating on corrected versus uncorrected contacts can lead to meaningfully different inferred epidemic parameters---a point we address next.

\subsection{Model calibration on corrected contacts improves parameter specification}\label{sec:debiasing_parameter_specification}

As we discussed earlier, biases from trajectory sparsity are often left unaddressed in epidemic modeling practice, yet the question remains whether model calibration could sufficiently debias the relevant outcomes on its own.
While our results are partially consistent with this hypothesis for key outcomes tied to the caseload curve, they also suggest that there is a critical benefit to correcting the trajectory sparsity in the contact network first.

Fig.~\ref{Fig:3}e clearly shows how the estimation of epidemic parameters and associated average reproduction number $\overline{R}_0$ computed from the ground-truth contacts ($\overline{R}_0$ formulation is introduced in Supplementary Section \ref{supp:epid_modeling}) are more robust to changes in sparsity level when using a corrected contact network.
The parameter misspecification of the model calibrated on sparse contacts increases with sparsity level, with estimated values of $\beta$ reaching values as high as $3.89 \pm 0.17$ $\times 10^{-3}$, and $\overline{R}_0$ as high as $4.77 \pm 0.12$ for 50--60\% sparsity, relative to ground truth $\beta^* = 0.84 \times 10^{-3}$ and $\overline{R}_0^* = 1.92$.
However, when calibration is applied after debiasing the network, the recovered values of $\beta$ and $\overline{R}_0$ in the high sparsity regime are $\beta=1.00 \pm 0.06$ $\times 10^{-3}$ and $\overline{R}_0=2.11 \pm 0.07$, respectively.
Indeed, the bulk of the estimated parameters result in an $\overline{R}_0$ between 1.5 and 3 across all sparsity ranges, while for uncorrected contacts, there are a considerable number of simulations where the recovered parameters induce an $\overline{R}_0$ greater than 4.5 and even 6 (Supplementary Tables \ref{tab:da_param_beta} and \ref{tab:da_param_R0}).
Namely, the recovered epidemic parameters are closer to the true parameters when using the corrected contacts, and consequently $\overline{R}_0$ is closer to the ground truth (Supplementary Fig.~\ref{fig:debiasing_contact_infectivity_reduction}c and Supplementary Table \ref{tab:da_avg_R0}).

The intuition behind this parameter misspecification is that, on average, epidemic incidence scales with the product of $\beta$ and the total volume of contacts.
When contact intensity is underestimated, model calibration compensates by increasing $\beta$ while keeping the product high enough to support the observed epidemic dynamics.
In other words, to obtain the same number of infections when using a sparser network, we must infer a more intense epidemic.
The misspecification is less relevant when the focus is on the predictions of models, but it is critical if the parameters are to be analyzed, compared with those in other studies, or used in simulations with different sparsity patterns.

\section{Discussion}
Recent studies have raised concerns about biases in human mobility data that can compromise scientific findings~\cite{barreras2024exciting, sanchez2026correcting, gallotti2024distorted}.
In epidemiological studies, mobility metrics constructed using different datasets often show disagreements~\cite{gallotti2024distorted, huang2021characteristics, noi2022assessing, chin2025bias, liu2024nonrepresentativeness}, even when they are expected to capture the same behavioral trends during COVID-19 lockdowns and reopening.
These differences could reflect demographic differences in panel composition \cite{Moro2021MobilityCities, coston2021leveraging, li_understanding_2024}, methodological choices in the construction of metrics~\cite{huang2021characteristics}, or other unobserved biases.
Our work contributes to the study of one such bias: sparsity in individual-level GPS records used for mobility-informed epidemic modeling.

Our experiments show that long gaps in the collection of mobility traces, owing to opportunistic sampling and technical limitations, introduce artificial sparsity in epidemic networks, leading to underestimates of the intensity of epidemic spread.
These effects from reduced network connectivity are consistent with prior work---both theoretical~\cite{newman1999scaling, chakrabarti2008epidemic,van2011decreasing, karrer2014percolation} and empirical~\cite{chang2021mobility, Chinazzi2020TheOutbreak, schlosser2020covid, brattig2023contact}, but the trajectory sparsity itself remains largely unaddressed.
When epidemic models start with aggregated inputs, it is challenging to even identify this missingness-induced sparsity as separate from the real sparsity of population-level contact patterns~\cite{chang2021mobility, kang2020multiscale, pei2020differential}.
We have implemented a framework that can isolate this effect, quantify the resulting bias rigorously, and evaluate the effectiveness of bias-mitigation strategies.
Our synthetic missingness assumes gaps are independent of visitation patterns---a simplification that enables a controlled counterfactual, but could impact our results.
However, using incomplete real data against external surveys risks conflating sparsity bias with panel composition effects~\cite{sanchez2026correcting}.

Addressing data-missingness bias is a pervasive challenge across disciplines where many possible solutions have emerged~\cite{rankin1985effects, tawn2020missing, Gelman2016TheVoter, Little1960ModellingSurveys, Sarndal2005EstimationNonresponse, barnett2020inferring, mercier2022effective}.
In practice, measurements of bias and its correction are difficult to validate without a counterfactual with complete data, knowledge of the true missingness process, rich auxiliary datasets, or simplifying assumptions about the missingness patterns.
We therefore extend an analysis framework from survey nonresponse in which complete data are ``amputated'' according to a realistic missingness mechanism, so that distorted estimates can be compared against a ground-truth baseline \cite{santos2019generating, schouten2018generating}.
Unlike prior analyses of mobility metrics where missingness has been added completely at random~\cite{song2010limits, barnett2020inferring}, we make the milder assumption that missingness patterns are independent of the individual.
This allows us to sample random sparsity patterns from an empirical distribution and construct an experimental setup in which trajectories with prescribed degrees of sparsity, between 10\% and 60\%, have a fully observed counterfactual.
By overlaying empirical gaps onto complete records, we preserve the real temporal distribution of missing data---including heavy tails in gap durations, and correlations with time (see Fig.~\ref{Fig:2}b--d).
These features produce larger estimated biases than those obtained when gaps are merely ``shuffled randomly'' (Fig.~\ref{Fig:2}e).
We note that experiments of this kind require complete mobility data, or at least a highly complete sub-sample, as in the DTU campus data.
This sets the DTU data apart from other datasets where even the most complete users have considerable sparsity~\cite{song2010limits, calabrese2013understanding, gonzalez2008understanding, yoo2020quality}.
Our equivalent analysis using data from Cuebiq lacked a sufficiently complete baseline of trajectories to fully replicate our findings (Supplementary Fig.~\ref{Fig:Spectus}a).
Nevertheless, the Cuebiq dataset still successfully demonstrated the practical utility of our proposed re-weighting and model calibration mechanisms.

At the core of our study is the finding that sparsity bias can be successfully mitigated through network re-weighting or epidemic model calibration alone.
While we illustrate with baseline approaches---IPW for network weights and Bayesian kernel fitting in a parsimonious SIR model (comparable to \cite{Chinazzi2020TheOutbreak, chang2021mobility, aleta2020modelling})---their success in recovering ground-truth epidemic outcomes is noteworthy.
It demonstrates that even straightforward, established debiasing methods are effective, increasing confidence that more sophisticated imputation or weighting procedures \cite{mercier2022effective, luca2021survey, chen2019complete, Sarndal2005EstimationNonresponse, coston2021leveraging, sanchez2026correcting, bang2005doubly, zhang2025survey} could yield strong results, although such evaluations are left for future work.
Furthermore, our IPW implementation offers immediate practical advantages: it requires no external auxiliary data, it directly models missingness as a function of time---possibly aligning with predictable circadian rhythms, battery lifecycles, and app-usage habits---and it avoids additional biases that could be introduced by methods joining Points-of-Interest (POI) data \cite{sanchez2026correcting}.

Two key limitations regarding model calibration warrant caution.
First, limitations in available near-complete data led to an in-sample evaluation, meaning the calibration scenarios require further out-of-sample validation.
Second, the finding that calibration alone can debias outcomes does not mean sparsity has been addressed in models implementing it.
Our results suggest models with downward-biased mobility inputs would end up with misspecification in meaningful transmission parameters---``absorbing'' the uncorrected reduction in network density during calibration (Fig.~\ref{Fig:3}e)---biasing downstream inferences and rendering alternative-scenario simulations less realistic.
Consequently, existing epidemiological conclusions drawn from uncorrected mobility inputs---for example, regarding reproduction numbers, infection hotspots, undocumented cases, or increased infectivity over time \cite{li_substantial_2020, chang2021mobility, aleta2020modelling, pei2020differential, aleta2022quantifying, yabe_behavioral_2023, schlosser2020covid, ross2021household}---likely reflect this underlying bias to some degree.

Network re-weighting can mitigate bias in epidemic dynamics, but it cannot recover edges erased by sparsity, which become indistinguishable from truly absent contacts.
Since seeded outbreaks require sufficient transmission paths from the seeds to the rest of the nodes, the elimination of a majority of edges between users (Supplementary Fig.~\ref{fig:mm_contact_infectivity_reduction}) likely explains why the probability $p_{\texttt{outbreak}}$ remains underestimated after contact-network correction in the oracle model (bottom left panel of Fig.~\ref{Fig:3}e).
Adding calibration partly compensates such structural loss, at the cost of misspecified transmission parameters, bringing $p_{\texttt{outbreak}}$ closer to ground truth for sparsity ranges below 40\% (Supplementary Table \ref{tab:da_realized_outbreaks}).

Beyond the debiasing procedures, our study is limited by how it constructs epidemic networks.
GPS data are ill-suited to approximate true co-location as well as other proximity sensors, like Bluetooth, can~\cite{stehle2011simulation, liu2013face, ozella2021using}, due to spatial inaccuracies, lack of altitude data, and the ``urban canyon'' effect~\cite{Crawford2022ImpactData, bierlaire2013probabilistic, olynik2002temporal}.
Our proxy for co-location, based on geohash-8 tiles and hourly intervals, is therefore only an approximation to true co-presence.
However, the bias pathway we identify should carry over to other methodologies~\cite{Pepe2020COVID-19Lockdown, chang2021mobility, aleta2020modelling, Crawford2022ImpactData}, as long as missing data can make transmission events shorter or missing.
Our framework also uses individual-level epidemic networks, whereas many epidemic models use aggregated POI-visitation tables to then construct networks~\cite{chang2021mobility, li_understanding_2024, safegraph_what_2019}.
Extending weights-based debiasing to such aggregated networks remains an important target for future work, especially because privacy-preserving safeguards often prevent direct work with individual-level data \cite{barreras2024exciting}.

The DTU-derived networks are also smaller and span a short period when compared with realistic city-level epidemic dynamics.
Our experiments use around 1,000 trajectories, only 363 of which form the near-complete baseline, from a population of university students with demographics and behaviors diverging from the city's whole population.
The search for a complete baseline also limited the simulation window to 28 days, constraining the parameter regimes that could produce non-trivial dynamics within the study period.
Future work could use datasets with larger near-complete sub-samples, that are more representative of city populations, where long-term epidemic dynamics and more complex network structure could emerge~\cite{watts2005multiscale, wu2008community, salathe2010dynamics}.

We hope that the work in this paper draws attention to the critical need for addressing data-missingness bias in applications that integrate sparse human mobility datasets, as well as further investigation into the nature and causes of this sparsity.
While our focus was co-location epidemiological networks, sparsity bias can distort other mobility metrics such as radius of gyration, unique locations visited, time spent at home, mobility entropy, among others, which underlie societally relevant findings in other sub-disciplines of human mobility science where, equivalently, the treatment of data sparsity falls short.
We hope the guidance presented here supports the development of debiasing methods for these other metrics as well, improving replicability and increasing confidence in findings derived from human mobility data.

\section{Methods}\label{sec:methods}

\textbf{DTU Campus data.}
We use a panel of GPS location data directly collected from $1,000$ students at the Technical University of Denmark between 2014 and 2015, who were given Android smartphone devices as part of the Copenhagen Network Study~\cite{stopczynski2014measuring}.
This dataset has been employed in several studies including the testing of stop detection algorithms~\cite{aslak2020infostop, cuttone2014inferring}, digital proximity tracing~\cite{cencetti2021digital}, and human social network modeling~\cite{sapiezynski2019interaction}.
The dataset has a 1-second temporal resolution and 1-meter spatial resolution; however, the sampling frequency is irregular and there can be gaps of several hours, and even days, in individual signals (Fig.~\ref{Fig:1}c), while the location has a mean horizontal accuracy of 20m~\cite{stopczynski2014measuring}.
For our experiments, we employ mobility data from a specific period spanning 28 days, from February 8, 2014 to March 7, 2014---dubbed the study period---over which a maximal cohort of users exhibited less than 5\% of hours without data.
However, the temporal sparsity patterns of all users in the sample were employed to simulate sparsity in the complete data.
The resulting sample consisted of 363 complete users.
For pre-processing the data, detecting stops and constructing contact networks, we first removed anomalous pings if the speed computed from the previous ping is higher than a maximum of 100 km/h using the scikit mobility library \cite{pappalardo2022scikit} and down-sampled the data to 1-minute resolution as it has little impact on stop-detection and reduces the sample size (the methodology for stop detection and contact estimation is detailed in Supplementary Sections \ref{supp:stop_detection} and \ref{supp:contact_estimation} respectively).
The choice of hourly bins for defining completeness is ad hoc and a compromise between seeking the highest possible granularity for analysis of contacts and visits, and getting a reasonable sample size of ``complete'' users---since completeness converges to zero as the bin width is reduced~\cite{cuttone2014inferring}.

\noindent \textbf{Epidemic modeling.}

The epidemic model used in our experiments is an individual-level stochastic Susceptible-Infected-Removed (SIR) model~\cite{Anderson1992InfectiousControl}.
Individuals in a population $P$ move through each of three compartments: Susceptible (S), their initial state indicating they can be infected by the disease; Infected (I), indicating they have contracted the disease and can infect others; and Removed (R), which is the residual compartment in which they can no longer contract or transmit the disease.
We model transitions in a timescale of days but the daily transition probability ($S\to I$) is obtained by integrating minutes in contact against a per-minute infection rate $\beta$.
The ($I \to R$) transitions are parameterized by a daily recovery rate $\gamma$.
Formally, if $C_{ij} (t)$ is the edge weight of the contact network of day $t$ (minutes in contact), and $X_i(t)$ is an indicator variable for whether individual $i$ is in compartment $X$ on day $t$, then individual-level transitions are given by equations \ref{eq:infection_prob} and \ref{eq:recovery_prob}.

\begin{align}
P(I_i(t+1)\hspace{3pt}|\hspace{3pt}S_i(t)) &=  1 - (1 - \beta)^{\sum_{j \in P} C_{ij}(t) I_j(t)}\label{eq:infection_prob} \\
P(R_i(t+1)\hspace{3pt}|\hspace{3pt}I_i(t)) &= \gamma\label{eq:recovery_prob}
\end{align}

\noindent and the compartments are initialized by assigning $n_0$ random users to $I$, and the rest to $S$.
Furthermore, we compute a daily reproduction number $\overline{R}_0$ indicating the expected number of infected individuals after a random individual was seeded as infected in an otherwise susceptible population on day $t$~\cite{stehle2011simulation} (derivation is detailed in the Supplementary Section \ref{supp:epid_modeling}).
Because this statistic indicates whether an epidemic on static network $C(t)$ will result in an outbreak ($R_0>1$) or be extinguished ($R_0<1$), we use the average value of $\overline{R}_0$ over the study period, namely $\overline{R}_0$, as an indicator of the severity of the epidemic which is directly proportional to the contact volume.

The simulation parameters for the ``ground-truth'' epidemic, $\Theta^* = (\beta^*, \gamma^*, n_0^*)$, were chosen ad hoc to produce an epidemic regime of moderate intensity---preventing immediate spread or extinction.
They result in a typical duration of the epidemic that aligns with the extent of the study period, and a value of $\overline{R}_0 = 1.92$.

We note that the initial day of the study period (February 8) falls on a Saturday.
Simulations seeded on weekends often fail to generate outbreaks, leading to trivial epidemic dynamics, due to lower contact frequencies relative to weekdays---an effect that is particularly pronounced in the student population captured by the DTU Campus data (see Supplementary Section~\ref{supp:mm_stats_contacts_infectivity}).
We therefore initialize the simulations on Monday, February 10, yielding an epidemic time window of 26 days.

\noindent \textbf{Experimental pipeline.}
Our simulations use the complete trajectories $T^*$ as a baseline to estimate the ``ground truth'' EMOs, while the sparsity patterns observed in the remaining incomplete trajectories are used to generate biased versions of such baselines.
Incomplete trajectories are grouped by their sparsity range $l$ (from 10--20\% to 50--60\% of missing hours), and each group is denoted $T^{(l)}$.
Complete trajectories undergo stop-detection using the Lachesis algorithm~\cite{hariharan2004project} (see parameters in Supplementary Table \ref{tab:method_stop-detection}).
Then stops are used to infer the total time individuals spend in close proximity, which is aggregated to daily individual contact networks $\{C^{(l)}(t)\}$.
We estimate the contacts between individuals through a grid-based approach using the geohash spatial index.
We choose the \texttt{geohash-8} resolution (approximately 21m x 19m at latitude 55.6761°N) which corresponds approximately to the area covered by a small building.
We define two individuals as in contact at minute $t$ when their stop locations fall in the same geohash-8 tile.
To mitigate the boundary problem, we also consider them in contact when their stops fall in the same tile after a small translation.
We aggregate the contact network at a daily level in order to provide it as an input to the epidemic SIR model.

For a given set of epidemic parameters $\Theta$, we simulate an ensemble of stochastic SIR epidemics on these networks and summarize the resulting distributions of daily infections.
Simulations using the true parameters $\Theta^{*}$ and complete networks ${C^{*}(t)}$ serve as the ground-truth baseline.
To quantify the bias introduced by missing data, we repeat the same process after introducing temporal gaps sampled from $T^{(l)}$ into $T^{*}$, producing sparse contact networks ${C^{(l)}(t)}$.
Note that these networks are random because the sparsification is stochastic.
For each sparsity range $(l)$, we generate 50 realizations of the missingness pattern and, within each, run 100 epidemic simulations, yielding an ensemble of 5,000 epidemic curves per configuration.
The same number of simulations is run for the ground truth for comparability.
We also repeat this pipeline based on two forms of correction: (i) calibration of epidemic parameters, denoted $\widehat{\Theta}$, obtained by fitting model outputs to case data; and (ii) debiasing of sparse networks using weights, yielding ${\widehat{C}^{(l)}(t)}$.
Our numerical results therefore compare five configurations summarized in Supplementary Table \ref{tab:method_model_scenarios}.

\noindent \textbf{Trajectory Sparsification}
The data-driven and the random-shuffling approaches sample the sparse trajectories with replacement from $T^{(l)}$.
The data-driven procedure matches each complete trajectory to a sparse one.
As an example, once a complete trajectory $T^*_{i}$ is matched to a sparse trajectory ${T}^{(l)}_i$, if ${T}^{(l)}_i$ has a gap from 9 a.m. to 12 p.m. on 10 Feb. 2014, we remove records in the same interval for $T_{i}$ and repeat this operation for all the gaps of ${T}^{(l)}_i$ obtaining the sparsified complete trajectory $T_i^{(*,l)}$.
Supplementary Section \ref{supp:traj_sparsification} details the procedure for selecting the trajectories for sparsification.

The random-shuffling differs from the data-driven approach by the randomization of the starting time of the gaps for each matched sparse trajectory.
Specifically, the trajectory is viewed as an alternating sequence of gaps and recorded intervals (inset of Fig.~\ref{Fig:2}a top panel).
All gap segments are randomly permuted among themselves, and all record segments are permuted among themselves.
The shuffled gaps and records are then reassembled in alternating order (inset of Fig.~\ref{Fig:2}a center panel).
This procedure preserves the original distribution of gap durations while randomizing their temporal placement.
Instead, the random-uniform approach consists of removing a set of hours uniformly at random from each complete trajectory such that the number of removed hours is within the desired range $l$.
Interestingly, these sparsification procedures can result in increases in the duration of contacts (see Supplementary Fig.~\ref{fig:mm_duration_contact_stops}).
While rare, this artifact can occur due to errors in the stop-detection procedure we implemented, which can detect a stop by clustering pings from different visits to the same location.

\noindent \textbf{Contact correction with inverse probability weighting.}
We apply inverse probability weighting (IPW) at the level of each contact in the co-location networks, rescaling each hourly contact duration by the inverse of an estimated probability of being observed~\cite{Sarndal2005EstimationNonresponse}.
While similar weights have been proposed in human mobility science \cite{liao2025effect, sanchez2026correcting}, this is, to the best of our knowledge, the first application in mobility-informed epidemic modeling.

Formally, for user $i$ and hour $h$ of day $t$, we estimate the probability that location records are present, namely $\hat{p}_{i}^{h}$, (where $\hat{p}_{i}^h = 1 - p_i^{h}$ is the complement of the sparsity probability $p_i^h$).
Assuming that the data-missingness processes of different users are independent, the probability of observing a potential contact between users $i$ and $j$ during hour $h$ is given by $\hat{p}_{i}^{h}\hat{p}_{j}^{h}$.
The corrected daily contact duration is then provided by equation \ref{eq:contact_rescaling}

\begin{align}
    \widehat{C}_{ij}(t) = \sum\limits_{h} C_{ij}(h) \cdot \frac{1}{1- p_i^{h}}  \cdot \frac{1}{1 - p_j^{h}} \label{eq:contact_rescaling}
\end{align}

\noindent where $C_{ij}(h)$ are the observed minutes in contact during hour $h$ (Supplementary Section \ref{supp:methods_debiasing} details the methodology for contact rescaling).
In our experiments, $\hat{p}_{i}^{h}$ is estimated via maximum likelihood, assuming a constant hourly probability conditional on the hour of the day and a weekday/weekend indicator.
The daily contact $\widehat{C}_{ij}(t)$ is then capped at 24 hours to avoid impossible durations.

\noindent \textbf{Calibration of epidemic modeling parameters.} In most modeling scenarios, either the network is not available at a high level of detail (that is individual-level with weights of contact duration at the minute resolution) and/or knowledge of the data quality of the trajectories from which the estimated network is derived is absent.
These conditions preclude applying the correction to the network.
An alternative common approach is calibration; that is, fitting the model to find the optimal parameters which, given the network as input, best align with a ground-truth reference curve.
We employ the median curve of infected as a ground-truth reference for calibration and define as objective function the RMSE between the ground truth and the biased daily median curves of infected.
The optimal parameters $\hat{\Theta} = (\hat{\beta}, \hat{\gamma}, \hat{n}_0)$ are obtained through Bayesian kernel fitting using the Optuna optimization package~\cite{akiba2019optuna} (Supplementary Section \ref{supp:methods_debiasing} details the methodology for calibration).

\section{Code \& Data Availability}

The DTU data are not publicly available due to privacy considerations, including European Union regulations and Danish Data Protection Agency rules.
To protect the privacy of participants, data cannot be shared freely, but are available to researchers who meet the criteria for access to confidential data, sign a confidentiality agreement and agree to work under supervision in Copenhagen.
Please direct your queries to L.A. (lauale@dtu.dk).
Data from Cuebiq are not publicly available due to privacy considerations, but can be accessed through Cuebiq's Social Impact program \url{https://cuebiq.com/social-impact/} for academic purposes.
Code for all our experiments and data to reproduce the figures is available at: \url{https://github.com/FedericoDelussu/Epidemic_modeling_on_sparse_GPS_trajectories.git}.

\section{Acknowledgments}
This work was made possible with the support of the National Science Foundation, NSF award 2341932.
\printbibliography
\end{refsection}

\newpage

\appendix

\section*{Supplementary Information}
\renewcommand{\thesubsection}{S.\arabic{subsection}}

\setcounter{figure}{0}
\renewcommand{\thefigure}{\arabic{figure}}
\renewcommand{\figurename}{Supplementary Figure}

\setcounter{table}{0}
\renewcommand{\thetable}{\arabic{table}}
\renewcommand{\tablename}{Supplementary Table}

\setcounter{equation}{0}

\begin{refsection}
\subsection{Experiments on a secondary dataset}\label{supp:Spectus}

\begin{figure}[H]
\centering
\includegraphics[width=0.7\textwidth]
{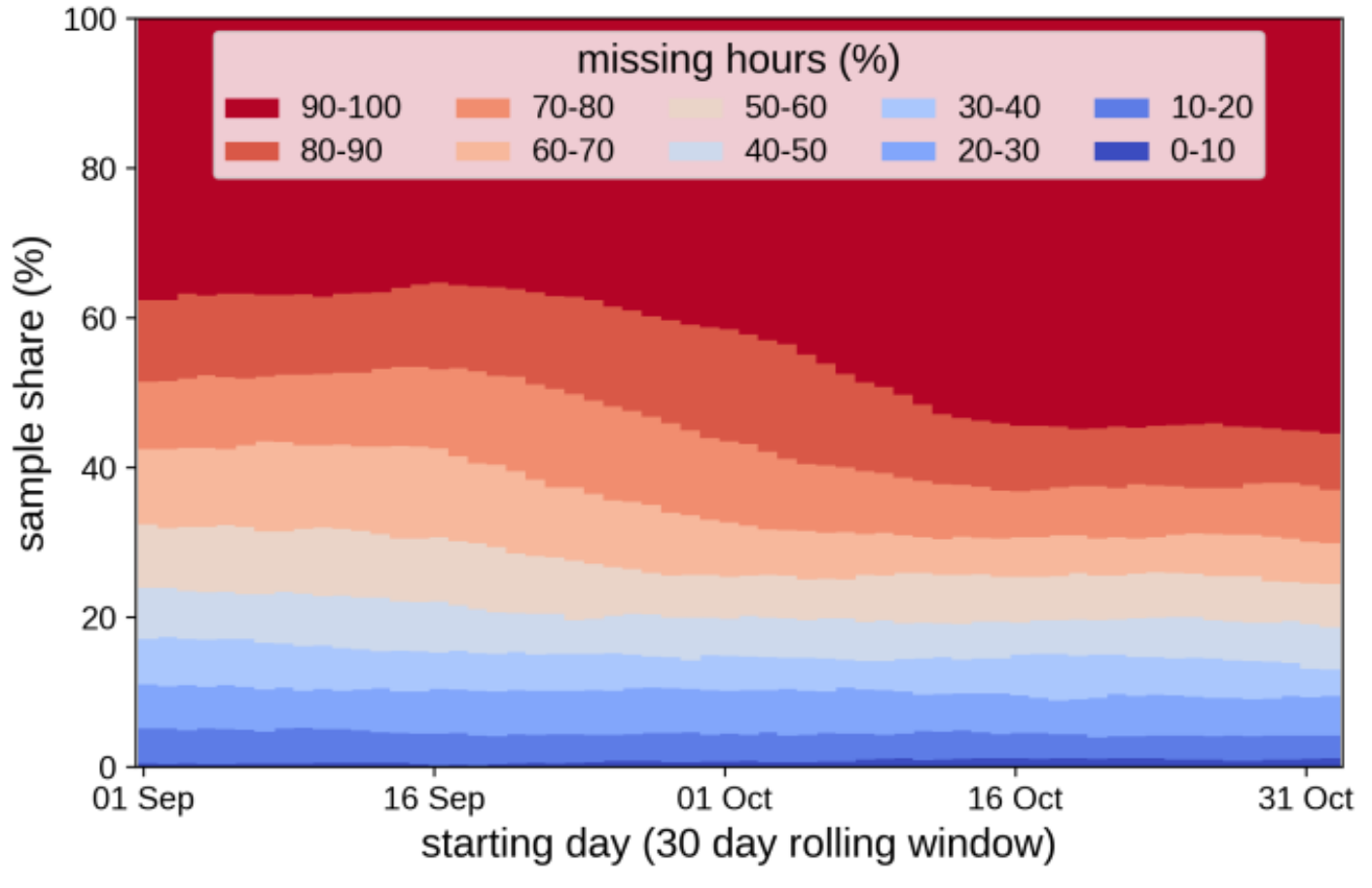}
\caption{\textbf{Time series of trajectory completeness for Cuebiq data.}
Given a sample of anonymized 1,356 GPS trajectories, sparsity is quantified by a 30-day rolling window with a step of 1 day.
The layered areas represent the daily fraction of trajectories with sparsity ranges across 0--100\% of missing hours over the 30-day period.}
\label{Fig:Spectus}
\end{figure}

We validate the debiasing methods (Section \ref{sec:debiasing-methods}) on a large-scale commercial US location dataset made available by Cuebiq, a leading location intelligence company that provides data to academic and humanitarian initiatives through its Data for Good program.
These data are collected through a CCPA- and GDPR-compliant framework from users who have opted in to share location data.
The data are anonymized and privacy-enhanced by removing records associated with sensitive locations and by obfuscating the locations of users' recurrently visited areas, such as homes and workplaces.
The sampling frame of this dataset is adult smartphone users in the United States, further filtered to devices active near the Cornell University Campus in Ithaca, NY---a college town chosen to capture the mobility of university students, similarly to the DTU dataset---ranging from September 1st, 2024 to November 30th, 2024. 
The period was chosen to balance trajectory completeness---at least 10 days observed in the campus area---with as large a number of devices as possible during 2024.
This resulted in a final sample of 852 trajectories for the experiments, which also include the pings outside the campus area.

Over this study period a very small fraction of trajectories had 0--10\% sparsity (Supplementary Fig.~\ref{Fig:Spectus}a), making the exact replication of our main findings infeasible.
Instead, we use this widely-used commercial dataset to demonstrate the practical applicability of our debiasing methods.
Our experiments compare EMOs on a co-location network constructed with the same processing pipeline, applying calibration on these biased contacts as well as on the rescaled contacts using IPW.

For these experiments, we use real caseload data from the COVID-19 pandemic for Tompkins County starting on 2020-09-10, obtained from the New York Times dataset \cite{nyt_covid19_data}\footnote{Available on \href{https://raw.githubusercontent.com/nytimes/covid-19-data/refs/heads/master/us-counties-2020.csv}{https://raw.githubusercontent.com/nytimes/covid-19-data/refs/heads/master/us-counties-2020.csv}}.
The curve is expressed as a percentage, dividing by 30{,}000, approximately the population of the city of Ithaca, to match the scale of simulated caseloads.
In particular, using a seed of 10 infected individuals corresponds to 1.2\% of our sample, which matches the rescaled target curve.

We note three differences in the experiment with respect to the DTU Campus data: (i) we calibrate to the cumulative infected and not to the daily infected as the latter signal is noisy and subject to high fluctuations; (ii) we use IPW rescaling coefficients computed from the overall trajectory sparsity without bucketing by hourly coefficients during weekday and weekend periods; and (iii) in the absence of a baseline to sparsify to different prescribed levels of sparsity, we consider only one set of calibrated parameters.

Consistent with our findings, calibration on corrected contacts reduces the variability in the prediction of the total infected (Fig.~\ref{Fig:Spectus_right_only}).
The 95\% CI of the total infected narrows from (1.2, 11.3)\% under sparse contacts to (1.3, 8.7)\% after contact correction while the IQR decreases from (4.4, 7.8)\% to (4.3, 6.8)\%.
Calibration on biased contacts leads to higher estimated infectivity, effectively absorbing the underestimation of contacts, with $\beta$ increasing from 0.005 when calibrating on corrected contacts to 0.70 when calibrating on sparse contacts, while $\gamma$ remains comparable (0.09 versus 0.08).

Since we calibrate to a true caseload of COVID-19, for which the daily infected peaks at the end of the selected time window, the biases on the dynamics, expressed by the predicted $t_{\texttt{peak}}$ and $I_{\texttt{peak}}$ cannot be easily quantified, yet the point estimates are fairly similar, with mean total infected (\%) just below 6\%.
However, likely due to sample size restrictions, simulated curves show convergence to that value early on followed by an extinction of the epidemic, while the real dynamics show exponential growth that would continue past the study period.

\subsection{Stop detection}\label{supp:stop_detection}

The procedure for contact estimation from the GPS trajectories involves the intermediate step of stop detection.
We use the sequential Lachesis stop detection algorithm \cite{hariharan2004project} which iteratively classifies each ping (lat, lon, t) as belonging to a stop or a trip.
A stop is defined as a temporally ordered collection of consecutive pings $\mathcal{P}_n$ which satisfies the conditions for stop assessment described in Supplementary Table \ref{tab:method_stop-detection}: (i) a minimum stop duration $\delta^{(stop)}_{min}$ of 10 minutes, comparable to the temporal frequency of the GPS trajectories in the DTU Campus data
(ii) a maximum temporal gap $\delta_{max}$ of 6 hours between consecutive pings, in order to classify sequences of pings with night-time gaps which are at most 6 hours long as stops, and (iii) a maximum diameter $D_{max}$ of 50 meters, comparable to the spatial extent of \texttt{geohash-8} cells used for contact estimation.
The diameter is computed as the maximum Euclidean distance across all pairwise distances between the pings in $P_n$.
Once the pings in $\mathcal{P}_n$ are classified as belonging to a stop, each stop location $s_{loc}$ is computed as the medoid of the sequence of pings.
That is the ping with the minimum average distance from the other pings in $\mathcal{P}_n$.

{\small\setlength{\tabcolsep}{4pt}
\begin{table}[H]
\centering

% ---- TOP SMALL TABULAR (2 columns, full width) ----
\begin{tabular*}{\textwidth}{@{\extracolsep{\fill}} l l}
\toprule
\textbf{spatio-temporal location data} & \textbf{description} \\
\midrule
$(lat,lon)$ & latitude and longitude coordinates \\
$m$ & time at minute resolution \\
\textbf{$p=(lat,lon,m)$} & individual ping \\
$\mathcal{P}_n = \{p_1, \ldots ,p_n\}$ & temporally ordered sequence of $n$ pings \\
\bottomrule
\end{tabular*}

% ---- MAIN TABULAR (same width) ----
\begin{tabular*}{\textwidth}{@{\extracolsep{\fill}} l l p{5cm} l}
\textbf{parameter} & \textbf{value} & \textbf{condition} & \textbf{formula} \\
\midrule
$\delta^{stop}_{min}$ & 10 minutes & minimum stop duration &
$m_n - m_1 \geq \delta^{(stop)}_{min}$ \\

$D_{max}$ & 50 meters & maximum roaming distance &
$\text{Diameter}(\mathcal{P}_n) < D_{max}$ \\

$\delta_{max}$ & 6 hours & minimum temporal distance between consecutive pings &
$m_{i+1} - m_i < \delta_{max}$ \\
\bottomrule
\end{tabular*}

\caption{\textbf{Lachesis algorithm for stop detection.}
(top) Individual trajectory data provided as input to the Lachesis algorithm.
(bottom) Parameters and conditions of stop assessment for a temporally ordered sequence of pings $\mathcal{P}_n$.}
\label{tab:method_stop-detection}

\end{table}

}

\subsection{Contact estimation from stop locations}\label{supp:contact_estimation}

After stop detection, each trajectory $T_i$ is converted to a sequence of stops $S_i = \{s_{i1}, .., s_{in}\}$ where each stop is expressed as $s := (s_{loc}, s_{start}, s_{end})$ denoting the stop location $s_{loc}$ and the stop start time $s_{start}$ and stop end time $s_{end}$ at the minute resolution.
For each pair of individuals $(i,j)$ we can estimate if they have been in contact during minute $m$ on geohash $q$ using:

\begin{equation}
    c_{ij}(m,q) = \delta_{i}^{(stop)}(m)\delta_j^{(stop)}(m)I[\text{gh}(s_i(m)) = q]I[\text{gh}(s_j(m)) = q],
\end{equation}\label{eq:contact_minute}

\noindent where $\delta_{i}^{(stop)}(m)$ is the indicator that $i$ is in a stop on minute $m$ and $s_i(m)$ is the corresponding stop location on minute $m$.
The stop location is mapped using the geohash spatial index to a cell $q = \text{gh}(s)$ at \texttt{geohash-8} resolution (21 x 19 m).
In order to account for spatial boundary effects we correct $c_{ij}(m,q)$ by adding a boundary term:

\begin{equation}
c^{\text{boundary}}_{ij}(m,q) =
\begin{cases}
0 & \text{if } c_{ij}(m,q) = 1, \\
1 & \text{if} \sum\limits_{\tau  \in \{\tau^{\uparrow}, \tau^{\nwarrow}, \tau^{\leftarrow}\}} \delta_{i}^{(stop)}(m)\delta_j^{(stop)}(m)I[\text{gh}(\tau(s_i(m))) = \text{gh}(\tau(s_j(m)))] > 0,
\end{cases}
\end{equation}\label{eq:contact_boundary}

\noindent where $\tau^{\uparrow}$ is a spatial translation operator which moves the stop location $l$ vertically upward by half of the height of the \texttt{geohash-8} cell ($\sim$ 10.5m), $\tau^{\leftarrow}$ is a spatial translation operator which moves the stop location $l$ horizontally leftward by half of the width of the \texttt{geohash-8} cell ($\sim$ 9.5m).
$\tau^{\nwarrow}$ is the combination of $\tau^{\uparrow}$ and $\tau^{\leftarrow}$ and accounts for neighboring geohashes which have one border in common.
The boundary term $c^{\text{boundary}}_{ij}(m,q)$ is different from 0 if two stop locations are in neighboring geohashes.

Given the activity space $Q$, defined as the set of geohashes where all individuals are located, we can define the indicator of contact between users $(i,j)$ on minute $m$ as:

\begin{equation}
    C_{ij}(m) = \sum_{q \in Q} c_{ij}(m,q) + c^{\text{boundary}}_{ij}(m,q)
\end{equation}\label{eq:contact_minute_geohashes}

We implement the $C_{ij}(m)$ computation and remove double counting of boundary contacts.
Thus $C_{ij}(m)$ can take values 0 or 1.
We then aggregate the estimated minute-level contacts at different temporal resolutions.
We use the hourly scale ($C_{ij}(h) = \sum_{m\in h } C_{ij}(m)$) to study intra-day temporal contact patterns.
We use the daily resolution ($C_{ij}(t) = \sum_{h \in t } C_{ij}(h)$) for running the SIR simulation on contact networks in which the edge weight represents the contact duration measured in minutes (Section \ref{supp:epid_modeling}).

\subsection{Epidemic modeling}\label{supp:epid_modeling}

Supplementary Table \ref{tab:method_SIR} describes the epidemic modeling parameters and the input contacts of the individual SIR compartmental model.

The model has parameters $\Theta = (\beta, \gamma, n_0)$ which govern the daily transitions over the different compartments, $S \rightarrow I$ and $I \rightarrow R$, described in the Methods Section in the main manuscript.

The probability that a Susceptible becomes Infected $P(S \rightarrow I)$ is obtained by compounding $\beta$---the probability of infection after a one-minute contact event with an infected---during day $t$.
The probability that an Infected transitions into the Removed compartment is constant and equal to $\gamma$ (see equations \ref{eq:infection_prob}, \ref{eq:recovery_prob} in the main manuscript).

{\scriptsize\setlength{\tabcolsep}{1pt}
\begin{table}[H]
\centering
\setlength{\abovecaptionskip}{4pt}
\setlength{\belowcaptionskip}{4pt}

\begin{tabular}{l l p{4cm} p{8cm}}
\toprule
\textbf{symbol} & \textbf{ground truth} & \textbf{name} & \textbf{description} \\
\midrule
$\Theta = (\beta, \gamma, n_0)$ & $\Theta^* = (\beta^*, \gamma^*, n_0^*)$  & SIR parameters & described below \\
$\beta$ & $\beta^{*} = 0.84 \cdot 10^{-3}$ & infection probability & prob. of infection given 1 minute contact with an infected. \\
$\gamma$ & $\gamma^{*} = 0.27$ & daily removal rate & prob. of transitioning to the Removed compartment over 1 day .\\
$n_0$ & $n_0^{*} = 3$ & seed size & initial number of infected individuals \\
\midrule
$P$ & $|P| = 363$ & population & individuals with complete trajectories \\
$D$ & $|D|=26$ & simulation days & period of 26 days starting from 2014-02-10 \\
$C_{ij}(t)$ & $C^*_{ij}(t)$ & contact duration & minutes in contact between $i$ and $j$ on day $t$  \\
$C(t)$ & $C^*(t)$ & daily contact network & $C(t) = \{C_{ij}(t) \quad;\quad i,j \in P\}$ for day $t$ \\
$C$ & $C^*$ & contacts & sequence of daily contact matrices \\ 
\midrule
\end{tabular}
\caption{\textbf{Individual SIR compartmental model.}
Summary of notation.}
\label{tab:method_SIR}
\end{table}

}

\noindent \textbf{Estimation of reproduction number $R_0$}

Given the contacts $C(t)$ on a given day $t$, we measure the severity of the epidemic spreading.
We introduce $R_0$ as the expected number of infected individuals after a random individual was seeded as infected in an otherwise susceptible population on day $t$~\cite{stehle2011simulation}.
Given an infected individual $i$ in a population $P$; we compute the expected number of infections $n_i$ generated by $i$:

\begin{equation}
    n_i(C(t), \beta,\gamma) = (1 - \gamma) + \sum\limits_{j \in P} [1 - (1-\beta)^{C_{ij}(t)}]
\end{equation}\label{eq:sir_number_infections}

$n_i$ depends on the epidemiological parameters $(\beta, \gamma)$ and the daily contact data $C(t)$ described in Supplementary Table \ref{tab:method_SIR}.
The formula is made up of two terms: (i) the probability that $i$ does not transition to the Removed compartment and (ii) the sum of the probabilities that $i$ infects its susceptible neighbors.

We then compute $\overline{R}_0$ as the average of $n_i$ computed for each individual of the population $P$ and obtain:

\begin{equation}
    R_0(C(t), \beta,\gamma) =  \frac{1}{|P|} \sum\limits_{i \in P} n_i(C(t), \beta,\gamma)
\end{equation}\label{eq:sir_R0}

Finally, in order to have a global metric of epidemic intensity, we compute a temporal average of $\overline{R}_0$ over the days $D$ of the study period:

\begin{equation}
  \overline{R}_0(C, \beta, \gamma) = \frac{1}{D} \sum_{t=1}^{D} R_0(C(t), \beta,\gamma)
\end{equation}\label{eq:sir_R0_avg}

In our implementation for computing $\overline{R}_0$ and the average $\overline{R}_0$ we use equation \ref{eq:sir_number_infections} without applying the approximation.

\subsection{Sparsification and selection of sparse trajectories}\label{supp:traj_sparsification}

\textbf{Sparsification}

We pre-process the data by selecting only trajectories whose longest temporal gap is shorter than one week, in order to focus on realistically sparse signals---characterized by gaps of the order of hours to days---while excluding trajectories dominated by extended inactivity longer than one week, which do not reflect the empirical missingness patterns of interest.

\begin{figure}[H]
\centering
\includegraphics[width=.8\textwidth]
{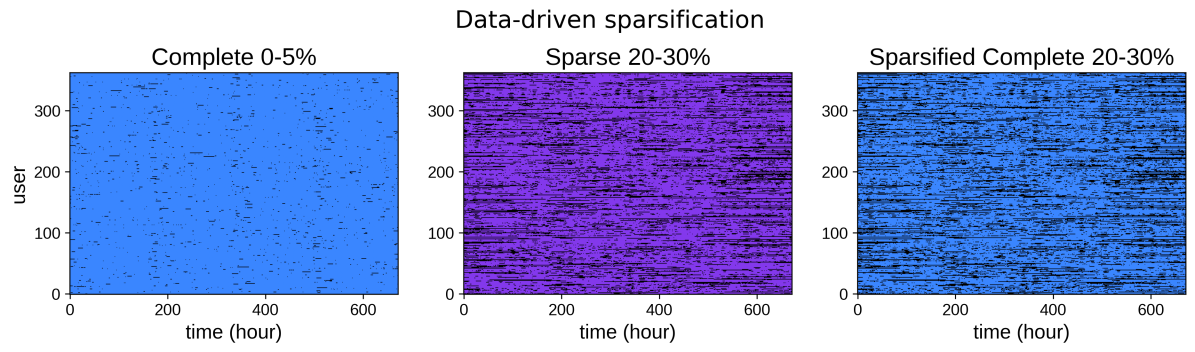}
\caption{\textbf{Instance of sparsification of the sample of complete trajectories for the 20--30\% range of missing hours.}
(left) Complete trajectories $T^*$, sample size is $N=363$.
(center) Sparse trajectories $T^{(l)}_{(N)}$ sampled with replacement from $T^{(l)}$ for $l= 20--30\%$.
(right) Sparsified complete trajectories $T^{(*,l)}$.}
\label{Fig:sparsification_instance}
\end{figure}

The procedure for the sparsification of the $N$ complete trajectories ($T^*$ sample) is described as follows: (i) The sample $T^{(l)}_{(N)}$ is collected from the set of sparse trajectories $T^{(l)}$ by sampling N times with replacement; (ii) Each complete trajectory $T^*_i$ is matched to a sparse trajectory $T^{(l)}_i$ in $T^{(l)}_{(N)}$; (iii) The sequence of temporal gaps of $T^{(l)}_i$ is applied to the complete trajectory $T^*_i$ by removing records with timestamps in those temporal gaps at the hour resolution, obtaining the sparsified trajectory $T^{(*,l)}_i$.
Supplementary Fig.~\ref{Fig:sparsification_instance} displays an instance of sparsification for the range of 20--30\% missing hours.
Supplementary Table \ref{tab:method_exp_pipeline} displays a summary of the notation for the sample of GPS trajectories and the corresponding estimated contacts.

{\small\setlength{\tabcolsep}{4pt}
\begin{table}[h]
    \centering 
    \begin{tabular}{llll}
        \toprule
        \textbf{GPS trajectory sample (N = 363)}  &  \textbf{Estimated contacts} \\
        \midrule
        $T^*$ : Complete (0-5\% missing hours) &  $C^*$ : ground truth \\ 
        $T^{(l)}_{(N)}$ : Sparse within range $l$ & none \\ 
        $T^{(*,l)}$ : Sparsified &  $C^{(l)}$ : biased contacts  \\
        \bottomrule
    \end{tabular}
\caption{\textbf{Summary of notation for trajectory sparsification and estimated contacts.}}
\label{tab:method_exp_pipeline}    
\end{table}

}

\noindent \textbf{Expansion of the sample of sparse trajectories}

\begin{figure}[H]
\centering
\includegraphics[width=.8\textwidth]
{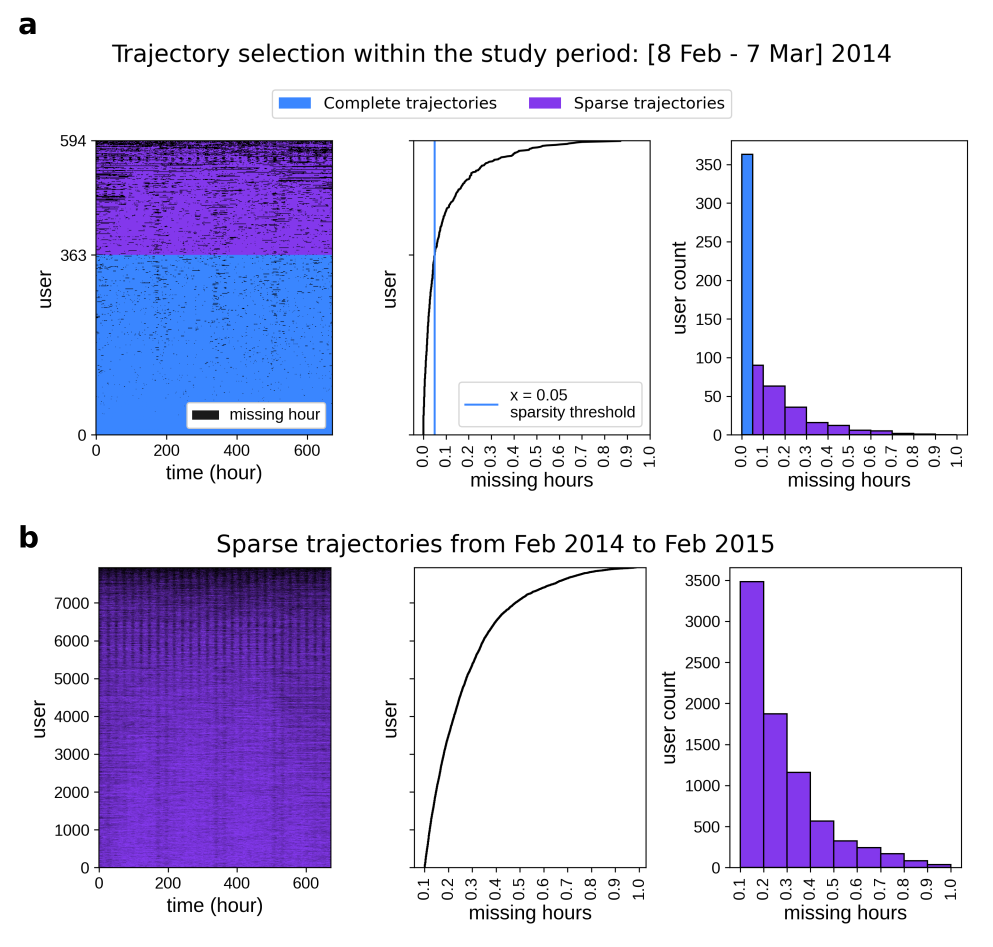}
\caption{\textbf{Selection of sparse trajectories for sparsification.}
\textbf{(a)} (left) Trajectory sparsity over the study period of 28 days from 8 Feb to 7 Mar 2014.
Rows indicate the users while columns indicate the time at the hour resolution.
Missing hours are shown in black and hours with data are shown in color.
Complete trajectories with less than 5\% of missing hours are marked in blue, while sparse trajectories are marked in purple.
(center) Individual-level sparsity quantified as the fraction of missing hours over the study period.
User ordering is the same as in the left panel.
(right) Count of users in each sparsity bin.
Sparse trajectories employed for sparsification are in ranges 10--20\%, 20--30\%, 30--40\%, 40--50\%, and 50--60\% of missing hours.
\textbf{(b)} Panels in row (a) are replicated for the sample of all sparse trajectories from Feb 2014 to Feb 2015 with the same duration as the study period and aligned by day of the week.}
\label{Fig:sparse_traj_count}
\end{figure}

The set of complete trajectories (sparsity between 0\% and 5\%) during the study period (from 8 Feb 2014 to 7 March 2014) consists of 363 sequences.
However, the number of devices in each sparsity range for that specific study period is considerably lower (Supplementary Fig.~\ref{Fig:sparse_traj_count}a, right panel), because this is a highly complete dataset.
This poses a problem when sampling with replacement, since it would result in users with redundant (perfectly correlated) data missingness, which might bias the results.
%Extension of the sparse sample
To overcome this issue, we expand the pool of gap sequences to be sampled from in the following way: for each of the non-complete users in the study period, we scan all the 28-day periods (aligned by day of the week) in the time frame between Feb 2014 and Feb 2015 to obtain many more gap sequences with varying sparsity levels.
With this approach, the number of gap sequences to sample from reaches a greater size than $N=363$ (Supplementary Table \ref{tab:count_sparse_trajectories} and Supplementary Fig.~\ref{Fig:sparse_traj_count}b, right panel).

\begin{table}[H]
\centering
{\scriptsize\setlength{\tabcolsep}{2pt}
\begin{tabular}{lrrrrrrrrrrr}
\toprule
missing hours (\%) & 0-5 & 5-10 & 10-20 & 20-30 & 30-40 & 40-50 & 50-60 & 60-70 & 70-80 & 80-90 & 90-100 \\
\midrule
\textbf{8 Feb 2014 - 7 Mar 2014}  & 363 & 90 & 63 & 36 & 16 & 12 & 6 & 5 & 2 & 1 & 0 \\
\textbf{2 Feb 2014 - 2 Feb 2015} & 11327 & 3972 & 3482 & 1873 & 1159 & 568 & 326 & 242 & 169 & 82 & 37 \\
\bottomrule
\end{tabular}

}
\caption{\textbf{Count of trajectories by sparsity level.}
Top row displays the count of individuals during the study period from 8 Feb 2014 to 7 Mar 2014 for each sparsity range.
The bottom row reports the counts of gap sequences for each sparsity range.
These counts reflect many gap sequences per user, corresponding to different 28-day sub-trajectories between Feb 2014 and Feb 2015 of the incomplete users (aligned by day of the week).}
\label{tab:count_sparse_trajectories}
\end{table}

\subsection{Debiasing approaches}\label{supp:methods_debiasing}

The different epidemic modeling approaches are described in Supplementary Table \ref{tab:method_model_scenarios}.

The ground-truth outcomes are obtained by simulating an SIR model on the ground-truth contact network $C^*$ estimated from the sample of complete trajectories using ground-truth parameters $\Theta^* = (\beta^*, \gamma^*, n_0^*)$ (Supplementary Table \ref{tab:method_SIR}).

In the first set of experiments we use the Oracle modeling which assumes knowledge of $\Theta^*$ both when simulating the epidemic on the sparse contacts $C^{(l)}$ and rescaled contacts $\hat{C}^{(l)}$ for a given trajectory sparsity level $l$ ranging from 10--20\% to 50--60\% of missing hours.

In the second set of experiments we use parameters $\hat{\Theta}$ obtained from calibration on the sparse contacts $C^{(l)}$ and rescaled contacts $\hat{C}^{(l)}$.

{\scriptsize\setlength{\tabcolsep}{2pt}
\begin{table}[H]
\centering
\begin{tabular}{lcc}
\toprule
\textbf{Epidemic modeling scenario} & \textbf{Network} & \textbf{Epidemic parameters} \\
\midrule
ground truth & ground truth $C^*$ & $\Theta^*$ (Table \ref{tab:method_SIR}) \\
oracle on sparse contacts & sparse contacts $C^{(l)}$  & $\Theta^*$ \\
oracle on corrected contacts & rescaled contacts $\hat{C}^{(l)}$ & $\Theta^*$ \\
calibration on sparse contacts & sparse contacts $C^{(l)}$ & $\hat{\Theta}$ calibrated \\
calibration on corrected contacts & rescaled contacts $\hat{C}^{(l)}$ & $\hat{\Theta}$ calibrated \\
\bottomrule
\end{tabular}
\caption{\textbf{Epidemic modeling scenarios.}
Summary of notation.}
\label{tab:method_model_scenarios}
\end{table}

}

\noindent \textbf{Contact correction at hourly level with inverse probability weighting}

Contact correction rescales the estimated contact duration by using inverse probability weighting (IPW).
As the contacts are estimated from the GPS trajectory data, we assume that the estimated duration is proportional to the level of completeness of such trajectories.
We model the biased contact duration $\widetilde{C}_{ij}(h)$ as proportional to the ground-truth duration $\hat{C}_{ij}(h)$---which is the target---and the probability $\hat{p}^h$ of observing a record during hour $h$ for both user $i$ and $j$.

\begin{equation}
\widetilde{C}_{ij}(h) = \hat{C}_{ij}(h) \cdot \hat{p}_i^{h} \cdot \hat{p}_j^{h}
\end{equation}\label{eq:contact_scaling}

By leveraging the heterogeneous temporal distribution of trajectory sparsity, we bucket each hour $h$ based on the hour of day (from 0 to 23) and the \textit{week period}: weekday (from Monday to Friday) or weekend (Saturday and Sunday).
We thus compute 48 hourly bucket probabilities for each user.
As an example, for 5 pm on a Friday (weekday), we evaluate the probability as the occurrence of record presence for the hours matching 5 pm on a weekday over the study period.
We express $\hat{p}^h$ as the complement of the sparsity probability; $\hat{p}^h = 1 - p^h$ and explicitly write the target $\hat{C}_{ij}(h)$ as:

\begin{equation}
\hat{C}_{ij}(h) = \widetilde{C}_{ij}(h) \cdot \frac{1}{1- p_i^{h}}  \cdot \frac{1}{1 - p_j^{h}}
\end{equation}\label{eq:contact_scaling_aggregation}

\noindent \textbf{Calibration of epidemic modeling parameters}

The calibration parameters $\hat{\Theta}$ are searched within the boundaries $(\beta, \gamma, n_0)$ of the parameter space specified by Supplementary Table \ref{tab:method_calibration_boundaries}.

The algorithm for parameter search is Bayesian kernel fitting \cite{akiba2019optuna}.
The algorithm starts from an initial condition and performs 100 trials.
The final parameter estimate corresponds to the minimum score.
The score computation is detailed in the Methods Section of the main article.

{\small\setlength{\tabcolsep}{4pt}
\begin{table}[H]
\centering
\begin{tabular}{lccc}
\toprule
 & $\beta$ & $\gamma$ & $n_0$ \\
\midrule
min & 5 $\cdot 10^{-5}$ & 0.117 & 1 \\
max & 1 $\cdot 10^{-2}$ & 0.632 & 10 \\
\bottomrule
\end{tabular}
\caption{\textbf{Boundaries for calibration parameter search.}}
\label{tab:method_calibration_boundaries}
\end{table}

}

\subsection{Changes in sparsity descriptive statistics}\label{supp:mm_stats_sparsity}

We consider the empirical sample of sparse trajectories $T^{(l)}$---employed for our proposed data-driven sparsification described in the main article---and compute descriptive statistics of sparsity for each range $l$ (from 10--20\% to 50--60\% missing hours over a period of 28 days).
We compute the same statistics for the random-uniform sample $T^{(l)}_{RU}$, having the same sample size of $T^{(l)}$ but differing by the randomization of the gap sequences (Trajectory sparsification is described in the Methods Section~\ref{sec:methods} in the main article).

We first compare the probability mass function of gap duration evaluated at the sample level:

\begin{equation}
    p_{gap}(d|T) = \frac{\# \text{gaps of d hours in } T}{\# \text{gaps in } T}\label{eq:gap_pmf}
\end{equation}

\begin{figure}[H]
\centering
\includegraphics[width=\textwidth]{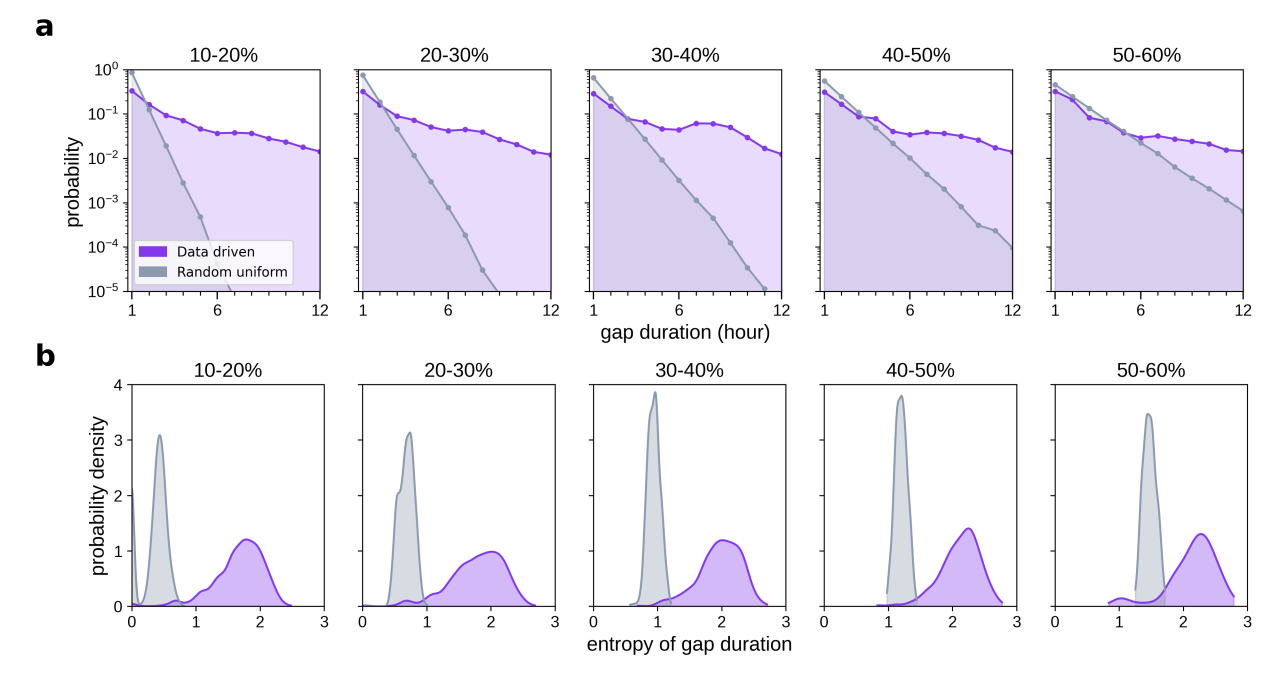}
\caption{\textbf{Variation in gap statistics.}
Each panel title denotes the range of missing hours in the sample of trajectories.
\textbf{(a)} Probability distribution of gap duration given the sample of trajectories used for sparsification under the data-driven and random-uniform approaches (described in Section \ref{main:random_sparsity} in the main article).
\textbf{(b)} Density of the trajectory-level entropy of gap duration (computed according to equation \ref{eq:gap_pmf_entropy}) from the samples described in (a).}
\label{fig:mm_gap_statistics}
\end{figure}

For example, we compute the probability of observing a gap of 2 hours as the occurrence of observations of 2-hour gaps over all observed gaps in all of the trajectories of the sample $T^{(l)}$.
From Supplementary Fig.~\ref{fig:mm_gap_statistics}, we observe that the data-driven probability has larger tails with respect to the random-uniform, the latter tends to have gaps of shorter duration.
Additionally, Supplementary Table \ref{tab:mm_gap_duration} shows that the average gap duration is less than 2 hours in the random-uniform case, consistently for all levels of sparsity.

We then compute the entropy of gap durations at the trajectory-level according to the Shannon formula ($H(X) = - \sum_{i=1}^{n} p_i \log p_i$).

\begin{equation}
    H_{gap}{(T_i)} = H(p_{gap}(d|T_i))\label{eq:gap_pmf_entropy}
\end{equation}

We gather a sample of entropy estimates evaluated for each trajectory in $T^{(l)}$ and evaluate the density of the entropy estimates using Gaussian kernel density estimation.
We observe that the entropy estimates increase with sparsity.
Consistently for all sparsity levels, the data-driven distributions are shifted toward larger entropy values and are more spread out than the random-uniform ones.
This result indicates the heterogeneity of gap duration patterns under empirical missingness (Supplementary Fig.~\ref{fig:mm_gap_statistics}b).

\begin{table}[H]
\centering
{\footnotesize\setlength{\tabcolsep}{4pt}
\begin{tabular}{lccccc}
\toprule
 & missing hours (\%) & avg. gap duration 
 (mean, mean std) & quartiles & 95\% CI \\
outcome &  &  &  &  \\
\midrule
\midrule
Data-driven & 10-20 & 3.67 $\pm$ 0.02 & (1, 2, 4) & (1, 15) \\
Data-driven & 20-30 & 6.21 $\pm$ 0.05 & (1, 3, 7) & (1, 37) \\
Data-driven & 30-40 & 7.30 $\pm$ 0.08 & (1, 3, 7) & (1, 51) \\
Data-driven & 40-50 & 7.10 $\pm$ 0.08 & (1, 3, 8) & (1, 39) \\
Data-driven & 50-60 & 7.13 $\pm$ 0.10 & (1, 3, 7) & (1, 42) \\
\midrule
Random uniform & 10-20 & 1.0569 $\pm$ 0.0004 & (1, 1, 1) & (1, 2) \\
Random uniform & 20-30 & 1.1734 $\pm$ 0.0008 & (1, 1, 1) & (1, 2) \\
Random uniform & 30-40 & 1.3277 $\pm$ 0.0014 & (1, 1, 1) & (1, 3) \\
Random uniform & 40-50 & 1.5275 $\pm$ 0.0022 & (1, 1, 2) & (1, 4) \\
Random uniform & 50-60 & 1.8017 $\pm$ 0.0040 & (1, 1, 2) & (1, 5) \\
\bottomrule
\end{tabular}
}
\caption{\textbf{Gap duration for the data-driven and random-uniform approaches.}}
\label{tab:mm_gap_duration}
\end{table}
\newpage

\subsection{Changes in the contact patterns}\label{supp:mm_stats_contacts_infectivity}
We compute the number of detected contacts (Supplementary Tables \ref{tab:mm_detected_contacts}) and the average duration of daily contacts (Supplementary Table \ref{tab:mm_contact_avg_duration}) over the study period.
We also compute the average reproduction number $\overline{R}_0$ (Supplementary Table \ref{tab:mm_contact_avg_duration}) according to Supplementary equation \ref{eq:contact_boundary}.

\begin{figure}[H]
\centering
\includegraphics[width=.8\textwidth]{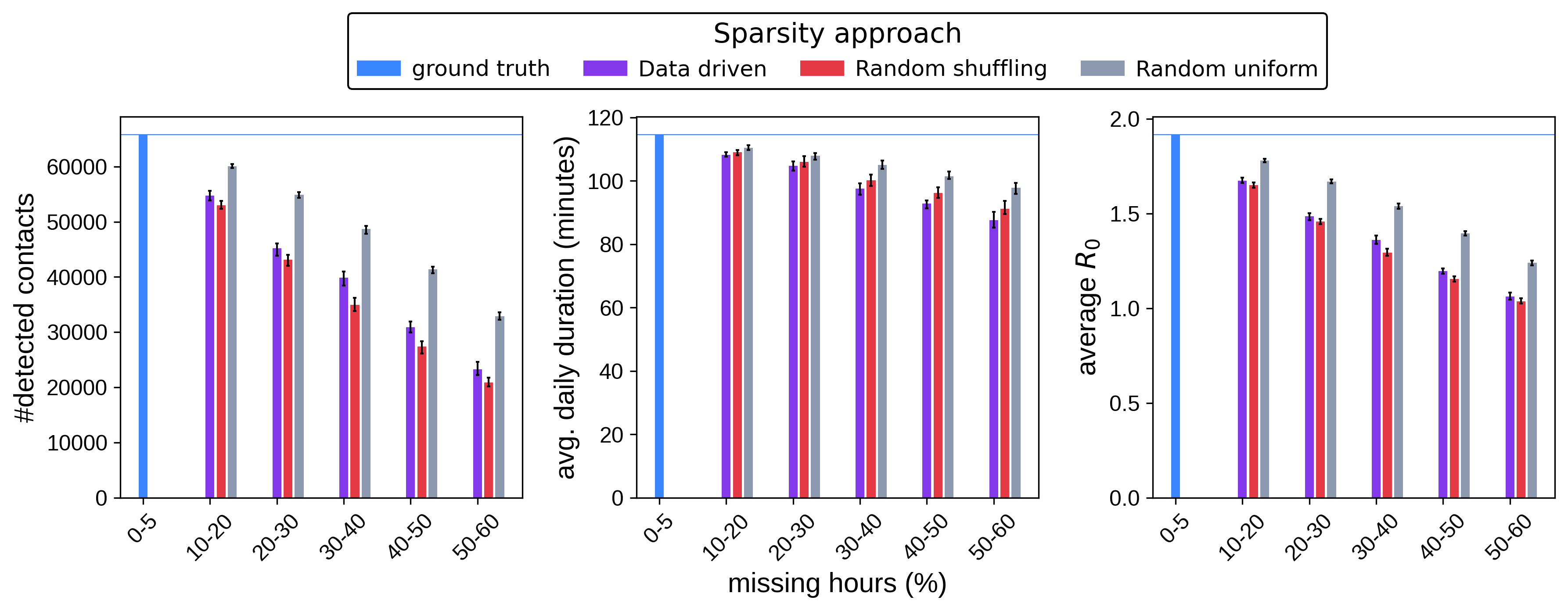}
\caption{\textbf{Variation in contact metrics and $\overline{R}_0$.}
Statistics derived from the contacts estimated over the study period.
Ground truth (blue) is compared to different sparsity scenarios, data-driven (purple) and random baselines (red and gray).
The bars represent the median value while the error bars denote the 95\% CI (reported in Supplementary Tables \ref{tab:mm_detected_contacts}, \ref{tab:mm_contact_avg_duration}, \ref{tab:mm_avg_R0}).
(left) Total number of detected daily contacts.
(center) Average duration of the daily detected contacts.
(right) Average $\overline{R}_0$ estimates computed from the contacts and the ground-truth epidemic parameters $(\beta, \gamma) = (0.84 \cdot 10^{-3}, 0.27)$.}
\label{fig:mm_contact_infectivity_reduction}
\end{figure}

We compare the ground-truth contacts obtained from the complete trajectories to the sparse contacts computed from the sparsified trajectories.
Supplementary Fig.~\ref{fig:mm_contact_infectivity_reduction} shows that both the number of contacts and the total contact time decrease steadily as sparsity increases.
The random-uniform model exhibits a weaker reduction as it is less affected by sparsity.

\begin{figure}[H]
\centering
\includegraphics[width=.7\textwidth]{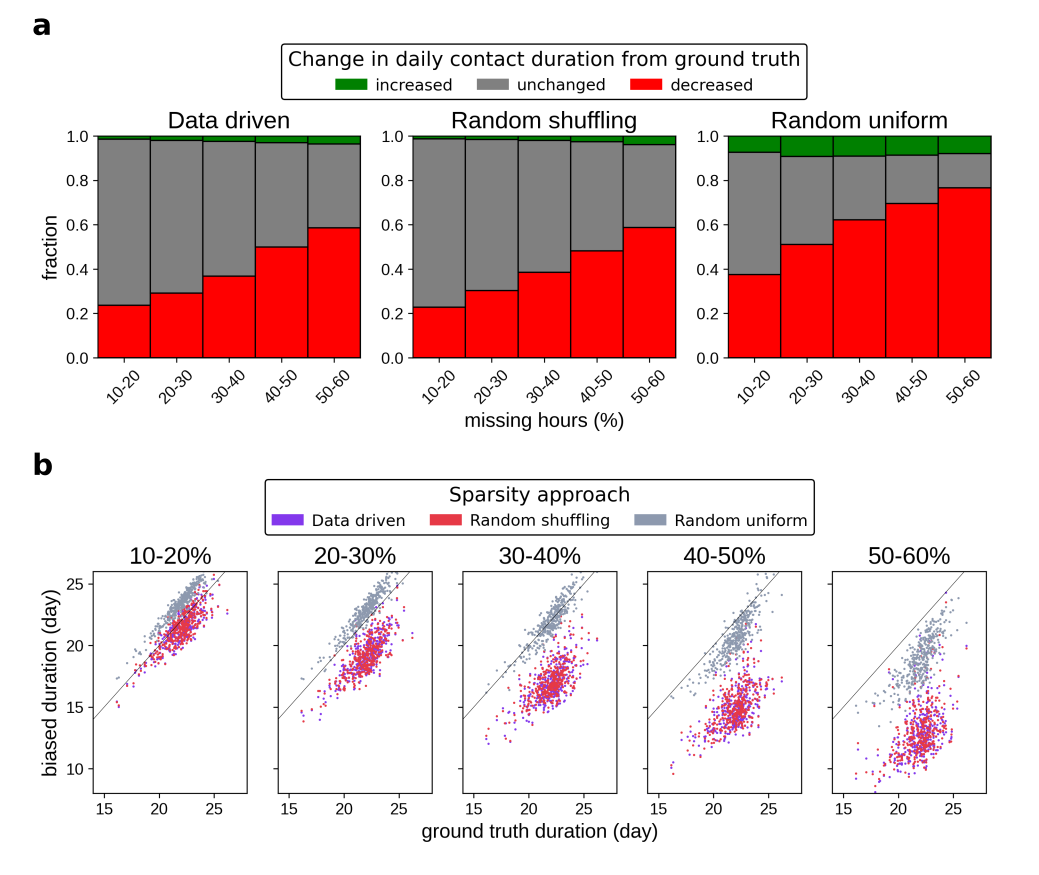}
\caption{\textbf{Variation of the estimated duration of detected contacts can be explained by the stop detection outcomes.}
\textbf{(a)} Share of estimated daily contacts according to the variation of duration with respect to the ground truth.
Each panel corresponds to a sparsification approach.
\textbf{(b)} Individual cumulative time spent in stops during [10 Feb--8 Mar].
Color denotes the sparsity approach for sparsification.
Each panel denotes the level of missing hours in the GPS location trajectories from 10--20\% to 50--60\%.
Each marker represents the individual cumulative time spent at stop locations over the study period.
This quantity is evaluated for the ground-truth scenario (x-coordinate) and for the sparse scenario (y-coordinate).
The line in each panel corresponds to the line $y=x$.}
\label{fig:mm_duration_contact_stops}
\end{figure}

Given the detected daily contacts, we compare the estimated biased daily contact durations to the ground-truth estimates.
From Supplementary Fig.~\ref{fig:mm_duration_contact_stops}a we observe that as sparsity increases there is an increasing fraction of contacts that have a decrease in duration with respect to the ground truth; while the fraction of contacts which are unchanged in duration shows an opposite trend.

Moreover, there is a small fraction of contacts (less than 10\%) that have an increase in duration with respect to the ground truth.
To explain this phenomenon in more detail we compute the total stop time for each individual over the study period.
In Supplementary Fig.~\ref{fig:mm_duration_contact_stops}b, we compare the stop durations (ground truth versus biased) and we observe that, as sparsity increases, the total time spent in stops tends to decrease---except for 10--20\% missing hours in the random-uniform case.
In this latter case, biased stop durations are greater than the ground truth.
This is observed for a minor fraction of the individuals also for the data-driven and random-shuffling approaches which have the same empirical gap duration distribution.

This result suggests that the Lachesis stop detection algorithm is robust to short-duration gaps as observed in the Random-uniform case; in contrast to the data-driven and random-shuffling cases for which the average gap duration is larger (Supplementary Table \ref{tab:mm_gap_duration}).
The robustness of stop detection can be explained by the tolerance parameter of the intertemporal distance between consecutive pings within the same stop; this is set to 6 hours in order to tolerate night-time gaps which occur with a larger frequency (Supplementary Table \ref{tab:method_stop-detection}).
Therefore, adding gaps of short duration might result---in some cases---in longer detected stops in the biased scenario.

\begin{figure}[H]
\centering
\includegraphics[width=\textwidth]{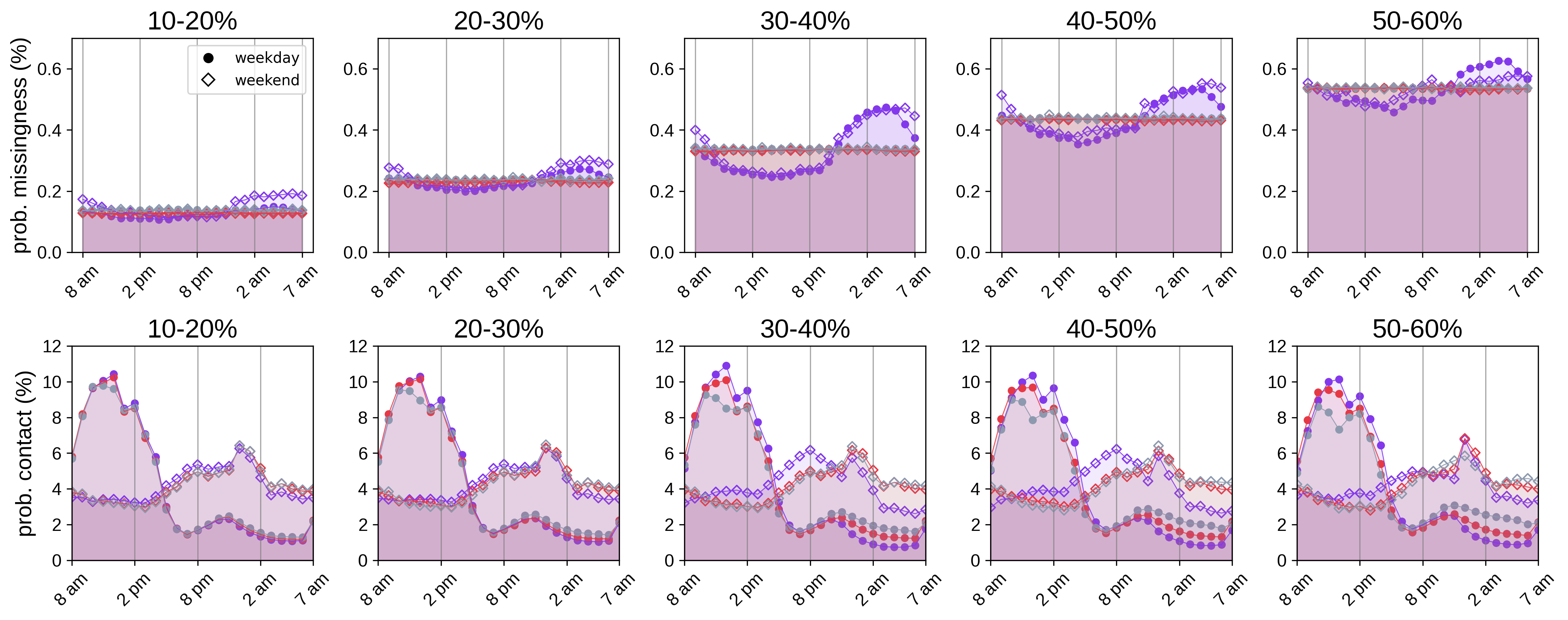}
\caption{\textbf{Probability of missingness and contact.}
Each panel denotes the range of missing hours (from 10--20\% to 50--60\%) and compares the sparsification approaches: data-driven (purple), random-shuffling (red), and random-uniform (gray).
\textbf{(a)} Probability that an hour has missing records during the weekday and the weekend.
\textbf{(b)} Probability that a contact is observed on a given hour during the weekday and the weekend.}
\label{fig:mm_prob_missingness_contact}
\end{figure}

\begin{figure}[H]
\centering
\includegraphics[width=.6\textwidth]{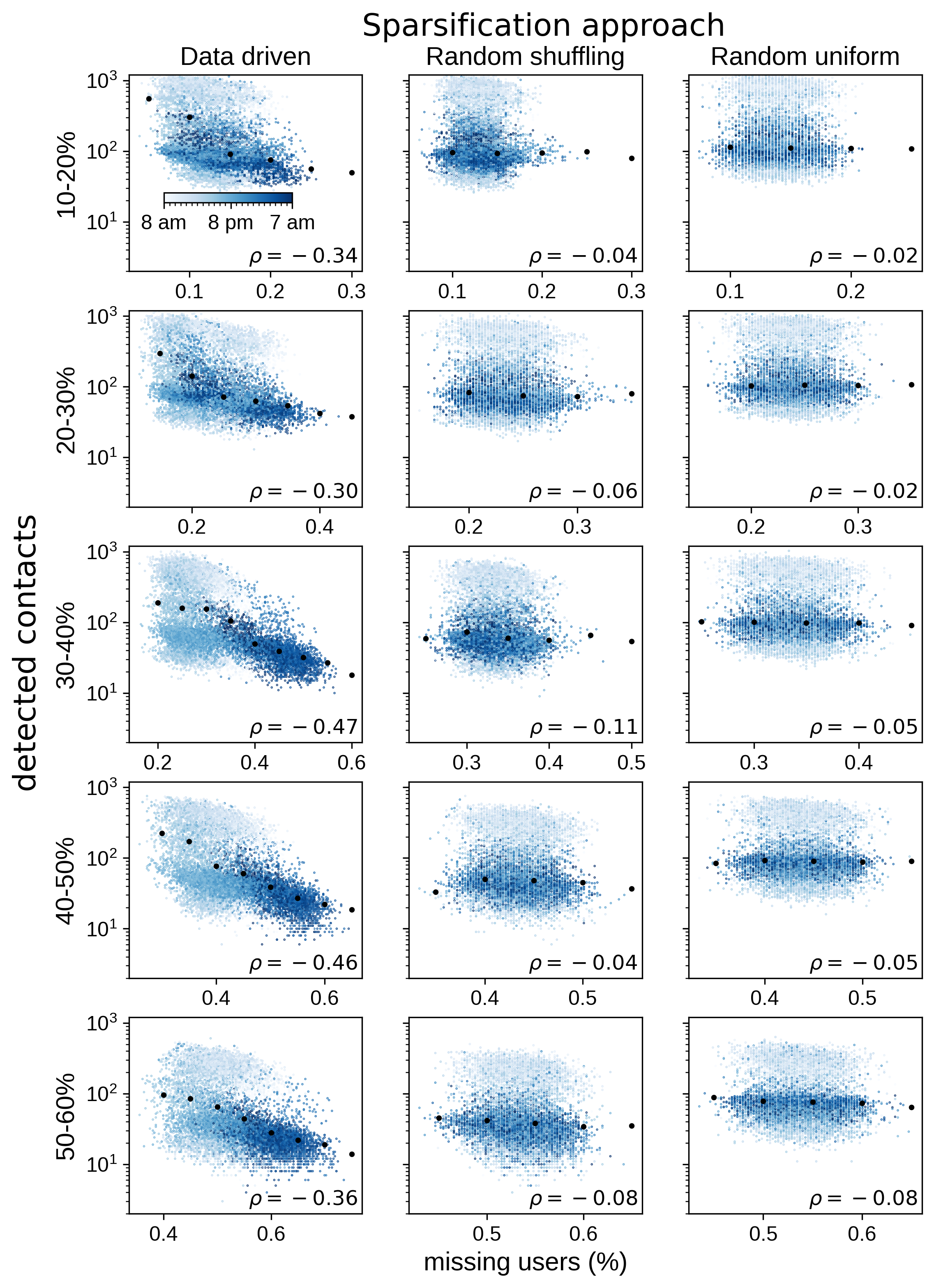}
\caption{\textbf{Correlation between missing users and detected contacts.}
Each row denotes the range of missing hours while the columns denote the sparsification approach: data-driven vs. random-shuffling and random-uniform baselines.}
\label{fig:mm_corr_missingness_contact}
\end{figure}

\begin{table}[H]
\centering
{\small\setlength{\tabcolsep}{3pt}
\begin{tabular}{lrrr}
\toprule
sparsity & Data driven & Random shuffling & Random uniform \\
\midrule
10-20 & -0.34 & -0.04 & -0.02 \\
20-30 & -0.30 & -0.06 & -0.02 \\
30-40 & -0.47 & -0.11 & -0.05 \\
40-50 & -0.46 & -0.04 & -0.05 \\
50-60 & -0.36 & -0.08 & -0.08 \\
\bottomrule
\end{tabular}

}
\caption{\textbf{Correlation between the detected contacts and the fraction of missing users.}
The correlation is reported for each range of missing hours and sparsification approach.}
\label{tab:mm_correlation_contacts_missingusers}
\end{table}

\clearpage
\begin{landscape}
\setlength{\intextsep}{4pt}
\captionsetup{skip=2pt}
\begin{table}[H]
\centering
{\scriptsize\setlength{\tabcolsep}{2pt}
\begin{tabular}{lccccc}
\toprule
 & missing hours (\%) & \textbf{detected contacts}
 (mean, mean std) & gt. rel. change (\%)  
 (mean, mean std) & quartiles & 95\% CI \\
outcome &  &  &  &  &  \\
\midrule
\midrule
Data-driven & 10-20 & 54729 $\pm$ 59 & -16.78 $\pm$ 0.09 & (54481, 54710, 55015) & (53875, 55597) \\
Data-driven & 20-30 & 45107 $\pm$ 79 & -31.41 $\pm$ 0.12 & (44745, 45164, 45474) & (43865, 46096) \\
Data-driven & 30-40 & 39861 $\pm$ 104 & -39.39 $\pm$ 0.16 & (39392, 39910, 40351) & (38417, 41009) \\
Data-driven & 40-50 & 30886 $\pm$ 76 & -53.04 $\pm$ 0.12 & (30499, 30933, 31220) & (29914, 31895) \\
Data-driven & 50-60 & 23309 $\pm$ 90 & -64.56 $\pm$ 0.14 & (22820, 23233, 23784) & (22213, 24648) \\
\midrule
Random shuffling & 10-20 & 53057 $\pm$ 61 & -19.32 $\pm$ 0.09 & (52771, 52960, 53417) & (52343, 53808) \\
Random shuffling & 20-30 & 43102 $\pm$ 69 & -34.46 $\pm$ 0.11 & (42806, 43112, 43455) & (42056, 43986) \\
Random shuffling & 30-40 & 35019 $\pm$ 92 & -46.75 $\pm$ 0.14 & (34620, 34923, 35480) & (33833, 36211) \\
Random shuffling & 40-50 & 27364 $\pm$ 73 & -58.39 $\pm$ 0.11 & (27142, 27370, 27652) & (26130, 28333) \\
Random shuffling & 50-60 & 20904 $\pm$ 65 & -68.22 $\pm$ 0.10 & (20606, 20857, 21236) & (20210, 21791) \\
\midrule
Random uniform & 10-20 & 60035 $\pm$ 28 & -8.71 $\pm$ 0.04 & (59892, 60032, 60172) & (59695, 60432) \\
Random uniform & 20-30 & 54869 $\pm$ 42 & -16.57 $\pm$ 0.06 & (54641, 54868, 55043) & (54303, 55404) \\
Random uniform & 30-40 & 48641 $\pm$ 54 & -26.04 $\pm$ 0.08 & (48384, 48678, 48854) & (47848, 49274) \\
Random uniform & 40-50 & 41336 $\pm$ 51 & -37.15 $\pm$ 0.08 & (41042, 41366, 41629) & (40677, 41863) \\
Random uniform & 50-60 & 32924 $\pm$ 56 & -49.94 $\pm$ 0.09 & (32636, 32886, 33148) & (32259, 33612) \\
\bottomrule
\end{tabular}
}
\caption{\textbf{Count of detected contacts.}
For each approach (data-driven, random-uniform, and random-shuffling) and range of missing hours, the point estimates, IQR, and 95\% CI are computed from 50 realizations of missingness.
The relative change is computed with respect to the ground-truth number of detected contacts, namely, $N = 65,766$.}
\label{tab:mm_detected_contacts}
\end{table}

\begin{table}[H]
\centering
{\scriptsize\setlength{\tabcolsep}{2pt}
\begin{tabular}{lccccc}
\toprule
 & missing hours (\%) & \textbf{avg. contact duration} 
 (mean, mean std) & gt. rel. change (\%) 
 (mean, mean std) & quartiles & 95\% CI \\
outcome &  &  &  &  &  \\
\midrule
\midrule
Data-driven & 10-20 & 108.27 $\pm$ 0.06 & -5.50 $\pm$ 0.46 & (108, 108, 109) & (108, 109) \\
Data-driven & 20-30 & 104.78 $\pm$ 0.11 & -8.55 $\pm$ 0.45 & (104, 105, 105) & (103, 106) \\
Data-driven & 30-40 & 97.55 $\pm$ 0.13 & -14.86 $\pm$ 0.42 & (97, 98, 98) & (96, 99) \\
Data-driven & 40-50 & 92.79 $\pm$ 0.09 & -19.02 $\pm$ 0.40 & (92, 93, 93) & (91, 94) \\
Data-driven & 50-60 & 87.71 $\pm$ 0.19 & -23.45 $\pm$ 0.40 & (87, 88, 89) & (85, 90) \\
\midrule
Random shuffling & 10-20 & 108.96 $\pm$ 0.08 & -4.90 $\pm$ 0.46 & (108, 109, 109) & (108, 110) \\
Random shuffling & 20-30 & 105.95 $\pm$ 0.13 & -7.53 $\pm$ 0.46 & (105, 106, 107) & (104, 108) \\
Random shuffling & 30-40 & 100.26 $\pm$ 0.13 & -12.49 $\pm$ 0.44 & (100, 100, 101) & (98, 102) \\
Random shuffling & 40-50 & 96.11 $\pm$ 0.12 & -16.12 $\pm$ 0.42 & (95, 96, 97) & (95, 98) \\
Random shuffling & 50-60 & 91.33 $\pm$ 0.17 & -20.29 $\pm$ 0.41 & (91, 91, 92) & (89, 94) \\
\midrule
Random uniform & 10-20 & 110.42 $\pm$ 0.06 & -3.63 $\pm$ 0.47 & (110, 110, 111) & (110, 111) \\
Random uniform & 20-30 & 107.79 $\pm$ 0.08 & -5.93 $\pm$ 0.46 & (107, 108, 108) & (107, 109) \\
Random uniform & 30-40 & 104.98 $\pm$ 0.10 & -8.37 $\pm$ 0.45 & (105, 105, 106) & (104, 106) \\
Random uniform & 40-50 & 101.56 $\pm$ 0.09 & -11.36 $\pm$ 0.43 & (101, 101, 102) & (101, 103) \\
Random uniform & 50-60 & 97.75 $\pm$ 0.11 & -14.69 $\pm$ 0.42 & (97, 98, 98) & (96, 99) \\
\bottomrule
\end{tabular}
}
\caption{\textbf{Average duration of daily detected contacts (minute).}
For each approach (data-driven, random-uniform, and random-shuffling) and range of missing hours, the point estimates, IQR, and 95\% CI are computed from 50 realizations of missingness.
The relative change is computed with respect to the ground-truth average duration, namely, 114 minutes.}
\label{tab:mm_contact_avg_duration}
\end{table}

\newpage
\begin{table}[H]
\centering
{\footnotesize\setlength{\tabcolsep}{2pt}\renewcommand{\arraystretch}{1.15}
\begin{tabular}{lccccc}
\toprule
 & missing hours (\%) & \textbf{avg. contact duration} 
 (mean, mean std) & gt. rel. change (\%) 
 (mean, mean std) & quartiles & 95\% CI \\
outcome &  &  &  &  &  \\
\midrule
\midrule
Data-driven & 10-20 & 1.676 $\pm$ 0.001 & -12.60 $\pm$ 0.05 & (1.67, 1.67, 1.68) & (1.66, 1.69) \\
Data-driven & 20-30 & 1.487 $\pm$ 0.001 & -22.42 $\pm$ 0.07 & (1.48, 1.49, 1.49) & (1.47, 1.50) \\
Data-driven & 30-40 & 1.362 $\pm$ 0.002 & -28.97 $\pm$ 0.09 & (1.35, 1.36, 1.37) & (1.34, 1.38) \\
Data-driven & 40-50 & 1.196 $\pm$ 0.001 & -37.61 $\pm$ 0.06 & (1.19, 1.20, 1.20) & (1.18, 1.21) \\
Data-driven & 50-60 & 1.064 $\pm$ 0.001 & -44.49 $\pm$ 0.07 & (1.06, 1.06, 1.07) & (1.05, 1.08) \\
\midrule
Random shuffling & 10-20 & 1.651 $\pm$ 0.001 & -13.91 $\pm$ 0.06 & (1.65, 1.65, 1.66) & (1.64, 1.66) \\
Random shuffling & 20-30 & 1.459 $\pm$ 0.001 & -23.91 $\pm$ 0.05 & (1.45, 1.46, 1.46) & (1.44, 1.47) \\
Random shuffling & 30-40 & 1.295 $\pm$ 0.001 & -32.48 $\pm$ 0.07 & (1.29, 1.29, 1.30) & (1.28, 1.31) \\
Random shuffling & 40-50 & 1.154 $\pm$ 0.001 & -39.79 $\pm$ 0.06 & (1.15, 1.15, 1.16) & (1.14, 1.17) \\
Random shuffling & 50-60 & 1.039 $\pm$ 0.001 & -45.81 $\pm$ 0.05 & (1.03, 1.04, 1.04) & (1.03, 1.05) \\
\midrule
Random uniform & 10-20 & 1.781 $\pm$ 0.001 & -7.11 $\pm$ 0.03 & (1.78, 1.78, 1.78) & (1.77, 1.79) \\
Random uniform & 20-30 & 1.669 $\pm$ 0.001 & -12.92 $\pm$ 0.04 & (1.67, 1.67, 1.67) & (1.66, 1.68) \\
Random uniform & 30-40 & 1.541 $\pm$ 0.001 & -19.61 $\pm$ 0.05 & (1.54, 1.54, 1.55) & (1.53, 1.55) \\
Random uniform & 40-50 & 1.397 $\pm$ 0.001 & -27.15 $\pm$ 0.04 & (1.39, 1.40, 1.40) & (1.38, 1.41) \\
Random uniform & 50-60 & 1.242 $\pm$ 0.001 & -35.22 $\pm$ 0.05 & (1.24, 1.24, 1.25) & (1.23, 1.25) \\
\bottomrule
\end{tabular}
}
\caption{\textbf{Average $\overline{R}_0$ over the study period.}
For each approach (data-driven, random-uniform, and random-shuffling) and range of missing hours, the point estimates, IQR, and 95\% CI are computed from 50 realizations of missingness.
The relative change is computed with respect to the ground-truth average $\overline{R}_0$, namely, 1.917.}
\label{tab:mm_avg_R0}
\end{table}
\end{landscape}
\clearpage

\newpage

\subsection{Changes in size and dynamics of the epidemic}\label{supp:mm_emobias}

We compute the variation of epidemic descriptive statistics---described in Section \ref{sec:bias-on-emos} of the main article---for different sparsity ranges (from 10--20\% to 50--60\% of missing hours).
For each sparsity range and sparsity approach (data-driven and random baselines---described in Section \ref{main:random_sparsity} in the main article) we gather an ensemble of 5,000 epidemic curves consisting of 100 epidemic simulations for each of the 50 missingness realizations.
We simulate an ensemble of 5,000 epidemic curves in the ground-truth scenario to match the sample size and ensure comparability.
We report the ground-truth value and the corresponding relative change and absolute change with respect to the ground truth for each sparse biased scenario.

Realized outbreaks are defined as the epidemic curves in the ensemble which result in more than 5\% of total infected.
The fraction of realized outbreaks $p_{\texttt{outbreak}}$ is computed for each missingness realization and its statistics are reported in Supplementary Table \ref{tab:mm_realized_outbreaks}.

The statistics for the infected peak $I_{\texttt{peak}}$ (Supplementary Table \ref{tab:mm_size_peak}) and the total infected $I_{\texttt{tot}}$ (Supplementary Table \ref{tab:mm_size_total}) are computed from the ensemble of realized outbreaks.

Given the realized outbreaks, we compute, for each day, the fraction of simulations in which the epidemic reaches its peak on that day $t_{\texttt{peak}}$.
This provides the distribution of peak-timing (Supplementary Table \ref{tab:mm_dynamic_peak_day}).
We then evaluate the occurrence of the day of last infection $t_{\texttt{tot}}$ which corresponds to the day in which the cumulative infected curve reaches a plateau (see bottom panel of Fig.~\ref{Fig:1}e).
This provides the distribution of last infection timing (Supplementary Table \ref{tab:mm_dynamic_lastcase_day}).

\begin{table}[H]
\centering
{\small\setlength{\tabcolsep}{4pt}
\begin{tabular}{lccc}
\toprule
outcome & missing hours (\%) & count \\
\midrule
\textbf{ground truth} & 0-5 & 4916 \\
\midrule
Data driven & 10-20 & 4832 \\
Data driven & 20-30 & 4637 \\
Data driven & 30-40 & 4507 \\
Data driven & 40-50 & 4066 \\
Data driven & 50-60 & 3253 \\
\midrule
Random shuffling & 10-20 & 4803 \\
Random shuffling & 20-30 & 4597 \\
Random shuffling & 30-40 & 4319 \\
Random shuffling & 40-50 & 3682 \\
Random shuffling & 50-60 & 2672 \\
\midrule
Random uniform & 10-20 & 4887 \\
Random uniform & 20-30 & 4862 \\
Random uniform & 30-40 & 4756 \\
Random uniform & 40-50 & 4546 \\
Random uniform & 50-60 & 4093 \\
\bottomrule
\end{tabular}

}
\caption{\textbf{Realized outbreaks sample size.}
Number of realized outbreaks for each sparsity approach and sparsity range.
The initial size of the ensemble for each range is 5,000.}
\label{tab:mm_realized_outbreaks_sample_size}
\end{table}

\begin{table}[H]
\centering
{\scriptsize\setlength{\tabcolsep}{0.5pt}
\begin{tabular}{lcccccc}
\toprule
 & missing hours (\%) & \makecell{Realized outbreaks frac. \\ (mean, mean std)} & \makecell{gt. abs. change \\ (mean, mean std)} & \makecell{gt. rel. change \\ (mean, mean std)} & quartiles & 95\% CI \\
outcome &  &  &  &  &  &  \\
\midrule
\textbf{ground truth} (gt.) & 0-5 & 98.32 $\pm$ 0.15 & - & - & (98.00, 98.00, 98.00) & (98.00, 98.00) \\
\midrule
Data-driven & 10-20 & 96.64 $\pm$ 0.24 & -1.68 $\pm$ 0.24 & -1.71 $\pm$ 0.24 & (95.00, 97.00, 98.00) & (94.00, 100.00) \\
Data-driven & 20-30 & 92.74 $\pm$ 0.40 & -5.58 $\pm$ 0.40 & -5.68 $\pm$ 0.41 & (91.00, 93.00, 95.00) & (87.00, 98.00) \\
Data-driven & 30-40 & 90.14 $\pm$ 0.39 & -8.18 $\pm$ 0.39 & -8.32 $\pm$ 0.40 & (89.00, 90.00, 92.00) & (84.00, 94.00) \\
Data-driven & 40-50 & 81.32 $\pm$ 0.56 & -17.00 $\pm$ 0.56 & -17.29 $\pm$ 0.57 & (79.00, 82.00, 84.00) & (72.00, 89.00) \\
Data-driven & 50-60 & 65.06 $\pm$ 0.77 & -33.26 $\pm$ 0.77 & -33.83 $\pm$ 0.79 & (61.00, 65.00, 69.00) & (55.00, 74.00) \\
\midrule
Random shuffling & 10-20 & 96.06 $\pm$ 0.28 & -2.26 $\pm$ 0.28 & -2.30 $\pm$ 0.29 & (95.00, 96.00, 98.00) & (93.00, 100.00) \\
Random shuffling & 20-30 & 91.94 $\pm$ 0.36 & -6.38 $\pm$ 0.36 & -6.49 $\pm$ 0.36 & (90.00, 91.00, 94.00) & (87.00, 96.00) \\
Random shuffling & 30-40 & 86.38 $\pm$ 0.40 & -11.94 $\pm$ 0.40 & -12.14 $\pm$ 0.41 & (84.00, 86.00, 89.00) & (82.00, 92.00) \\
Random shuffling & 40-50 & 73.64 $\pm$ 0.74 & -24.68 $\pm$ 0.74 & -25.10 $\pm$ 0.75 & (69.00, 74.00, 77.00) & (64.00, 83.00) \\
Random shuffling & 50-60 & 53.44 $\pm$ 0.87 & -44.88 $\pm$ 0.87 & -45.65 $\pm$ 0.88 & (50.00, 53.00, 57.00) & (41.00, 65.00) \\
\midrule
Random uniform & 10-20 & 97.74 $\pm$ 0.23 & -0.58 $\pm$ 0.23 & -0.59 $\pm$ 0.23 & (96.00, 98.00, 99.00) & (95.00, 100.00) \\
Random uniform & 20-30 & 97.24 $\pm$ 0.27 & -1.08 $\pm$ 0.27 & -1.10 $\pm$ 0.28 & (96.00, 97.00, 99.00) & (93.00, 100.00) \\
Random uniform & 30-40 & 95.12 $\pm$ 0.30 & -3.20 $\pm$ 0.30 & -3.25 $\pm$ 0.30 & (94.00, 95.00, 97.00) & (91.00, 99.00) \\
Random uniform & 40-50 & 90.92 $\pm$ 0.38 & -7.40 $\pm$ 0.38 & -7.53 $\pm$ 0.38 & (89.00, 91.00, 93.00) & (86.00, 95.00) \\
Random uniform & 50-60 & 81.86 $\pm$ 0.57 & -16.46 $\pm$ 0.57 & -16.74 $\pm$ 0.58 & (79.00, 81.00, 85.00) & (76.00, 89.00) \\
\bottomrule
\end{tabular}
}
\caption{\textbf{Fraction of realized outbreaks $p_{\texttt{outbreak}}$.}
For each approach (data-driven, random-uniform, and random-shuffling) and range of missing hours, the point estimates, IQR, and 95\% CI are computed from a sample of 50 measurements of $p_{\texttt{outbreak}}$.
Each measurement is computed from the ensemble of 100 epidemic curves obtained for each of the 50 missingness realizations.}
\label{tab:mm_realized_outbreaks}
\end{table}

\begin{table}[H]
\centering
{\scriptsize\setlength{\tabcolsep}{2pt}
\begin{tabular}{lcccccc}
\toprule
 & missing hours (\%) & \makecell{I\_peak frac. \\ (mean, mean std)} & \makecell{gt. abs. change \\ (mean, mean std)} & \makecell{gt. rel. change \\ (mean, mean std)} & quartiles & 95\% CI \\
outcome &  &  &  &  &  &  \\
\midrule
\textbf{ground truth} (gt.) & 0-5 & 33.86 $\pm$ 0.06 & - & - & (31.00, 33.00, 37.00) & (27.00, 42.00) \\
\midrule
Data-driven & 10-20 & 29.44 $\pm$ 0.05 & -4.42 $\pm$ 0.08 & -13.05 $\pm$ 0.21 & (27.00, 29.00, 32.00) & (23.00, 36.00) \\
Data-driven & 20-30 & 25.43 $\pm$ 0.05 & -8.42 $\pm$ 0.08 & -24.88 $\pm$ 0.20 & (23.00, 25.00, 28.00) & (19.00, 32.00) \\
Data-driven & 30-40 & 21.02 $\pm$ 0.05 & -12.84 $\pm$ 0.08 & -37.93 $\pm$ 0.19 & (19.00, 21.00, 23.00) & (14.00, 28.00) \\
Data-driven & 40-50 & 14.01 $\pm$ 0.07 & -19.84 $\pm$ 0.09 & -58.61 $\pm$ 0.21 & (12.00, 14.00, 17.00) & (4.00, 22.00) \\
Data-driven & 50-60 & 6.87 $\pm$ 0.05 & -26.99 $\pm$ 0.08 & -79.72 $\pm$ 0.16 & (4.00, 7.00, 9.00) & (2.00, 13.00) \\
\midrule
Random shuffling & 10-20 & 28.23 $\pm$ 0.05 & -5.63 $\pm$ 0.08 & -16.62 $\pm$ 0.20 & (26.00, 28.00, 31.00) & (22.00, 35.00) \\
Random shuffling & 20-30 & 23.31 $\pm$ 0.05 & -10.55 $\pm$ 0.08 & -31.15 $\pm$ 0.19 & (21.00, 23.00, 25.00) & (17.00, 30.00) \\
Random shuffling & 30-40 & 17.35 $\pm$ 0.06 & -16.51 $\pm$ 0.08 & -48.76 $\pm$ 0.19 & (15.00, 17.00, 20.00) & (9.00, 25.00) \\
Random shuffling & 40-50 & 10.89 $\pm$ 0.07 & -22.96 $\pm$ 0.09 & -67.83 $\pm$ 0.21 & (8.00, 11.00, 14.00) & (2.00, 18.00) \\
Random shuffling & 50-60 & 5.49 $\pm$ 0.05 & -28.37 $\pm$ 0.08 & -83.80 $\pm$ 0.15 & (3.00, 5.00, 7.00) & (2.00, 11.00) \\
\midrule
Random uniform & 10-20 & 30.79 $\pm$ 0.05 & -3.07 $\pm$ 0.08 & -9.06 $\pm$ 0.21 & (28.00, 31.00, 33.00) & (24.00, 38.00) \\
Random uniform & 20-30 & 28.10 $\pm$ 0.05 & -5.75 $\pm$ 0.07 & -17.00 $\pm$ 0.20 & (26.00, 28.00, 30.00) & (22.00, 35.00) \\
Random uniform & 30-40 & 24.46 $\pm$ 0.05 & -9.39 $\pm$ 0.08 & -27.74 $\pm$ 0.19 & (22.00, 24.00, 27.00) & (18.00, 31.00) \\
Random uniform & 40-50 & 19.63 $\pm$ 0.05 & -14.22 $\pm$ 0.08 & -42.01 $\pm$ 0.19 & (17.00, 20.00, 22.00) & (13.00, 27.00) \\
Random uniform & 50-60 & 12.77 $\pm$ 0.06 & -21.08 $\pm$ 0.08 & -62.27 $\pm$ 0.19 & (10.00, 13.00, 15.00) & (4.00, 20.00) \\
\bottomrule
\end{tabular}
}
\caption{\textbf{Variation of peak of infected $I_{\texttt{peak}}$.}
For each approach (data-driven, random-uniform, and random-shuffling) and range of missing hours, the point estimates, IQR, and 95\% CI are computed from the sample of realized outbreaks.}
\label{tab:mm_size_peak}
\end{table}

\begin{table}[H]
\centering
{\scriptsize\setlength{\tabcolsep}{2pt}
\begin{tabular}{lcccccc}
\toprule
 & missing hours (\%) & \makecell{I\_tot frac. \\ (mean, mean std)} & \makecell{gt. abs. change \\ (mean, mean std)} & \makecell{gt. rel. change \\ (mean, mean std)} & quartiles & 95\% CI \\
outcome &  &  &  &  &  &  \\
\midrule
\textbf{ground truth} (gt.) & 0-5 & 85.29 $\pm$ 0.03 & - & - & (84.00, 85.00, 87.00) & (81.00, 89.00) \\
\midrule
Data-driven & 10-20 & 81.13 $\pm$ 0.04 & -4.16 $\pm$ 0.05 & -4.88 $\pm$ 0.05 & (80.00, 81.00, 83.00) & (76.00, 85.00) \\
Data-driven & 20-30 & 75.68 $\pm$ 0.06 & -9.61 $\pm$ 0.07 & -11.27 $\pm$ 0.08 & (74.00, 76.00, 78.00) & (67.00, 82.00) \\
Data-driven & 30-40 & 67.97 $\pm$ 0.11 & -17.32 $\pm$ 0.11 & -20.31 $\pm$ 0.13 & (66.00, 69.00, 72.00) & (50.00, 77.00) \\
Data-driven & 40-50 & 48.16 $\pm$ 0.20 & -37.13 $\pm$ 0.20 & -43.54 $\pm$ 0.23 & (42.00, 51.00, 57.00) & (13.00, 64.00) \\
Data-driven & 50-60 & 24.92 $\pm$ 0.19 & -60.37 $\pm$ 0.20 & -70.78 $\pm$ 0.23 & (16.00, 25.00, 33.00) & (6.00, 45.00) \\
\midrule
Random shuffling & 10-20 & 80.09 $\pm$ 0.04 & -5.20 $\pm$ 0.05 & -6.10 $\pm$ 0.06 & (79.00, 80.00, 82.00) & (75.00, 85.00) \\
Random shuffling & 20-30 & 73.13 $\pm$ 0.07 & -12.16 $\pm$ 0.08 & -14.26 $\pm$ 0.09 & (71.00, 74.00, 76.00) & (64.00, 79.00) \\
Random shuffling & 30-40 & 59.56 $\pm$ 0.15 & -25.73 $\pm$ 0.16 & -30.17 $\pm$ 0.18 & (56.00, 62.00, 66.00) & (29.00, 71.00) \\
Random shuffling & 40-50 & 39.14 $\pm$ 0.22 & -46.15 $\pm$ 0.22 & -54.11 $\pm$ 0.25 & (32.00, 42.00, 49.00) & (8.00, 58.00) \\
Random shuffling & 50-60 & 19.56 $\pm$ 0.18 & -65.73 $\pm$ 0.19 & -77.06 $\pm$ 0.22 & (11.00, 19.00, 27.00) & (6.00, 39.00) \\
\midrule
Random uniform & 10-20 & 82.59 $\pm$ 0.03 & -2.70 $\pm$ 0.04 & -3.16 $\pm$ 0.05 & (81.00, 83.00, 84.00) & (78.00, 87.00) \\
Random uniform & 20-30 & 79.66 $\pm$ 0.04 & -5.63 $\pm$ 0.05 & -6.61 $\pm$ 0.06 & (78.00, 80.00, 82.00) & (74.00, 84.00) \\
Random uniform & 30-40 & 74.80 $\pm$ 0.06 & -10.49 $\pm$ 0.07 & -12.30 $\pm$ 0.08 & (73.00, 75.00, 77.00) & (67.00, 81.00) \\
Random uniform & 40-50 & 65.47 $\pm$ 0.11 & -19.82 $\pm$ 0.11 & -23.23 $\pm$ 0.13 & (63.00, 67.00, 70.00) & (49.00, 74.00) \\
Random uniform & 50-60 & 46.46 $\pm$ 0.19 & -38.83 $\pm$ 0.19 & -45.53 $\pm$ 0.22 & (42.00, 49.00, 55.00) & (13.00, 62.00) \\
\bottomrule
\end{tabular}
}
\caption{\textbf{Variation of total of infected $I_{\texttt{tot}}$.}
For each approach (data-driven, random-uniform, and random-shuffling) and range of missing hours, the point estimates, IQR, and 95\% CI are computed from the sample of realized outbreaks.}
\label{tab:mm_size_total}
\end{table}

\clearpage
\begin{landscape}
\setlength{\intextsep}{4pt}
\captionsetup{skip=2pt}
\begin{table}[H]
\centering
{\scriptsize\setlength{\tabcolsep}{2pt}
\begin{tabular}{lccccccccccccccccccccccccccccc}
\toprule
 & missing hours (\%) & 1 & 2 & 3 & 4 & 5 & 6 & 7 & 8 & 9 & 10 & 11 & 12 & 13 & 14 & 15 & 16 & 17 & 18 & 19 & 20 & 21 & 22 & 23 & 24 & 25 & 26 & 27 & 28 \\
outcome &  &  &  &  &  &  &  &  &  &  &  &  &  &  &  &  &  &  &  &  &  &  &  &  &  &  &  &  &  \\
\midrule
\textbf{ground truth} (gt.) & 0-5 &  &  &  &  &  &  & 44.5 & 0.1 &  & 5.3 & 6.9 & 24.5 & 16.2 & 1.9 &  &  &  &  & 0.2 & 0.2 & 0.2 &  &  &  &  &  &  &  \\
\midrule
Data-driven & 10-20 &  &  &  &  &  &  & 18.1 & 0.1 &  & 1.6 & 3.7 & 22.7 & 43.3 & 6.3 &  &  &  & 0.2 & 0.5 & 1.3 & 2.1 &  &  &  &  &  &  &  \\
Data-driven & 20-30 &  &  &  &  &  &  & 5.6 &  &  & 0.7 & 1.3 & 10.3 & 53.2 & 14.0 &  &  &  & 0.2 & 0.6 & 3.1 & 10.4 &  &  &  &  &  & 0.2 & 0.3 \\
Data-driven & 30-40 &  &  &  &  &  &  & 3.5 &  &  & 0.2 & 0.6 & 3.8 & 43.9 & 13.9 &  &  & 0.1 & 0.3 & 0.3 & 4.8 & 27.1 &  &  &  & 0.1 & 0.2 & 0.4 & 0.8 \\
Data-driven & 40-50 &  &  & 0.1 &  &  & 0.2 & 3.1 & 0.4 &  & 0.3 & 0.5 & 2.7 & 22.1 & 6.7 &  &  &  & 0.2 & 0.2 & 1.9 & 53.4 & 0.2 &  & 0.1 & 0.7 & 1.1 & 2.8 & 3.0 \\
Data-driven & 50-60 &  &  &  & 0.3 & 1.0 & 2.5 & 11.8 & 1.8 & 0.1 & 1.6 & 1.9 & 4.6 & 16.7 & 8.2 & 0.3 & 0.1 & 0.3 & 0.3 & 0.5 & 2.8 & 30.5 & 0.7 & 0.1 & 0.5 & 0.9 & 1.9 & 4.5 & 6.1 \\
\midrule
Random shuffling & 10-20 &  &  &  &  &  &  & 14.6 & 0.1 &  & 2.3 & 3.2 & 20.0 & 45.2 & 8.5 &  &  &  & 0.3 & 0.4 & 2.1 & 3.4 &  &  &  &  &  &  &  \\
Random shuffling & 20-30 &  &  &  &  &  &  & 4.7 & 0.1 &  & 1.2 & 1.5 & 9.7 & 48.9 & 13.3 &  &  &  & 0.2 & 0.5 & 3.3 & 15.6 &  &  &  &  & 0.2 & 0.2 & 0.5 \\
Random shuffling & 30-40 &  &  &  &  &  & 0.1 & 2.2 &  &  & 0.2 & 0.4 & 3.3 & 29.2 & 11.8 & 0.1 &  & 0.1 & 0.3 & 0.3 & 4.5 & 41.1 &  &  & 0.1 & 0.3 & 0.6 & 2.3 & 3.0 \\
Random shuffling & 40-50 &  &  &  & 0.1 & 0.5 & 0.7 & 3.3 & 0.8 &  & 0.6 & 0.6 & 2.7 & 15.1 & 7.1 &  &  & 0.1 & 0.1 & 0.4 & 3.1 & 50.4 & 0.5 &  & 0.3 & 1.0 & 1.5 & 4.3 & 6.7 \\
Random shuffling & 50-60 &  &  & 0.3 & 1.0 & 2.6 & 3.3 & 11.5 & 2.2 & 0.4 & 2.4 & 2.5 & 4.5 & 11.7 & 6.1 & 0.4 & 0.1 & 0.6 & 0.9 & 0.9 & 4.0 & 28.6 & 0.6 & 0.1 & 0.9 & 1.9 & 2.1 & 4.6 & 5.9 \\
\midrule
Random uniform & 10-20 &  &  &  &  &  &  & 27.2 & 0.1 &  & 3.7 & 5.6 & 26.3 & 30.9 & 3.9 &  &  &  & 0.1 & 0.3 & 0.8 & 0.9 &  &  &  &  &  & 0.1 &  \\
Random uniform & 20-30 &  &  &  &  &  &  & 14.5 & 0.2 &  & 3.0 & 4.2 & 23.8 & 43.8 & 5.7 &  &  &  & 0.2 & 0.4 & 1.4 & 2.6 &  &  &  &  &  & 0.1 &  \\
Random uniform & 30-40 &  &  &  &  &  &  & 5.9 & 0.1 &  & 1.3 & 2.5 & 14.9 & 51.6 & 10.2 &  &  & 0.1 & 0.3 & 0.4 & 2.8 & 9.1 &  &  &  & 0.1 & 0.1 & 0.3 & 0.3 \\
Random uniform & 40-50 &  &  &  &  &  &  & 1.8 & 0.2 &  & 0.6 & 1.1 & 7.3 & 40.9 & 10.2 &  &  & 0.1 & 0.3 & 0.6 & 4.4 & 29.3 &  &  &  & 0.3 & 0.5 & 1.0 & 1.2 \\
Random uniform & 50-60 &  &  &  & 0.1 & 0.2 & 0.3 & 2.2 & 0.4 &  & 1.0 & 0.9 & 3.8 & 19.6 & 5.8 & 0.1 &  & 0.1 & 0.3 & 0.5 & 5.0 & 47.1 & 0.4 &  & 0.3 & 1.2 & 1.9 & 4.2 & 4.4 \\
\bottomrule
\end{tabular}
}
\caption{\textbf{Peak timing.}
Occurrence of the peak day over the ensemble of realized outbreaks and over all 50 missingness realizations.
Columns refer to the day of the simulation.
Reported values are rounded to one decimal place.}
\label{tab:mm_dynamic_peak_day}
\end{table}

\begin{table}[H]
\centering
{\scriptsize\setlength{\tabcolsep}{2pt}
\begin{tabular}{lccccccccccccccccccccccccccccc}
\toprule
 & missing hours (\%) & 1 & 2 & 3 & 4 & 5 & 6 & 7 & 8 & 9 & 10 & 11 & 12 & 13 & 14 & 15 & 16 & 17 & 18 & 19 & 20 & 21 & 22 & 23 & 24 & 25 & 26 & 27 & 28 \\
outcome &  &  &  &  &  &  &  &  &  &  &  &  &  &  &  &  &  &  &  &  &  &  &  &  &  &  &  &  &  \\
\midrule
\textbf{ground truth} (gt.) & 0-5 &  &  &  &  &  &  &  &  &  &  &  &  &  &  &  &  &  & 0.2 & 0.6 & 2.2 & 13.7 & 3.2 & 1.8 & 10.8 & 15.7 & 15.8 & 18.5 & 17.5 \\
\midrule
Data-driven & 10-20 &  &  &  &  &  &  &  &  &  &  &  &  &  &  &  &  &  &  & 0.1 & 0.4 & 5.4 & 1.7 & 0.9 & 6.2 & 12.1 & 16.1 & 24.1 & 32.9 \\
Data-driven & 20-30 &  &  &  &  &  &  &  &  &  &  &  &  &  &  &  &  &  &  &  &  & 1.9 & 0.6 & 0.2 & 2.9 & 6.5 & 11.6 & 26.0 & 50.2 \\
Data-driven & 30-40 &  &  &  &  &  &  &  &  &  &  &  &  &  &  &  &  &  &  &  &  & 0.1 & 0.1 & 0.1 & 1.0 & 2.6 & 5.6 & 17.8 & 72.6 \\
Data-driven & 40-50 &  &  &  &  &  &  &  &  &  &  &  & 0.1 & 0.1 & 0.3 &  &  &  & 0.1 & 0.1 & 0.2 & 0.2 & 0.1 &  & 0.4 & 0.7 & 1.8 & 9.3 & 86.4 \\
Data-driven & 50-60 &  &  &  &  &  &  & 0.1 & 0.1 &  & 0.3 & 0.3 & 0.3 & 1.0 & 1.5 & 0.3 & 0.3 & 0.8 & 0.8 & 0.7 & 1.0 & 2.9 & 0.6 & 0.3 & 1.4 & 1.8 & 2.6 & 10.5 & 72.3 \\
\midrule
Random shuffling & 10-20 &  &  &  &  &  &  &  &  &  &  &  &  &  &  &  &  &  &  & 0.1 & 0.3 & 3.1 & 1.4 & 0.5 & 4.7 & 10.3 & 14.2 & 25.9 & 39.6 \\
Random shuffling & 20-30 &  &  &  &  &  &  &  &  &  &  &  &  &  &  &  &  &  &  &  &  & 0.4 & 0.3 & 0.2 & 1.5 & 3.3 & 7.6 & 21.5 & 65.2 \\
Random shuffling & 30-40 &  &  &  &  &  &  &  &  &  &  &  &  &  & 0.1 &  &  &  &  &  &  &  &  &  & 0.2 & 0.7 & 2.4 & 8.9 & 87.4 \\
Random shuffling & 40-50 &  &  &  &  &  &  &  &  & 0.1 & 0.1 & 0.1 & 0.2 & 0.3 & 0.5 & 0.1 & 0.1 & 0.3 & 0.2 & 0.1 & 0.2 & 0.6 & 0.2 & 0.1 & 0.2 & 0.6 & 1.0 & 5.4 & 89.6 \\
Random shuffling & 50-60 &  &  &  &  &  &  & 0.1 & 0.1 & 0.1 & 0.3 & 0.2 & 0.8 & 1.5 & 2.4 & 0.9 & 0.4 & 1.0 & 1.2 & 0.9 & 1.7 & 3.3 & 1.1 & 0.7 & 1.8 & 2.2 & 3.8 & 10.0 & 65.3 \\
\midrule
Random uniform & 10-20 &  &  &  &  &  &  &  &  &  &  &  &  &  &  &  &  &  & 0.1 & 0.1 & 0.7 & 7.7 & 2.1 & 1.2 & 8.2 & 13.5 & 16.4 & 22.4 & 27.6 \\
Random uniform & 20-30 &  &  &  &  &  &  &  &  &  &  &  &  &  &  &  &  &  &  & 0.1 & 0.2 & 3.7 & 1.6 & 0.6 & 5.2 & 10.2 & 14.7 & 25.0 & 38.8 \\
Random uniform & 30-40 &  &  &  &  &  &  &  &  &  &  &  &  &  &  &  &  &  &  &  &  & 1.0 & 0.5 & 0.2 & 2.4 & 5.8 & 10.1 & 24.4 & 55.6 \\
Random uniform & 40-50 &  &  &  &  &  &  &  &  &  &  &  &  &  &  &  &  &  &  &  &  & 0.2 & 0.1 & 0.1 & 0.7 & 1.5 & 3.6 & 15.1 & 78.7 \\
Random uniform & 50-60 &  &  &  &  &  &  &  &  &  &  & 0.1 &  & 0.2 & 0.3 &  &  & 0.1 & 0.1 & 0.1 & 0.2 & 0.2 & 0.1 &  & 0.1 & 0.5 & 1.7 & 6.6 & 89.4 \\
\bottomrule
\end{tabular}
}
\caption{\textbf{Last infection timing.}
Occurrence of the last infection day over the ensemble of realized outbreaks for all 50 missingness realizations.
Columns refer to the day of the simulation.
Reported values are rounded to one decimal place.}
\label{tab:mm_dynamic_lastcase_day}
\end{table}
\end{landscape}
\clearpage

\newpage

\subsection{Debiasing impact on epidemic predictions and parameter estimation}\label{supp:debiasing}

We compare the ground-truth contacts to the sparse and rescaled contacts.
We compute the contact metrics and the average reproduction number $\overline{R}_0$ as in Supplementary Section \ref{supp:mm_stats_contacts_infectivity}.

\begin{figure}[H]
\centering
\includegraphics[width=.7\textwidth]{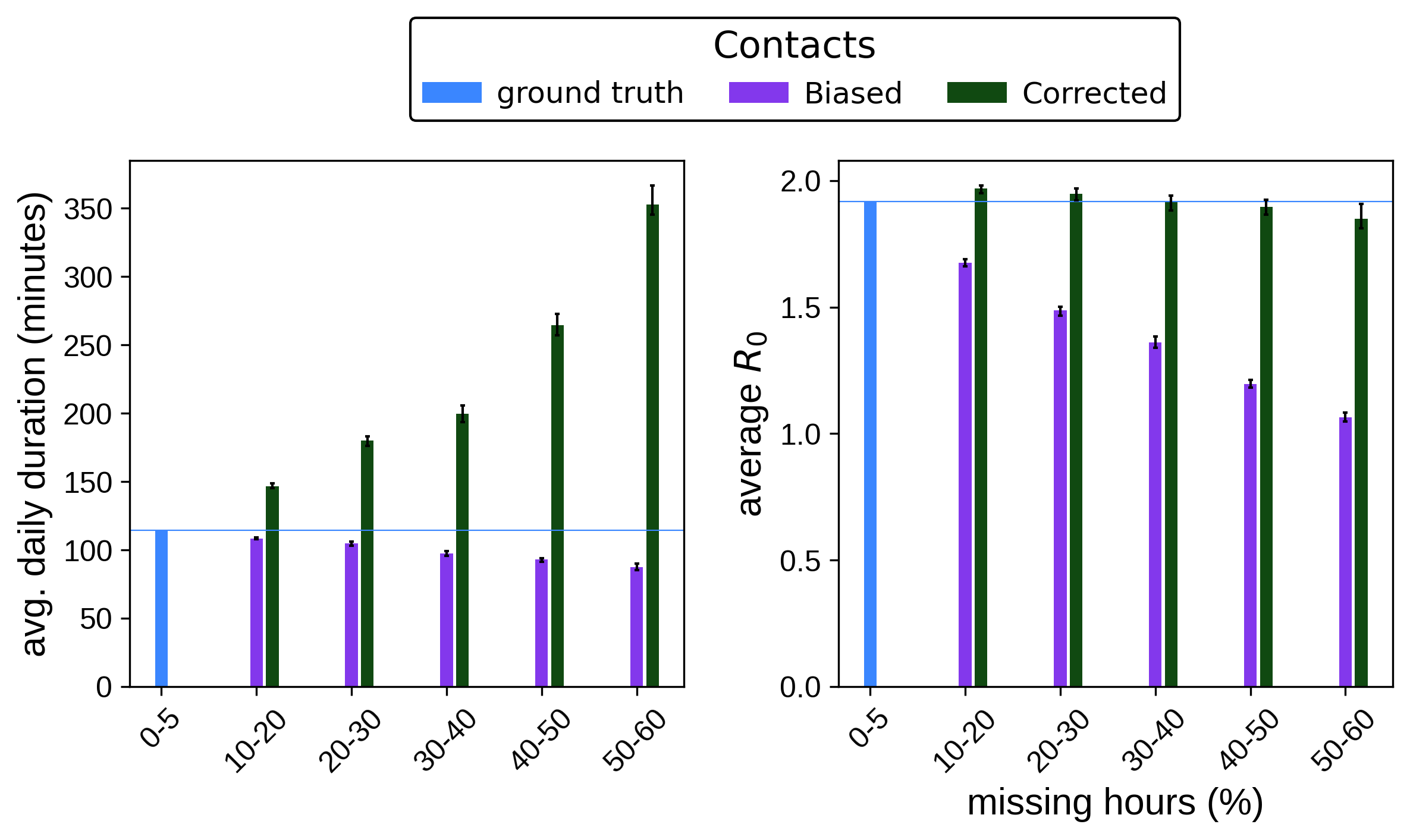}
\caption{\textbf{Variation in contact metrics and $\overline{R}_0$.}
Statistics derived from the contacts estimated over the study period.
Ground truth (blue) is compared to biased (purple) and rescaled contacts (green) under the data-driven sparsity scenario.
The bars represent the median value while the error bars denote the 95\% CI (reported in Supplementary Tables \ref{tab:da_contact_avg_duration} and \ref{tab:da_avg_R0}).
(left) The total number of detected daily contacts remains the same for both the biased and rescaled contacts as the rescaling changes only the duration of the detected contacts.
(center) Average duration of the daily detected contacts.
(right) Average $\overline{R}_0$ estimates from the contacts and the ground-truth epidemic parameters $(\beta^*, \gamma^*) = (0.84 \cdot 10^{-3}, 0.27)$.}
\label{fig:debiasing_contact_infectivity_reduction}
\end{figure}

Supplementary Fig.~\ref{fig:debiasing_contact_infectivity_reduction}a shows that, by design, contact rescaling preserves the number of biased detected contacts, while Supplementary Fig.~\ref{fig:debiasing_contact_infectivity_reduction}b shows that it increases the average contact duration, reaching up to three times the ground-truth value for 50--60\% of missing hours.
Supplementary Fig.~\ref{fig:debiasing_contact_infectivity_reduction}c shows that the estimated average reproduction number $\overline{R}_0$ recovers values close to the ground truth.

We compute the variation of epidemic descriptive statistics under the debiasing approaches (detailed in Supplementary Table \ref{tab:method_model_scenarios}) following the same procedure reported in Supplementary Section \ref{supp:mm_emobias}.
We report the biased and ground-truth values for the fraction of realized outbreaks, $p_{\texttt{outbreak}}$ (Supplementary Table \ref{tab:da_realized_outbreaks}), peak of infected individuals, $I_{\texttt{peak}}$ (Supplementary Table \ref{tab:da_size_peak}), total of infected, $I_{\texttt{tot}}$ (Supplementary Table \ref{tab:da_size_total}), timing of the peak, $t_{\texttt{peak}}$ (Supplementary Table \ref{tab:da_dynamic_peak_day}), and last case $t_{\texttt{tot}}$ (Supplementary Table \ref{tab:da_dynamic_lastcase_day}).
These results show how calibration, contact rescaling and the combination of the two mitigate bias in the predictions of the epidemic size and dynamics.

Finally, for both sparse and rescaled contacts, we report the statistics for the calibrated parameters $\beta$ and $\gamma$, in Supplementary Tables \ref{tab:da_param_beta} and \ref{tab:da_param_gamma} respectively.
We provide the associated average reproduction number $\overline{R}_0$ (Supplementary Table \ref{tab:da_param_R0}) and the scores of the calibration parameter search (Supplementary Table \ref{tab:da_param_score}).

\clearpage
\begin{landscape}
\noindent \textbf{Changes in contact metrics and average $\overline{R}_0$}
\setlength{\intextsep}{4pt}
\captionsetup{skip=2pt}
\begin{table}[H]
\centering
{\footnotesize\setlength{\tabcolsep}{2pt}
\begin{tabular}{lccccc}
\toprule
 & missing hours (\%) & \textbf{avg. contact duration} 
 (mean, mean std) & gt. rel. change (\%) 
 (mean, mean std) & quartiles & 95\% CI \\
outcome &  &  &  &  &  \\
\midrule
\midrule
biased & 10-20 & 108.27 $\pm$ 0.06 & -5.50 $\pm$ 0.46 & (108, 108, 109) & (108, 109) \\
biased & 20-30 & 104.78 $\pm$ 0.11 & -8.55 $\pm$ 0.45 & (104, 105, 105) & (103, 106) \\
biased & 30-40 & 97.55 $\pm$ 0.13 & -14.86 $\pm$ 0.42 & (97, 98, 98) & (96, 99) \\
biased & 40-50 & 92.79 $\pm$ 0.09 & -19.02 $\pm$ 0.40 & (92, 93, 93) & (91, 94) \\
biased & 50-60 & 87.71 $\pm$ 0.19 & -23.45 $\pm$ 0.40 & (87, 88, 89) & (85, 90) \\
\midrule
rescaled & 10-20 & 146.84 $\pm$ 0.15 & 28.16 $\pm$ 0.63 & (146, 147, 147) & (145, 149) \\
rescaled & 20-30 & 179.77 $\pm$ 0.25 & 56.90 $\pm$ 0.78 & (179, 180, 181) & (176, 183) \\
rescaled & 30-40 & 199.79 $\pm$ 0.48 & 74.37 $\pm$ 0.93 & (198, 200, 202) & (194, 206) \\
rescaled & 40-50 & 264.76 $\pm$ 0.57 & 131.08 $\pm$ 1.21 & (263, 265, 267) & (257, 273) \\
rescaled & 50-60 & 354.05 $\pm$ 0.94 & 209.01 $\pm$ 1.69 & (349, 353, 358) & (345, 367) \\
\midrule
\end{tabular}
}
\caption{\textbf{Average duration of daily detected contacts (minute).}
The point estimates, IQR, and 95\% CI are computed from 50 realizations of missingness.
The relative change is computed with respect to the ground-truth average duration, namely, 114 minutes.}
\label{tab:da_contact_avg_duration}
\end{table}

\begin{table}[H]
\centering
{\small\setlength{\tabcolsep}{3pt}
\begin{tabular}{lccccc}
\toprule
 & missing hours (\%) & \textbf{avg. R0} 
 (mean, mean std) & gt. rel. change (\%) 
 (mean, mean std) & quartiles & 95\% CI \\
outcome &  &  &  &  &  \\
\midrule
\midrule
biased & 10-20 & 1.676 $\pm$ 0.001 & -12.60 $\pm$ 0.05 & (1.67, 1.67, 1.68) & (1.66, 1.69) \\
biased & 20-30 & 1.487 $\pm$ 0.001 & -22.42 $\pm$ 0.07 & (1.48, 1.49, 1.49) & (1.47, 1.50) \\
biased & 30-40 & 1.362 $\pm$ 0.002 & -28.97 $\pm$ 0.09 & (1.35, 1.36, 1.37) & (1.34, 1.38) \\
biased & 40-50 & 1.196 $\pm$ 0.001 & -37.61 $\pm$ 0.06 & (1.19, 1.20, 1.20) & (1.18, 1.21) \\
biased & 50-60 & 1.064 $\pm$ 0.001 & -44.49 $\pm$ 0.07 & (1.06, 1.06, 1.07) & (1.05, 1.08) \\
\midrule
rescaled & 10-20 & 1.967 $\pm$ 0.001 & 2.57 $\pm$ 0.06 & (1.96, 1.97, 1.97) & (1.95, 1.98) \\
rescaled & 20-30 & 1.947 $\pm$ 0.002 & 1.57 $\pm$ 0.09 & (1.94, 1.95, 1.96) & (1.92, 1.97) \\
rescaled & 30-40 & 1.912 $\pm$ 0.002 & -0.28 $\pm$ 0.12 & (1.90, 1.91, 1.92) & (1.88, 1.94) \\
rescaled & 40-50 & 1.897 $\pm$ 0.002 & -1.03 $\pm$ 0.12 & (1.88, 1.90, 1.91) & (1.87, 1.93) \\
rescaled & 50-60 & 1.853 $\pm$ 0.004 & -3.35 $\pm$ 0.20 & (1.84, 1.85, 1.87) & (1.81, 1.91) \\
\midrule
\end{tabular}
}
\caption{\textbf{Average $\overline{R}_0$ over the study period.}
The point estimates, IQR, and 95\% CI are computed from 50 realizations of missingness.
The relative change is computed with respect to the ground-truth average $\overline{R}_0$, namely, 1.917.}
\label{tab:da_avg_R0}
\end{table}
\end{landscape}
\clearpage

\noindent \textbf{Changes in size and dynamics of the epidemic}

\begin{table}[H]
\centering
{\small\setlength{\tabcolsep}{4pt}
\begin{tabular}{lcc}
\toprule
 missing hours (\%) &  & count \\
\midrule
0-5 & ground truth & 4916 \\
\midrule
10-20 & Oracle on biased contacts & 4832 \\
20-30 & Oracle on biased contacts & 4637 \\
30-40 & Oracle on biased contacts & 4507 \\
40-50 & Oracle on biased contacts & 4066 \\
50-60 & Oracle on biased contacts & 3253 \\
\toprule
 missing hours (\%) & \textbf{Debiasing approach} & count \\
\midrule
10-20 & Oracle on rescaled contacts & 4889 \\
20-30 & Oracle on rescaled contacts & 4840 \\
30-40 & Oracle on rescaled contacts & 4842 \\
40-50 & Oracle on rescaled contacts & 4800 \\
50-60 & Oracle on rescaled contacts & 4718 \\
\midrule
10-20 & Calib. on biased contacts & 4775 \\
20-30 & Calib. on biased contacts & 4838 \\
30-40 & Calib. on biased contacts & 4832 \\
40-50 & Calib. on biased contacts & 4604 \\
50-60 & Calib. on biased contacts & 4660 \\
\midrule
10-20 & Calib. on rescaled contacts & 4819 \\
20-30 & Calib. on rescaled contacts & 4766 \\
30-40 & Calib. on rescaled contacts & 4683 \\
40-50 & Calib. on rescaled contacts & 4594 \\
50-60 & Calib. on rescaled contacts & 4550 \\
\bottomrule
\end{tabular}

}
\caption{\textbf{Realized outbreaks sample size for each debiasing approach.}
Number of realized outbreaks for each debiasing approach and sparsity range.}
\label{tab:debiasing_realized_outbreaks_sample_size}
\end{table}

\clearpage
\begin{landscape}
\begin{table}[H]
\centering
{\footnotesize\setlength{\tabcolsep}{2pt}
\begin{tabular}{lcccccc}
\toprule
 & missing hours (\%) & \makecell{Realized outbreaks frac. \\ (mean, mean std)} & \makecell{gt. abs. change \\ (mean, mean std)} & \makecell{gt. rel. change \\ (mean, mean std)} & quartiles & 95\% CI \\
outcome &  &  &  &  &  &  \\
\midrule
\textbf{ground truth} (gt.) & 0-5 & 98.32 $\pm$ 0.00 & - & - & (98.00, 98.00, 98.00) & (98.00, 98.00) \\
\midrule
Oracle on sparse contacts & 10-20 & 96.64 $\pm$ 0.24 & -1.68 $\pm$ 0.24 & -1.71 $\pm$ 0.24 & (95.00, 97.00, 98.00) & (94.00, 100.00) \\
Oracle on sparse contacts & 20-30 & 92.74 $\pm$ 0.40 & -5.58 $\pm$ 0.40 & -5.68 $\pm$ 0.41 & (91.00, 93.00, 95.00) & (87.00, 98.00) \\
Oracle on sparse contacts & 30-40 & 90.14 $\pm$ 0.39 & -8.18 $\pm$ 0.39 & -8.32 $\pm$ 0.40 & (89.00, 90.00, 92.00) & (84.00, 94.00) \\
Oracle on sparse contacts & 40-50 & 81.32 $\pm$ 0.56 & -17.00 $\pm$ 0.56 & -17.29 $\pm$ 0.57 & (79.00, 82.00, 84.00) & (72.00, 89.00) \\
Oracle on sparse contacts & 50-60 & 65.06 $\pm$ 0.77 & -33.26 $\pm$ 0.77 & -33.83 $\pm$ 0.79 & (61.00, 65.00, 69.00) & (55.00, 74.00) \\
\midrule
Oracle on rescaled contacts & 10-20 & 97.78 $\pm$ 0.22 & -0.54 $\pm$ 0.22 & -0.55 $\pm$ 0.23 & (97.00, 98.00, 99.00) & (94.00, 100.00) \\
Oracle on rescaled contacts & 20-30 & 96.80 $\pm$ 0.29 & -1.52 $\pm$ 0.29 & -1.55 $\pm$ 0.29 & (96.00, 97.00, 98.00) & (92.00, 100.00) \\
Oracle on rescaled contacts & 30-40 & 96.84 $\pm$ 0.23 & -1.48 $\pm$ 0.23 & -1.51 $\pm$ 0.23 & (96.00, 97.00, 98.00) & (94.00, 100.00) \\
Oracle on rescaled contacts & 40-50 & 96.00 $\pm$ 0.29 & -2.32 $\pm$ 0.29 & -2.36 $\pm$ 0.30 & (95.00, 96.00, 98.00) & (91.00, 99.00) \\
Oracle on rescaled contacts & 50-60 & 94.36 $\pm$ 0.34 & -3.96 $\pm$ 0.34 & -4.03 $\pm$ 0.35 & (93.00, 94.00, 96.00) & (90.00, 99.00) \\
\midrule
Calib. on sparse contacts & 10-20 & 95.50 $\pm$ 1.02 & -2.82 $\pm$ 1.02 & -2.87 $\pm$ 1.04 & (94.00, 99.00, 100.00) & (76.00, 100.00) \\
Calib. on sparse contacts & 20-30 & 96.76 $\pm$ 0.61 & -1.56 $\pm$ 0.61 & -1.59 $\pm$ 0.62 & (95.00, 99.00, 100.00) & (88.00, 100.00) \\
Calib. on sparse contacts & 30-40 & 96.64 $\pm$ 0.53 & -1.68 $\pm$ 0.53 & -1.71 $\pm$ 0.54 & (94.00, 98.00, 100.00) & (89.00, 100.00) \\
Calib. on sparse contacts & 40-50 & 92.08 $\pm$ 1.25 & -6.24 $\pm$ 1.25 & -6.35 $\pm$ 1.27 & (89.00, 96.00, 98.00) & (69.00, 100.00) \\
Calib. on sparse contacts & 50-60 & 93.20 $\pm$ 1.08 & -5.12 $\pm$ 1.08 & -5.21 $\pm$ 1.10 & (90.00, 96.00, 98.00) & (72.00, 100.00) \\
\midrule
Calib. on rescaled contacts & 10-20 & 96.38 $\pm$ 0.89 & -1.94 $\pm$ 0.89 & -1.97 $\pm$ 0.91 & (95.00, 99.00, 100.00) & (77.00, 100.00) \\
Calib. on rescaled contacts & 20-30 & 95.32 $\pm$ 1.04 & -3.00 $\pm$ 1.04 & -3.05 $\pm$ 1.06 & (95.00, 98.00, 100.00) & (73.00, 100.00) \\
Calib. on rescaled contacts & 30-40 & 93.66 $\pm$ 1.09 & -4.66 $\pm$ 1.09 & -4.74 $\pm$ 1.11 & (91.00, 96.00, 99.00) & (75.00, 100.00) \\
Calib. on rescaled contacts & 40-50 & 91.88 $\pm$ 1.35 & -6.44 $\pm$ 1.35 & -6.55 $\pm$ 1.37 & (90.00, 94.00, 100.00) & (68.00, 100.00) \\
Calib. on rescaled contacts & 50-60 & 91.00 $\pm$ 1.36 & -7.32 $\pm$ 1.36 & -7.45 $\pm$ 1.38 & (86.00, 94.00, 99.00) & (70.00, 100.00) \\
\bottomrule
\end{tabular}
}
\caption{\textbf{Fraction of realized outbreaks $p_{\texttt{outbreak}}$.}
For each debiasing approach and range of missing hours, the point estimates, IQR, and 95\% CI are computed from a sample of 50 measurements of $p_{\texttt{outbreak}}$.
Each measurement is computed from the ensemble of 100 epidemic curves obtained for each of the 50 missingness realizations.}
\label{tab:da_realized_outbreaks}
\end{table}
\end{landscape}
\clearpage

{\setlength{\intextsep}{4pt}\captionsetup{skip=2pt}
\begin{table}[H]
\centering
{\scriptsize\setlength{\tabcolsep}{0.5pt}
\begin{tabular}{lcccccc}
\toprule
 & missing hours (\%) & \makecell{I\_max frac. \\ (mean, mean std)} & \makecell{gt. abs. change \\ (mean, mean std)} & \makecell{gt. rel. change \\ (mean, mean std)} & quartiles & 95\% CI \\
outcome &  &  &  &  &  &  \\
\midrule
\textbf{ground truth} (gt.) & 0-5 & 33.86 $\pm$ 0.06 & - & - & (31.00, 33.00, 37.00) & (27.00, 42.00) \\
\midrule
Oracle on sparse contacts & 10-20 & 29.44 $\pm$ 0.05 & -4.42 $\pm$ 0.08 & -13.05 $\pm$ 0.21 & (27.00, 29.00, 32.00) & (23.00, 36.00) \\
Oracle on sparse contacts & 20-30 & 25.43 $\pm$ 0.05 & -8.42 $\pm$ 0.08 & -24.88 $\pm$ 0.20 & (23.00, 25.00, 28.00) & (19.00, 32.00) \\
Oracle on sparse contacts & 30-40 & 21.02 $\pm$ 0.05 & -12.84 $\pm$ 0.08 & -37.93 $\pm$ 0.19 & (19.00, 21.00, 23.00) & (14.00, 28.00) \\
Oracle on sparse contacts & 40-50 & 14.01 $\pm$ 0.07 & -19.84 $\pm$ 0.09 & -58.61 $\pm$ 0.21 & (12.00, 14.00, 17.00) & (4.00, 22.00) \\
Oracle on sparse contacts & 50-60 & 6.87 $\pm$ 0.05 & -26.99 $\pm$ 0.08 & -79.72 $\pm$ 0.16 & (4.00, 7.00, 9.00) & (2.00, 13.00) \\
\midrule
Oracle on rescaled contacts & 10-20 & 34.89 $\pm$ 0.06 & 1.04 $\pm$ 0.08 & 3.07 $\pm$ 0.24 & (32.00, 35.00, 38.00) & (28.00, 43.00) \\
Oracle on rescaled contacts & 20-30 & 34.71 $\pm$ 0.06 & 0.85 $\pm$ 0.08 & 2.52 $\pm$ 0.24 & (32.00, 34.00, 37.00) & (28.00, 43.00) \\
Oracle on rescaled contacts & 30-40 & 33.42 $\pm$ 0.05 & -0.44 $\pm$ 0.08 & -1.29 $\pm$ 0.23 & (31.00, 33.00, 36.00) & (27.00, 41.00) \\
Oracle on rescaled contacts & 40-50 & 32.77 $\pm$ 0.06 & -1.08 $\pm$ 0.08 & -3.19 $\pm$ 0.23 & (30.00, 33.00, 36.00) & (26.00, 41.00) \\
Oracle on rescaled contacts & 50-60 & 32.15 $\pm$ 0.06 & -1.70 $\pm$ 0.08 & -5.03 $\pm$ 0.24 & (29.00, 32.00, 35.00) & (25.00, 41.00) \\
\midrule
Calib. on sparse contacts & 10-20 & 34.18 $\pm$ 0.08 & 0.33 $\pm$ 0.10 & 0.97 $\pm$ 0.29 & (30.00, 34.00, 38.00) & (25.00, 47.00) \\
Calib. on sparse contacts & 20-30 & 34.24 $\pm$ 0.07 & 0.39 $\pm$ 0.09 & 1.14 $\pm$ 0.27 & (31.00, 34.00, 37.00) & (26.00, 45.00) \\
Calib. on sparse contacts & 30-40 & 34.04 $\pm$ 0.07 & 0.18 $\pm$ 0.09 & 0.54 $\pm$ 0.26 & (31.00, 34.00, 37.00) & (26.00, 44.00) \\
Calib. on sparse contacts & 40-50 & 34.66 $\pm$ 0.07 & 0.81 $\pm$ 0.09 & 2.39 $\pm$ 0.28 & (31.00, 34.00, 38.00) & (26.00, 46.00) \\
Calib. on sparse contacts & 50-60 & 33.87 $\pm$ 0.07 & 0.02 $\pm$ 0.09 & 0.06 $\pm$ 0.27 & (30.00, 33.00, 37.00) & (25.00, 45.00) \\
\midrule
Calib. on rescaled contacts & 10-20 & 33.65 $\pm$ 0.07 & -0.20 $\pm$ 0.09 & -0.60 $\pm$ 0.28 & (30.00, 33.00, 37.00) & (25.00, 45.00) \\
Calib. on rescaled contacts & 20-30 & 34.50 $\pm$ 0.08 & 0.65 $\pm$ 0.10 & 1.91 $\pm$ 0.30 & (31.00, 34.00, 37.00) & (26.00, 49.00) \\
Calib. on rescaled contacts & 30-40 & 34.34 $\pm$ 0.07 & 0.49 $\pm$ 0.09 & 1.44 $\pm$ 0.27 & (31.00, 34.00, 37.00) & (26.00, 45.00) \\
Calib. on rescaled contacts & 40-50 & 34.55 $\pm$ 0.08 & 0.69 $\pm$ 0.10 & 2.04 $\pm$ 0.29 & (31.00, 34.00, 38.00) & (26.00, 47.00) \\
Calib. on rescaled contacts & 50-60 & 34.28 $\pm$ 0.08 & 0.42 $\pm$ 0.10 & 1.25 $\pm$ 0.30 & (30.00, 34.00, 37.00) & (25.00, 47.00) \\
\bottomrule
\end{tabular}
}
\caption{\textbf{Variation of peak of infected $I_{\texttt{peak}}$.}
For each debiasing approach and range of missing hours, the point estimates, IQR, and 95\% CI are computed from the sample of realized outbreaks.}
\label{tab:da_size_peak}
\end{table}

\begin{table}[H]
\centering
{\scriptsize\setlength{\tabcolsep}{0.5pt}
\begin{tabular}{lcccccc}
\toprule
 & missing hours (\%) & \makecell{I\_tot frac. \\ (mean, mean std)} & \makecell{gt. abs. change \\ (mean, mean std)} & \makecell{gt. rel. change \\ (mean, mean std)} & quartiles & 95\% CI \\
outcome &  &  &  &  &  &  \\
\midrule
\textbf{ground truth} (gt.) & 0-5 & 85.29 $\pm$ 0.03 & - & - & (84.00, 85.00, 87.00) & (81.00, 89.00) \\
\midrule
Oracle on sparse contacts & 10-20 & 81.13 $\pm$ 0.04 & -4.16 $\pm$ 0.05 & -4.88 $\pm$ 0.05 & (80.00, 81.00, 83.00) & (76.00, 85.00) \\
Oracle on sparse contacts & 20-30 & 75.68 $\pm$ 0.06 & -9.61 $\pm$ 0.07 & -11.27 $\pm$ 0.08 & (74.00, 76.00, 78.00) & (67.00, 82.00) \\
Oracle on sparse contacts & 30-40 & 67.97 $\pm$ 0.11 & -17.32 $\pm$ 0.11 & -20.31 $\pm$ 0.13 & (66.00, 69.00, 72.00) & (50.00, 77.00) \\
Oracle on sparse contacts & 40-50 & 48.16 $\pm$ 0.20 & -37.13 $\pm$ 0.20 & -43.54 $\pm$ 0.23 & (42.00, 51.00, 57.00) & (13.00, 64.00) \\
Oracle on sparse contacts & 50-60 & 24.92 $\pm$ 0.19 & -60.37 $\pm$ 0.20 & -70.78 $\pm$ 0.23 & (16.00, 25.00, 33.00) & (6.00, 45.00) \\
\midrule
Oracle on rescaled contacts & 10-20 & 86.17 $\pm$ 0.03 & 0.88 $\pm$ 0.04 & 1.03 $\pm$ 0.05 & (85.00, 86.00, 87.00) & (82.00, 89.00) \\
Oracle on rescaled contacts & 20-30 & 86.37 $\pm$ 0.03 & 1.08 $\pm$ 0.04 & 1.27 $\pm$ 0.04 & (85.00, 87.00, 88.00) & (83.00, 90.00) \\
Oracle on rescaled contacts & 30-40 & 85.67 $\pm$ 0.03 & 0.38 $\pm$ 0.04 & 0.44 $\pm$ 0.05 & (85.00, 86.00, 87.00) & (82.00, 89.00) \\
Oracle on rescaled contacts & 40-50 & 85.07 $\pm$ 0.03 & -0.22 $\pm$ 0.04 & -0.26 $\pm$ 0.05 & (84.00, 85.00, 87.00) & (81.00, 89.00) \\
Oracle on rescaled contacts & 50-60 & 83.64 $\pm$ 0.03 & -1.65 $\pm$ 0.04 & -1.94 $\pm$ 0.05 & (82.00, 84.00, 85.00) & (79.00, 88.00) \\
\midrule
Calib. on sparse contacts & 10-20 & 85.00 $\pm$ 0.05 & -0.29 $\pm$ 0.06 & -0.34 $\pm$ 0.07 & (83.00, 85.00, 88.00) & (77.00, 91.00) \\
Calib. on sparse contacts & 20-30 & 85.54 $\pm$ 0.04 & 0.25 $\pm$ 0.05 & 0.29 $\pm$ 0.06 & (83.00, 86.00, 88.00) & (79.00, 91.00) \\
Calib. on sparse contacts & 30-40 & 85.53 $\pm$ 0.04 & 0.24 $\pm$ 0.05 & 0.28 $\pm$ 0.06 & (84.00, 86.00, 88.00) & (79.00, 90.00) \\
Calib. on sparse contacts & 40-50 & 86.11 $\pm$ 0.05 & 0.82 $\pm$ 0.06 & 0.97 $\pm$ 0.06 & (84.00, 86.00, 88.00) & (80.00, 92.00) \\
Calib. on sparse contacts & 50-60 & 84.57 $\pm$ 0.05 & -0.72 $\pm$ 0.06 & -0.84 $\pm$ 0.07 & (83.00, 85.00, 87.00) & (77.00, 91.00) \\
\midrule
Calib. on rescaled contacts & 10-20 & 84.69 $\pm$ 0.05 & -0.60 $\pm$ 0.06 & -0.70 $\pm$ 0.07 & (82.00, 85.00, 87.00) & (77.00, 91.00) \\
Calib. on rescaled contacts & 20-30 & 85.69 $\pm$ 0.05 & 0.40 $\pm$ 0.06 & 0.47 $\pm$ 0.06 & (84.00, 86.00, 88.00) & (79.00, 93.00) \\
Calib. on rescaled contacts & 30-40 & 86.04 $\pm$ 0.04 & 0.75 $\pm$ 0.05 & 0.88 $\pm$ 0.06 & (84.00, 86.00, 88.00) & (80.00, 91.00) \\
Calib. on rescaled contacts & 40-50 & 86.00 $\pm$ 0.05 & 0.71 $\pm$ 0.06 & 0.84 $\pm$ 0.07 & (83.00, 86.00, 89.00) & (79.00, 92.00) \\
Calib. on rescaled contacts & 50-60 & 85.13 $\pm$ 0.06 & -0.16 $\pm$ 0.06 & -0.19 $\pm$ 0.07 & (83.00, 85.00, 88.00) & (77.00, 92.00) \\
\bottomrule
\end{tabular}
}
\caption{\textbf{Variation of total of infected $I_{\texttt{tot}}$.}
For each debiasing approach and range of missing hours, the point estimates, IQR, and 95\% CI are computed from the sample of realized outbreaks.}
\label{tab:da_size_total}
\end{table}
}
\clearpage

\begin{landscape}
\begin{table}[H]
\centering
{\scriptsize\setlength{\tabcolsep}{2pt}
\begin{tabular}{lccccccccccccccccccccccccccccc}
\toprule
 & missing hours (\%) & 1 & 2 & 3 & 4 & 5 & 6 & 7 & 8 & 9 & 10 & 11 & 12 & 13 & 14 & 15 & 16 & 17 & 18 & 19 & 20 & 21 & 22 & 23 & 24 & 25 & 26 & 27 & 28 \\
outcome &  &  &  &  &  &  &  &  &  &  &  &  &  &  &  &  &  &  &  &  &  &  &  &  &  &  &  &  &  \\
\midrule
\textbf{ground truth} (gt.) & 0-5 &  &  &  &  &  &  & 44.5 & 0.1 &  & 5.3 & 6.9 & 24.5 & 16.2 & 1.9 &  &  &  &  & 0.2 & 0.2 & 0.2 &  &  &  &  &  &  &  \\
\midrule
Calib. on sparse contacts & 10-20 &  &  &  &  &  &  & 52.1 & 0.1 &  & 3.6 & 5.7 & 21.9 & 14.9 & 0.9 &  &  &  & 0.1 & 0.2 & 0.3 & 0.1 &  &  &  &  &  &  &  \\
Calib. on sparse contacts & 20-30 &  &  &  &  &  &  & 48.9 &  &  & 3.5 & 7.4 & 22.5 & 15.3 & 1.5 &  &  &  & 0.1 & 0.2 & 0.4 & 0.1 &  &  &  &  &  &  &  \\
Calib. on sparse contacts & 30-40 &  &  &  &  &  & 0.1 & 51.7 &  &  & 3.3 & 6.6 & 20.9 & 14.4 & 1.6 &  &  &  & 0.2 & 0.3 & 0.3 & 0.3 &  &  &  &  &  &  &  \\
Calib. on sparse contacts & 40-50 &  &  &  &  &  & 0.2 & 52.0 &  &  & 4.5 & 6.2 & 21.5 & 12.5 & 1.7 &  &  & 0.1 & 0.2 & 0.3 & 0.5 & 0.3 &  &  &  &  &  &  &  \\
Calib. on sparse contacts & 50-60 &  &  &  &  &  & 0.1 & 55.9 & 0.1 &  & 3.0 & 7.1 & 16.6 & 12.9 & 2.2 &  &  &  & 0.2 & 0.4 & 0.9 & 0.5 &  &  &  &  &  &  &  \\
\midrule
Calib. on rescaled contacts & 10-20 &  &  &  &  &  &  & 49.0 & 0.1 &  & 3.8 & 6.3 & 22.2 & 16.0 & 1.6 &  &  &  & 0.2 & 0.2 & 0.4 & 0.2 &  &  &  &  &  &  &  \\
Calib. on rescaled contacts & 20-30 &  &  &  &  &  &  & 49.7 & 0.1 &  & 3.9 & 7.5 & 21.9 & 14.6 & 1.6 &  &  &  & 0.1 & 0.1 & 0.3 & 0.2 &  &  &  &  &  &  &  \\
Calib. on rescaled contacts & 30-40 &  &  &  &  &  & 0.2 & 52.4 & 0.1 &  & 3.4 & 7.3 & 21.1 & 12.9 & 1.6 &  &  &  & 0.2 & 0.4 & 0.3 & 0.3 &  &  &  &  &  &  &  \\
Calib. on rescaled contacts & 40-50 &  &  &  &  &  & 0.4 & 54.8 &  &  & 3.7 & 5.8 & 19.7 & 13.2 & 1.4 &  &  &  & 0.1 & 0.2 & 0.3 & 0.4 &  &  &  &  &  &  &  \\
Calib. on rescaled contacts & 50-60 &  &  &  &  &  &  & 55.6 & 0.1 &  & 2.8 & 7.0 & 16.9 & 13.4 & 2.4 &  &  &  & 0.1 & 0.4 & 0.7 & 0.7 &  &  &  &  &  &  &  \\
\midrule
Oracle on sparse contacts & 10-20 &  &  &  &  &  &  & 18.1 & 0.1 &  & 1.6 & 3.7 & 22.7 & 43.3 & 6.3 &  &  &  & 0.2 & 0.5 & 1.3 & 2.1 &  &  &  &  &  &  &  \\
Oracle on sparse contacts & 20-30 &  &  &  &  &  &  & 5.6 &  &  & 0.7 & 1.3 & 10.3 & 53.2 & 14.0 &  &  &  & 0.2 & 0.6 & 3.1 & 10.4 &  &  &  &  &  & 0.2 & 0.3 \\
Oracle on sparse contacts & 30-40 &  &  &  &  &  &  & 3.5 &  &  & 0.2 & 0.6 & 3.8 & 43.9 & 13.9 &  &  & 0.1 & 0.3 & 0.3 & 4.8 & 27.1 &  &  &  & 0.1 & 0.2 & 0.4 & 0.8 \\
Oracle on sparse contacts & 40-50 &  &  & 0.1 &  &  & 0.2 & 3.1 & 0.4 &  & 0.3 & 0.5 & 2.7 & 22.1 & 6.7 &  &  &  & 0.2 & 0.2 & 1.9 & 53.4 & 0.2 &  & 0.1 & 0.7 & 1.1 & 2.8 & 3.0 \\
Oracle on sparse contacts & 50-60 &  &  &  & 0.3 & 1.0 & 2.5 & 11.8 & 1.8 & 0.1 & 1.6 & 1.9 & 4.6 & 16.7 & 8.2 & 0.3 & 0.1 & 0.3 & 0.3 & 0.5 & 2.8 & 30.5 & 0.7 & 0.1 & 0.5 & 0.9 & 1.9 & 4.5 & 6.1 \\
\midrule
Oracle on rescaled contacts & 10-20 &  &  &  &  &  &  & 48.4 & 0.1 &  & 4.0 & 6.7 & 25.3 & 13.4 & 1.3 &  &  &  & 0.1 & 0.2 & 0.3 & 0.1 &  &  &  &  &  &  &  \\
Oracle on rescaled contacts & 20-30 &  &  &  &  &  &  & 45.6 & 0.1 &  & 4.6 & 7.1 & 25.5 & 14.1 & 1.9 &  &  &  & 0.2 & 0.2 & 0.5 & 0.2 &  &  &  &  &  &  &  \\
Oracle on rescaled contacts & 30-40 &  &  &  &  &  & 0.1 & 45.7 &  &  & 3.3 & 7.0 & 23.8 & 16.7 & 2.1 &  &  &  & 0.2 & 0.2 & 0.4 & 0.4 &  &  &  &  &  &  &  \\
Oracle on rescaled contacts & 40-50 &  &  &  &  &  & 0.1 & 48.6 &  &  & 3.0 & 4.8 & 22.2 & 18.0 & 1.7 &  &  &  &  & 0.4 & 0.7 & 0.5 &  &  &  &  &  &  &  \\
Oracle on rescaled contacts & 50-60 &  &  &  &  &  &  & 52.8 & 0.1 &  & 2.4 & 6.4 & 18.1 & 15.4 & 2.5 &  &  &  & 0.1 & 0.3 & 0.8 & 1.0 &  &  &  &  &  &  &  \\
\bottomrule
\end{tabular}
}
\caption{\textbf{Peak timing.}
Occurrence of the peak day over the ensemble of realized outbreaks and over all 50 missingness realizations.
Columns refer to the day of the simulation.
Reported values are rounded to one decimal place.}
\label{tab:da_dynamic_peak_day}
\end{table}

\newpage
\begin{table}[H]
\centering
{\scriptsize\setlength{\tabcolsep}{2pt}
\begin{tabular}{lccccccccccccccccccccccccccccc}
\toprule
 & missing hours (\%) & 1 & 2 & 3 & 4 & 5 & 6 & 7 & 8 & 9 & 10 & 11 & 12 & 13 & 14 & 15 & 16 & 17 & 18 & 19 & 20 & 21 & 22 & 23 & 24 & 25 & 26 & 27 & 28 \\
outcome &  &  &  &  &  &  &  &  &  &  &  &  &  &  &  &  &  &  &  &  &  &  &  &  &  &  &  &  &  \\
\midrule
\textbf{ground truth} (gt.) & 0-5 &  &  &  &  &  &  &  &  &  &  &  &  &  &  &  &  &  & 0.2 & 0.6 & 2.2 & 13.7 & 3.2 & 1.8 & 10.8 & 15.7 & 15.8 & 18.5 & 17.5 \\
\midrule
Calib. on sparse contacts & 10-20 &  &  &  &  &  &  &  &  &  &  &  &  &  &  &  &  &  & 0.6 & 0.9 & 3.0 & 14.4 & 3.4 & 1.5 & 11.1 & 15.2 & 14.4 & 17.1 & 18.3 \\
Calib. on sparse contacts & 20-30 &  &  &  &  &  &  &  &  &  &  &  &  &  &  &  &  & 0.1 & 0.2 & 0.7 & 3.0 & 16.4 & 3.7 & 1.7 & 9.9 & 14.7 & 15.1 & 18.2 & 16.5 \\
Calib. on sparse contacts & 30-40 &  &  &  &  &  &  &  &  &  &  &  &  &  &  &  &  &  & 0.2 & 0.6 & 1.7 & 13.6 & 3.0 & 1.4 & 9.7 & 15.3 & 15.6 & 18.7 & 20.1 \\
Calib. on sparse contacts & 40-50 &  &  &  &  &  &  &  &  &  &  &  &  &  &  &  &  &  & 0.2 & 0.3 & 1.6 & 11.7 & 2.9 & 1.3 & 9.3 & 14.0 & 15.4 & 20.8 & 22.4 \\
Calib. on sparse contacts & 50-60 &  &  &  &  &  &  &  &  &  &  &  &  &  &  &  &  &  & 0.1 & 0.1 & 0.5 & 6.5 & 1.7 & 0.5 & 6.4 & 11.1 & 15.5 & 24.2 & 33.4 \\
\midrule
Calib. on rescaled contacts & 10-20 &  &  &  &  &  &  &  &  &  &  &  &  &  &  &  &  & 0.1 & 0.4 & 0.7 & 2.7 & 14.2 & 3.1 & 1.6 & 10.5 & 15.4 & 15.0 & 18.2 & 18.1 \\
Calib. on rescaled contacts & 20-30 &  &  &  &  &  &  &  &  &  &  &  &  &  &  &  &  & 0.2 & 0.4 & 1.1 & 3.3 & 17.1 & 3.9 & 1.8 & 11.0 & 14.7 & 14.2 & 17.3 & 15.0 \\
Calib. on rescaled contacts & 30-40 &  &  &  &  &  &  &  &  &  &  &  &  &  &  &  &  & 0.1 & 0.3 & 0.8 & 2.1 & 15.7 & 3.4 & 1.3 & 10.0 & 14.8 & 15.0 & 18.5 & 17.9 \\
Calib. on rescaled contacts & 40-50 &  &  &  &  &  &  &  &  &  &  &  &  &  &  &  &  &  & 0.2 & 0.3 & 1.7 & 12.1 & 3.6 & 1.2 & 8.3 & 13.5 & 15.4 & 20.0 & 23.8 \\
Calib. on rescaled contacts & 50-60 &  &  &  &  &  &  &  &  &  &  &  &  &  &  &  &  &  & 0.1 &  & 0.7 & 6.3 & 1.8 & 0.6 & 5.7 & 9.9 & 15.7 & 24.5 & 34.5 \\
\midrule
Oracle on sparse contacts & 10-20 &  &  &  &  &  &  &  &  &  &  &  &  &  &  &  &  &  &  & 0.1 & 0.4 & 5.4 & 1.7 & 0.9 & 6.2 & 12.1 & 16.1 & 24.1 & 32.9 \\
Oracle on sparse contacts & 20-30 &  &  &  &  &  &  &  &  &  &  &  &  &  &  &  &  &  &  &  &  & 1.9 & 0.6 & 0.2 & 2.9 & 6.5 & 11.6 & 26.0 & 50.2 \\
Oracle on sparse contacts & 30-40 &  &  &  &  &  &  &  &  &  &  &  &  &  &  &  &  &  &  &  &  & 0.1 & 0.1 & 0.1 & 1.0 & 2.6 & 5.6 & 17.8 & 72.6 \\
Oracle on sparse contacts & 40-50 &  &  &  &  &  &  &  &  &  &  &  & 0.1 & 0.1 & 0.3 &  &  &  & 0.1 & 0.1 & 0.2 & 0.2 & 0.1 &  & 0.4 & 0.7 & 1.8 & 9.3 & 86.4 \\
Oracle on sparse contacts & 50-60 &  &  &  &  &  &  & 0.1 & 0.1 &  & 0.3 & 0.3 & 0.3 & 1.0 & 1.5 & 0.3 & 0.3 & 0.8 & 0.8 & 0.7 & 1.0 & 2.9 & 0.6 & 0.3 & 1.4 & 1.8 & 2.6 & 10.5 & 72.3 \\
\midrule
Oracle on rescaled contacts & 10-20 &  &  &  &  &  &  &  &  &  &  &  &  &  &  &  &  & 0.1 & 0.2 & 0.7 & 3.2 & 16.9 & 3.4 & 2.0 & 12.4 & 16.1 & 14.1 & 16.3 & 14.5 \\
Oracle on rescaled contacts & 20-30 &  &  &  &  &  &  &  &  &  &  &  &  &  &  &  &  & 0.1 & 0.2 & 0.8 & 2.7 & 18.1 & 4.0 & 1.9 & 11.1 & 14.7 & 15.0 & 15.7 & 15.6 \\
Oracle on rescaled contacts & 30-40 &  &  &  &  &  &  &  &  &  &  &  &  &  &  &  &  &  &  & 0.3 & 1.7 & 12.4 & 3.3 & 1.4 & 10.6 & 15.2 & 15.9 & 19.9 & 19.2 \\
Oracle on rescaled contacts & 40-50 &  &  &  &  &  &  &  &  &  &  &  &  &  &  &  &  &  & 0.1 & 0.3 & 0.8 & 9.6 & 3.2 & 1.1 & 8.2 & 13.7 & 16.9 & 19.9 & 26.2 \\
Oracle on rescaled contacts & 50-60 &  &  &  &  &  &  &  &  &  &  &  &  &  &  &  &  &  & 0.1 & 0.2 & 0.8 & 7.1 & 1.5 & 0.8 & 5.6 & 10.4 & 15.5 & 23.3 & 34.9 \\
\bottomrule
\end{tabular}
}
\caption{\textbf{Last infection timing.}
Occurrence of the last infection day over the ensemble of realized outbreaks and over all 50 missingness realizations.
Columns refer to the day of the simulation.
Reported values are rounded to one decimal place.}
\label{tab:da_dynamic_lastcase_day}
\end{table}
\end{landscape}
\clearpage

\noindent \textbf{Changes in parameter estimation}
\setlength{\intextsep}{2pt}
\captionsetup{skip=1pt}
\begin{table}[H]
\centering
{\small\setlength{\tabcolsep}{4pt}\renewcommand{\arraystretch}{1.0}
\begin{tabular}{lccc}
\toprule
\multicolumn{4}{c}{\textbf{sparse contacts}} \\
\toprule
 & (mean, mean std) & quartiles & 95\% CI \\
missing hours (\%) &  &  &  \\
\midrule
10-20 & (1.115 $\pm$ 0.043) & [0.924, 1.029, 1.296] & [0.702, 1.782] \\
20-30 & (1.387 $\pm$ 0.043) & [1.159, 1.404, 1.560] & [0.917, 2.032] \\
30-40 & (1.770 $\pm$ 0.058) & [1.421, 1.729, 2.134] & [1.168, 2.601] \\
40-50 & (2.839 $\pm$ 0.123) & [2.284, 2.626, 3.265] & [1.586, 5.043] \\
50-60 & (3.898 $\pm$ 0.175) & [3.102, 3.688, 4.422] & [2.240, 6.834] \\
\bottomrule
\end{tabular}
\par\smallskip
\begin{tabular}{lccc}
\toprule
\multicolumn{4}{c}{\textbf{rescaled contacts}} \\
\toprule
 & (mean, mean std) & quartiles & 95\% CI \\
missing hours (\%) &  &  &  \\
\midrule
10-20 & (0.789 $\pm$ 0.028) & [0.632, 0.791, 0.934] & [0.530, 1.207] \\
20-30 & (0.870 $\pm$ 0.041) & [0.715, 0.769, 0.959] & [0.547, 1.764] \\
30-40 & (0.931 $\pm$ 0.031) & [0.771, 0.907, 1.024] & [0.626, 1.432] \\
40-50 & (1.017 $\pm$ 0.055) & [0.699, 0.949, 1.235] & [0.544, 1.849] \\
50-60 & (1.000 $\pm$ 0.060) & [0.756, 0.903, 1.192] & [0.562, 1.634] \\
\bottomrule
\end{tabular}
}
\caption{\textbf{$\beta$ infection probability at minute resolution.}
Parameter estimation from calibration on sparse and corrected contacts.
Values are reported in units of $\times 10^{-3}$.
The ground-truth value is $0.84 \times 10^{-3}$.}
\label{tab:da_param_beta}
\end{table}

\begin{table}[H]
\centering
{\small\setlength{\tabcolsep}{4pt}\renewcommand{\arraystretch}{1.0}
\begin{tabular}{lccc}
\toprule
\multicolumn{4}{c}{\textbf{sparse contacts}} \\
\toprule
 & (mean, mean std) & quartiles & 95\% CI \\
missing hours (\%) &  &  &  \\
\midrule
10-20 & (0.264 $\pm$ 0.002) & [0.254, 0.264, 0.274] & [0.233, 0.292] \\
20-30 & (0.262 $\pm$ 0.002) & [0.254, 0.262, 0.270] & [0.235, 0.288] \\
30-40 & (0.262 $\pm$ 0.002) & [0.252, 0.261, 0.271] & [0.238, 0.290] \\
40-50 & (0.258 $\pm$ 0.002) & [0.250, 0.258, 0.268] & [0.230, 0.284] \\
50-60 & (0.251 $\pm$ 0.002) & [0.242, 0.251, 0.259] & [0.232, 0.268] \\
\bottomrule
\end{tabular}
\par\smallskip
\begin{tabular}{lccc}
\toprule
\multicolumn{4}{c}{\textbf{rescaled contacts}} \\
\toprule
 & (mean, mean std) & quartiles & 95\% CI \\
missing hours (\%) &  &  &  \\
\midrule
10-20 & (0.263 $\pm$ 0.003) & [0.248, 0.259, 0.279] & [0.223, 0.306] \\
20-30 & (0.266 $\pm$ 0.003) & [0.255, 0.265, 0.276] & [0.239, 0.310] \\
30-40 & (0.268 $\pm$ 0.002) & [0.257, 0.267, 0.276] & [0.234, 0.294] \\
40-50 & (0.259 $\pm$ 0.002) & [0.247, 0.262, 0.269] & [0.230, 0.290] \\
50-60 & (0.250 $\pm$ 0.003) & [0.235, 0.246, 0.264] & [0.225, 0.283] \\
\bottomrule
\end{tabular}
}
\caption{\textbf{$\gamma$ recovery probability at daily resolution.}
Parameter estimation from calibration on sparse and corrected contacts.
The ground-truth value is 0.27.}
\label{tab:da_param_gamma}
\end{table}

\clearpage
\begin{table}[H]
\centering
{\small\setlength{\tabcolsep}{4pt}\renewcommand{\arraystretch}{1.0}
\begin{tabular}{lccc}
\toprule
\multicolumn{4}{c}{\textbf{sparse contacts}} \\
\toprule
 & (mean, mean std) & quartiles & 95\% CI \\
missing hours (\%) &  &  &  \\
\midrule
10-20 & (5.69 $\pm$ 0.30) & [4.02, 5.51, 6.98] & [2.60, 8.82] \\
20-30 & (5.05 $\pm$ 0.30) & [3.38, 4.43, 6.40] & [2.39, 8.73] \\
30-40 & (5.03 $\pm$ 0.27) & [3.62, 5.00, 6.39] & [2.06, 8.62] \\
40-50 & (5.02 $\pm$ 0.27) & [3.54, 4.68, 6.59] & [2.31, 8.92] \\
50-60 & (5.37 $\pm$ 0.25) & [3.97, 4.89, 6.61] & [3.25, 8.90] \\
\bottomrule
\end{tabular}
\par\smallskip
\begin{tabular}{lccc}
\toprule
\multicolumn{4}{c}{\textbf{rescaled contacts}} \\
\toprule
 & (mean, mean std) & quartiles & 95\% CI \\
missing hours (\%) &  &  &  \\
\midrule
10-20 & (6.09 $\pm$ 0.38) & [3.97, 6.06, 7.30] & [2.05, 12.15] \\
20-30 & (5.84 $\pm$ 0.38) & [3.85, 5.36, 7.31] & [2.49, 11.90] \\
30-40 & (5.47 $\pm$ 0.32) & [3.75, 5.29, 6.70] & [2.25, 10.47] \\
40-50 & (6.24 $\pm$ 0.34) & [4.27, 5.94, 7.92] & [2.83, 10.84] \\
50-60 & (7.15 $\pm$ 0.32) & [5.42, 7.06, 8.43] & [4.27, 10.92] \\
\bottomrule
\end{tabular}
}
\caption{\textbf{Score of parameter estimation.}
Score is computed as the RMSE of the median of the ensemble of daily infected.
The RMSE is evaluated between the ground truth and the sparse scenario (using either sparse or rescaled contacts for calibration).}
\label{tab:da_param_score}
\end{table}

\begin{table}[H]
\centering
{\small\setlength{\tabcolsep}{4pt}\renewcommand{\arraystretch}{1.0}
\begin{tabular}{lccc}
\toprule
\multicolumn{4}{c}{\textbf{sparse contacts}} \\
\toprule
 & (mean, mean std) & quartiles & 95\% CI \\
missing hours (\%) &  &  &  \\
\midrule
10-20 & (2.24 $\pm$ 0.05) & [2.02, 2.15, 2.46] & [1.76, 2.99] \\
20-30 & (2.56 $\pm$ 0.05) & [2.33, 2.58, 2.76] & [2.01, 3.24] \\
30-40 & (2.96 $\pm$ 0.06) & [2.62, 2.95, 3.33] & [2.32, 3.76] \\
40-50 & (3.94 $\pm$ 0.10) & [3.51, 3.80, 4.34] & [2.82, 5.60] \\
50-60 & (4.77 $\pm$ 0.12) & [4.23, 4.68, 5.20] & [3.47, 6.61] \\
\bottomrule
\end{tabular}
\par\smallskip
\begin{tabular}{lccc}
\toprule
\multicolumn{4}{c}{\textbf{rescaled contacts}} \\
\toprule
 & (mean, mean std) & quartiles & 95\% CI \\
missing hours (\%) &  &  &  \\
\midrule
10-20 & (1.84 $\pm$ 0.03) & [1.64, 1.83, 2.04] & [1.53, 2.34] \\
20-30 & (1.94 $\pm$ 0.05) & [1.75, 1.82, 2.04] & [1.54, 3.02] \\
30-40 & (2.02 $\pm$ 0.04) & [1.82, 1.98, 2.14] & [1.64, 2.61] \\
40-50 & (2.12 $\pm$ 0.06) & [1.75, 2.05, 2.38] & [1.56, 3.08] \\
50-60 & (2.11 $\pm$ 0.07) & [1.83, 2.01, 2.36] & [1.58, 2.86] \\
\bottomrule
\end{tabular}
}
\caption{\textbf{$\overline{R}_0$ computed from estimated parameters and ground-truth contacts.}
$\overline{R}_0$ is computed from the ground-truth contacts and the parameters calibrated on the biased or rescaled contacts.
Ground-truth simulation parameters provide an estimate of $\overline{R}_0 = 1.92$.}
\label{tab:da_param_R0}
\end{table}
\clearpage

\printbibliography[
  heading=bibintoc,
  title={Supplementary References},
  resetnumbers=true
]

@article{lucchini2021living,
  title={Living in a pandemic: changes in mobility routines, social activity and adherence to COVID-19 protective measures},
  author={Lucchini, Lorenzo and Centellegher, Simone and Pappalardo, Luca and Gallotti, Riccardo and Privitera, Filippo and Lepri, Bruno and De Nadai, Marco},
  journal={Scientific reports},
  volume={11},
  number={1},
  pages={24452},
  year={2021},
  publisher={Nature Publishing Group UK London}
}

@article{liao2026urban,
  title={Urban mobility enables deprivation bubble breaking in Indian and Mexican cities},
  author={Liao, Yuan and Delussu, Federico and de Sojo, S{\'\i}lvia and Alessandretti, Laura and Desiderio, Antonio},
  journal={arXiv preprint arXiv:2603.29782},
  year={2026}
}

@article{gauvin2020gender,
  title={Gender gaps in urban mobility},
  author={Gauvin, Laetitia and Tizzoni, Michele and Piaggesi, Simone and Young, Andrew and Adler, Natalia and Verhulst, Stefaan and Ferres, Leo and Cattuto, Ciro},
  journal={Humanities and Social Sciences Communications},
  volume={7},
  number={1},
  pages={11},
  year={2020},
  publisher={Palgrave}
}

@article{xu2018planning,
  title={Planning for electric vehicle needs by coupling charging profiles with urban mobility},
  author={Xu, Yanyan and {\c{C}}olak, Serdar and Kara, Emre C and Moura, Scott J and Gonz{\'a}lez, Marta C},
  journal={Nature Energy},
  volume={3},
  number={6},
  pages={484--493},
  year={2018},
  publisher={Nature Publishing Group UK London}
}

@article{mercier2022effective,
  title={Effective resistance against pandemics: Mobility network sparsification for high-fidelity epidemic simulations},
  author={Mercier, Alexander and Scarpino, Samuel and Moore, Cristopher},
  journal={PLOS Computational Biology},
  volume={18},
  number={11},
  pages={e1010650},
  year={2022},
  publisher={Public Library of Science San Francisco, CA USA}
}

@inproceedings{spielman2008graph,
  title={Graph sparsification by effective resistances},
  author={Spielman, Daniel A and Srivastava, Nikhil},
  booktitle={Proceedings of the fortieth annual ACM symposium on Theory of computing},
  pages={563--568},
  year={2008}
}

@article{newman1999scaling,
  title={Scaling and percolation in the small-world network model},
  author={Newman, Mark EJ and Watts, Duncan J},
  journal={Physical review E},
  volume={60},
  number={6},
  pages={7332},
  year={1999},
  publisher={APS}
}

@article{olynik2002temporal,
  title={Temporal characteristics of GPS error sources and their impact on relative positioning},
  author={Olynik, Michael and others},
  journal={University of Calgary: Calgary, Canada},
  year={2002}
}

@article{bierlaire2013probabilistic,
  title={A probabilistic map matching method for smartphone GPS data},
  author={Bierlaire, Michel and Chen, Jingmin and Newman, Jeffrey},
  journal={Transportation Research Part C: Emerging Technologies},
  volume={26},
  pages={78--98},
  year={2013},
  publisher={Elsevier}
}

@article{zhang2025survey,
  title={A survey on point-of-interest recommendation: Models, architectures, and security},
  author={Zhang, Qianru and Yang, Peng and Yu, Junliang and Wang, Haixin and He, Xingwei and Yiu, Siu-Ming and Yin, Hongzhi},
  journal={IEEE Transactions on Knowledge and Data Engineering},
  year={2025},
  publisher={IEEE}
}

@article{bang2005doubly,
  title={Doubly robust estimation in missing data and causal inference models},
  author={Bang, Heejung and Robins, James M},
  journal={Biometrics},
  volume={61},
  number={4},
  pages={962--973},
  year={2005},
  publisher={Oxford University Press}
}

@article{gonzalez2008understanding,
  title={Understanding individual human mobility patterns},
  author={Gonzalez, Marta C and Hidalgo, Cesar A and Barabasi, Albert-Laszlo},
  journal={nature},
  volume={453},
  number={7196},
  pages={779--782},
  year={2008},
  publisher={Nature Publishing Group UK London}
}

@article{calabrese2013understanding,
  title={Understanding individual mobility patterns from urban sensing data: A mobile phone trace example},
  author={Calabrese, Francesco and Diao, Mi and Di Lorenzo, Giusy and Ferreira Jr, Joseph and Ratti, Carlo},
  journal={Transportation research part C: emerging technologies},
  volume={26},
  pages={301--313},
  year={2013},
  publisher={Elsevier}
}

@inproceedings{zhou2020demystifying,
  title={Demystifying diehard android apps},
  author={Zhou, Hao and Wang, Haoyu and Zhou, Yajin and Luo, Xiapu and Tang, Yutian and Xue, Lei and Wang, Ting},
  booktitle={Proceedings of the 35th IEEE/ACM International Conference on Automated Software Engineering},
  pages={187--198},
  year={2020}
}

@article{pastor2015epidemic,
  title={Epidemic processes in complex networks},
  author={Pastor-Satorras, Romualdo and Castellano, Claudio and Van Mieghem, Piet and Vespignani, Alessandro},
  journal={Reviews of Modern Physics},
  volume={87},
  number={3},
  pages={925--979},
  year={2015},
  publisher={American Physical Society},
  doi={10.1103/RevModPhys.87.925}
}

@article{valdano2015analytical,
  title={Analytical computation of the epidemic threshold on temporal networks},
  author={Valdano, Eugenio and Ferreri, Luca and Poletto, Chiara and Colizza, Vittoria},
  journal={Physical Review X},
  volume={5},
  number={2},
  pages={021005},
  year={2015},
  publisher={American Physical Society},
  doi={10.1103/PhysRevX.5.021005}
}

@article{schlosser2020covid,
  title={COVID-19 lockdown induces disease-mitigating structural changes in mobility networks},
  author={Schlosser, Frank and Maier, Benjamin F and Jack, Olivia and Hinrichs, David and Zachariae, Adrian and Brockmann, Dirk},
  journal={Proceedings of the National Academy of Sciences},
  volume={117},
  number={52},
  pages={32883--32890},
  year={2020},
  publisher={National Academy of Sciences}
}

@article{van2011decreasing,
  title={Decreasing the spectral radius of a graph by link removals},
  author={Van Mieghem, Piet and Stevanovi{\'c}, Dragan and Kuipers, Fernando and Li, Cong and Van De Bovenkamp, Ruud and Liu, Daijie and Wang, Huijuan},
  journal={Physical Review E—Statistical, Nonlinear, and Soft Matter Physics},
  volume={84},
  number={1},
  pages={016101},
  year={2011},
  publisher={APS}
}

@article{brattig2023contact,
  title={Contact networks have small metric backbones that maintain community structure and are primary transmission subgraphs},
  author={Brattig Correia, Rion and Barrat, Alain and Rocha, Luis M},
  journal={PLOS Computational Biology},
  volume={19},
  number={2},
  pages={e1010854},
  year={2023},
  publisher={Public Library of Science San Francisco, CA USA}
}

@article{chakrabarti2008epidemic,
  title={Epidemic thresholds in real networks},
  author={Chakrabarti, Deepayan and Wang, Yang and Wang, Chenxi and Leskovec, Jurij and Faloutsos, Christos},
  journal={ACM Transactions on Information and System Security (TISSEC)},
  volume={10},
  number={4},
  pages={1--26},
  year={2008},
  publisher={ACM New York, NY, USA}
}

@article{noi2022assessing,
  title={Assessing COVID-induced changes in spatiotemporal structure of mobility in the United States in 2020: a multi-source analytical framework},
  author={Noi, Evgeny and Rudolph, Alexander and Dodge, Somayeh},
  journal={International Journal of Geographical Information Science},
  volume={36},
  number={3},
  pages={585--616},
  year={2022},
  publisher={Taylor \& Francis}
}

@article{van2008virus,
  title={Virus spread in networks},
  author={Van Mieghem, Piet and Omic, Jasmina and Kooij, Robert},
  journal={IEEE/ACM Transactions On Networking},
  volume={17},
  number={1},
  pages={1--14},
  year={2008},
  publisher={IEEE}
}

@article{Pepe2020COVID-19Lockdown,
    title = {{COVID-19 outbreak response, a dataset to assess mobility changes in Italy following national lockdown}},
    year = {2020},
    journal = {Scientific data},
    author = {Pepe, Emanuele and Bajardi, Paolo and Gauvin, Laetitia and Privitera, Filippo and Lake, Brennan and Cattuto, Ciro and Tizzoni, Michele},
    number = {1},
    pages = {1--7},
    volume = {7},
    publisher = {Nature Publishing Group}
}

@article{huang2021characteristics,
  title        = {The characteristics of multi-source mobility datasets and how they reveal the luxury nature of social distancing in the US during the COVID-19 pandemic},
  author       = {Huang, Xiao and Li, Zhenlong and Jiang, Yuqin and Ye, Xinyue and Deng, Chengbin and Zhang, Jiajia and Li, Xiaoming},
  year         = 2021,
  journal      = {International Journal of Digital Earth},
  publisher    = {Taylor \& Francis},
  volume       = 14,
  number       = 4,
  pages        = {424--442}
}

@article{Crawford2022ImpactData,
    title = {{Impact of close interpersonal contact on COVID-19 incidence: Evidence from 1 year of mobile device data}},
    year = {2022},
    journal = {Science Advances},
    author = {Crawford, Forrest W. and Jones, Sydney A. and Cartter, Matthew and Dean, Samantha G. and Warren, Joshua L. and Li, Zehang Richard and Barbieri, Jacqueline and Campbell, Jared and Kenney, Patrick and Valleau, Thomas and Morozova, Olga},
    number = {1},
    month = {1},
    pages = {5499},
    volume = {8},
    publisher = {American Association for the Advancement of Science},
    url = {https://www.science.org/doi/10.1126/sciadv.abi5499},
    doi = {10.1126/SCIADV.ABI5499/SUPPL{\_}FILE/SCIADV.ABI5499{\_}SM.PDF},
    issn = {23752548},
    pmid = {34995121}
}

@article{Painter2021PoliticalMandates,
    title = {{Political beliefs affect compliance with government mandates}},
    year = {2021},
    journal = {Journal of Economic Behavior {\&} Organization},
    author = {Painter, Marcus and Qiu, Tian},
    pages = {688--701},
    volume = {185},
    publisher = {Elsevier}
}

@article{savi2023standardised,
  title={A standardised differential privacy framework for epidemiological modeling with mobile phone data},
  author={Savi, Merveille Koissi and Yadav, Akash and Zhang, Wanrong and Vembar, Navin and Schroeder, Andrew and Balsari, Satchit and Buckee, Caroline O and Vadhan, Salil and Kishore, Nishant},
  journal={PLOS Digital Health},
  volume={2},
  number={10},
  pages={e0000233},
  year={2023},
  publisher={Public Library of Science San Francisco, CA USA}
}

@article{chen2022strategic,
  title={Strategic COVID-19 vaccine distribution can simultaneously elevate social utility and equity},
  author={Chen, Lin and Xu, Fengli and Han, Zhenyu and Tang, Kun and Hui, Pan and Evans, James and Li, Yong},
  journal={Nature Human Behaviour},
  volume={6},
  number={11},
  pages={1503--1514},
  year={2022},
  publisher={Nature Publishing Group UK London}
}

@article{kang2020multiscale,
  title        = {Multiscale dynamic human mobility flow dataset in the US during the COVID-19 epidemic},
  author       = {Kang, Yuhao and Gao, Song and Liang, Yunlei and Li, Mingxiao and Rao, Jinmeng and Kruse, Jake},
  year         = 2020,
  journal      = {Scientific data},
  publisher    = {Nature Publishing Group},
  volume       = 7,
  number       = 1,
  pages        = {1--13}
}

@article{gozzi2023estimating,
  title={Estimating the impact of COVID-19 vaccine inequities: a modeling study},
  author={Gozzi, Nicol{\`o} and Chinazzi, Matteo and Dean, Natalie E and Longini Jr, Ira M and Halloran, M Elizabeth and Perra, Nicola and Vespignani, Alessandro},
  journal={Nature Communications},
  volume={14},
  number={1},
  pages={3272},
  year={2023},
  publisher={Nature Publishing Group UK London}
}

@article{chang2021mobility,
  title={Mobility network models of COVID-19 explain inequities and inform reopening},
  author={Chang, Serina and Pierson, Emma and Koh, Pang Wei and Gerardin, Jaline and Redbird, Beth and Grusky, David and Leskovec, Jure},
  journal={Nature},
  volume={589},
  number={7840},
  pages={82--87},
  year={2021},
  publisher={Nature Publishing Group UK London}
}

@article{barreras2024exciting,
  title={The exciting potential and daunting challenge of using GPS human-mobility data for epidemic modeling},
  author={Barreras, Francisco and Watts, Duncan J},
  journal={Nature Computational Science},
  pages={1--14},
  year={2024},
  publisher={Nature Publishing Group US New York}
}

@article{buckee2020aggregated,
  title={Aggregated mobility data could help fight COVID-19},
  author={Buckee, Caroline O and Balsari, Satchit and Chan, Jennifer and Crosas, Merc{\`e} and Dominici, Francesca and Gasser, Urs and Grad, Yonatan H and Grenfell, Bryan and Halloran, M Elizabeth and Kraemer, Moritz UG and others},
  journal={Science},
  volume={368},
  number={6487},
  pages={145--146},
  year={2020},
  publisher={American Association for the Advancement of Science}
}

@article{yabe2024enhancing,
  title={Enhancing human mobility research with open and standardized datasets},
  author={Yabe, Takahiro and Luca, Massimiliano and Tsubouchi, Kota and Lepri, Bruno and Gonzalez, Marta C and Moro, Esteban},
  journal={Nature Computational Science},
  volume={4},
  number={7},
  pages={469--472},
  year={2024},
  publisher={Nature Publishing Group US New York}
}

@article{graser2021exploratory,
  title={An exploratory data analysis protocol for identifying problems in continuous movement data},
  author={Graser, Anita},
  journal={Journal of Location Based Services},
  volume={15},
  number={2},
  pages={89--117},
  year={2021},
  publisher={Taylor \& Francis}
}

@article{schlosser2021biases,
  title={Biases in human mobility data impact epidemic modeling},
  author={Schlosser, Frank and Sekara, Vedran and Brockmann, Dirk and Garcia-Herranz, Manuel},
  journal={arXiv preprint arXiv:2112.12521},
  year={2021}
}

@article{wesolowski2016connecting,
  title={Connecting mobility to infectious diseases: the promise and limits of mobile phone data},
  author={Wesolowski, Amy and Buckee, Caroline O and Eng{\o}-Monsen, Kenth and Metcalf, Charlotte Jessica Eland},
  journal={The Journal of infectious diseases},
  volume={214},
  number={suppl\_4},
  pages={S414--S420},
  year={2016},
  publisher={Oxford University Press}
}

@article{tizzoni2014use,
  title={On the use of human mobility proxies for modeling epidemics},
  author={Tizzoni, Michele and Bajardi, Paolo and Decuyper, Adeline and Kon Kam King, Guillaume and Schneider, Christian M and Blondel, Vincent and Smoreda, Zbigniew and Gonz{\'a}lez, Marta C and Colizza, Vittoria},
  journal={PLoS computational biology},
  volume={10},
  number={7},
  pages={e1003716},
  year={2014},
  publisher={Public Library of Science San Francisco, USA}
}

@article{wellenius2021impacts,
  title={Impacts of social distancing policies on mobility and COVID-19 case growth in the US},
  author={Wellenius, Gregory A and Vispute, Swapnil and Espinosa, Valeria and Fabrikant, Alex and Tsai, Thomas C and Hennessy, Jonathan and Dai, Andrew and Williams, Brian and Gadepalli, Krishna and Boulanger, Adam and others},
  journal={Nature communications},
  volume={12},
  number={1},
  pages={3118},
  year={2021},
  publisher={Nature Publishing Group UK London}
}

@article{aleta2020modelling,
  title={Modelling the impact of testing, contact tracing and household quarantine on second waves of COVID-19},
  author={Aleta, Alberto and Martin-Corral, David and Pastore y Piontti, Ana and Ajelli, Marco and Litvinova, Maria and Chinazzi, Matteo and Dean, Natalie E and Halloran, M Elizabeth and Longini Jr, Ira M and Merler, Stefano and others},
  journal={Nature human behaviour},
  volume={4},
  number={9},
  pages={964--971},
  year={2020},
  publisher={Nature Publishing Group UK London}
}

@article{weill2020social,
  title={Social distancing responses to COVID-19 emergency declarations strongly differentiated by income},
  author={Weill, Joakim A and Stigler, Matthieu and Deschenes, Olivier and Springborn, Michael R},
  journal={Proceedings of the national academy of sciences},
  volume={117},
  number={33},
  pages={19658--19660},
  year={2020},
  publisher={National Academy of Sciences}
}

@article{aleta2022quantifying,
  title={Quantifying the importance and location of SARS-CoV-2 transmission events in large metropolitan areas},
  author={Aleta, Alberto and Mart{\'\i}n-Corral, David and Bakker, Michiel A and Pastore y Piontti, Ana and Ajelli, Marco and Litvinova, Maria and Chinazzi, Matteo and Dean, Natalie E and Halloran, M Elizabeth and Longini Jr, Ira M and others},
  journal={Proceedings of the National Academy of Sciences},
  volume={119},
  number={26},
  pages={e2112182119},
  year={2022},
  publisher={National Academy of Sciences}
}

@article{stehle2011simulation,
  title={Simulation of an SEIR infectious disease model on the dynamic contact network of conference attendees},
  author={Stehl{\'e}, Juliette and Voirin, Nicolas and Barrat, Alain and Cattuto, Ciro and Colizza, Vittoria and Isella, Lorenzo and R{\'e}gis, Corinne and Pinton, Jean-Fran{\c{c}}ois and Khanafer, Nagham and Van den Broeck, Wouter and others},
  journal={BMC medicine},
  volume={9},
  pages={1--15},
  year={2011},
  publisher={Springer}
}

@article{couture2022jue,
  title={JUE Insight: Measuring movement and social contact with smartphone data: a real-time application to COVID-19},
  author={Couture, Victor and Dingel, Jonathan I and Green, Allison and Handbury, Jessie and Williams, Kevin R},
  journal={Journal of Urban Economics},
  volume={127},
  pages={103328},
  year={2022},
  publisher={Elsevier}
}

@article{davis2020estimating,
  title={Estimating the establishment of local transmission and the cryptic phase of the COVID-19 pandemic in the USA},
  author={Davis, Jessica T and Chinazzi, Matteo and Perra, Nicola and Mu, Kunpeng and y Piontti, Ana Pastore and Ajelli, Marco and Dean, Natalie E and Gioannini, Corrado and Litvinova, Maria and Merler, Stefano and others},
  journal={medRxiv},
  year={2020}
}

@article{bhouri2021covid,
  title={COVID-19 dynamics across the US: A deep learning study of human mobility and social behavior},
  author={Bhouri, Mohamed Aziz and Sahli Costabal, F. and Wang, Hanwen and Linka, Kevin and Peirlinck, Mathias and Kuhl, Ellen and Perdikaris, Paris},
  journal={Computer Methods in Applied Mechanics and Engineering},
  volume={382},
  pages={113891},
  year={2021},
  publisher={Elsevier},
  doi={10.1016/j.cma.2021.113891}
}

@article{salathe2010dynamics,
  title={Dynamics and control of diseases in networks with community structure},
  author={Salath{\'e}, Marcel and Jones, James H},
  journal={PLoS computational biology},
  volume={6},
  number={4},
  pages={e1000736},
  year={2010},
  publisher={Public Library of Science San Francisco, USA}
}

@article{simini2012universal,
  title={A universal model for mobility and migration patterns},
  author={Simini, Filippo and Gonz{\'a}lez, Marta C and Maritan, Amos and Barab{\'a}si, Albert-L{\'a}szl{\'o}},
  journal={Nature},
  volume={484},
  number={7392},
  pages={96--100},
  year={2012},
  publisher={Nature Publishing Group UK London}
}

@article{grantz2020use,
  title={The use of mobile phone data to inform analysis of COVID-19 pandemic epidemiology},
  author={Grantz, Kyra H and Meredith, Hannah R and Cummings, Derek AT and Metcalf, C Jessica E and Grenfell, Bryan T and Giles, John R and Mehta, Shruti and Solomon, Sunil and Labrique, Alain and Kishore, Nishant and others},
  journal={Nature communications},
  volume={11},
  number={1},
  pages={4961},
  year={2020},
  publisher={Nature Publishing Group UK London}
}

@article{Moro2021MobilityCities,
    title = {{Mobility patterns are associated with experienced income segregation in large US cities}},
    year = {2021},
    journal = {Nature communications},
    author = {Moro, Esteban and Calacci, Dan and Dong, Xiaowen and Pentland, Alex},
    number = {1},
    pages = {4633},
    volume = {12},
    publisher = {Nature Publishing Group UK London}
}

@article{gallotti2024distorted,
  title={Distorted insights from human mobility data},
  author={Gallotti, Riccardo and Maniscalco, Davide and Barthelemy, Marc and De Domenico, Manlio},
  journal={Communications Physics},
  volume={7},
  number={1},
  pages={421},
  year={2024},
  publisher={Nature Publishing Group UK London}
}

@article{alessandretti2020scales,
  title={The scales of human mobility},
  author={Alessandretti, Laura and Aslak, Ulf and Lehmann, Sune},
  journal={Nature},
  volume={587},
  number={7834},
  pages={402--407},
  year={2020},
  publisher={Nature Publishing Group}
}

@article{Gelman2016TheVoter,
    title = {{The Mythical Swing Voter}},
    year = {2016},
    journal = {Quarterly Journal of Political Science},
    author = {Gelman, Andrew and Goel, Sharad and Rivers, Douglas and Rothschild, David and Gelman, A and Goel, S and Rivers, D and Rothschild, D},
    pages = {103--130},
    volume = {11},
    url = {http://dx.doi.org/10.1561/100.00015031_supp},
    doi = {10.1561/100.00015031{\_}supp},
    issn = {1554-0626}
}

@article{Little1960ModellingSurveys,
    title = {{Modelling Differential Nonresponse in Sample Surveys}},
    year = {1960},
    journal = {Source: Sankhy{\={a}}: The Indian Journal of Statistics, Series B},
    author = {Little, Thomas C and Gelman, Andrew},
    number = {1},
    pages = {101--126},
    volume = {60},
    url = {https://www.jstor.org/stable/25053025}
}

@article{santos2019generating,
  title={Generating synthetic missing data: A review by missing mechanism},
  author={Santos, Miriam Seoane and Pereira, Ricardo Cardoso and Costa, Adriana Fonseca and Soares, Jastin Pompeu and Santos, Jo{\~a}o and Abreu, Pedro Henriques},
  journal={IEEE Access},
  volume={7},
  pages={11651--11667},
  year={2019},
  publisher={IEEE}
}

@article{schouten2018generating,
  title={Generating missing values for simulation purposes: a multivariate amputation procedure},
  author={Schouten, Rianne Margaretha and Lugtig, Peter and Vink, Gerko},
  journal={Journal of Statistical Computation and Simulation},
  volume={88},
  number={15},
  pages={2909--2930},
  year={2018},
  publisher={Taylor \& Francis}
}

@article{Sarndal2005EstimationNonresponse,
    title = {{Estimation in surveys with nonresponse}},
    year = {2005},
    journal = {Estimation in Surveys with Nonresponse},
    author = {S{\"{a}}rndal, Carl Erik and Lundstr{\"{o}}m, Sixten},
    month = {6},
    pages = {1--199},
    publisher = {Wiley Blackwell},
    url = {https://onlinelibrary.wiley.com/doi/book/10.1002/0470011351},
    isbn = {9780470011355},
    doi = {10.1002/0470011351}
}

@article{liao2025effect,
  title={The effect of limited mobility on the experienced segregation of foreign-born minorities},
  author={Liao, Yuan and Gil, Jorge and Yeh, Sonia and Pereira, Rafael HM and Alessandretti, Laura},
  journal={npj Sustainable Mobility and Transport},
  volume={2},
  number={1},
  pages={29},
  year={2025},
  publisher={Nature Publishing Group UK London}
}

@article{wang2018urban,
  author  = {Wang, Qi and Phillips, Nolan E. and Small, Mario L. and Sampson, Robert J.},
  title   = {Urban mobility and neighborhood isolation in America’s 50 largest cities},
  journal = {Proceedings of the National Academy of Sciences},
  year    = {2018},
  volume  = {115},
  number  = {30},
  pages   = {7735--7740},
  doi     = {10.1073/pnas.1802537115}
}

@article{song2010limits,
  title={Limits of predictability in human mobility},
  author={Song, Chaoming and Qu, Zehui and Blumm, Nicholas and Barab{\'a}si, Albert-L{\'a}szl{\'o}},
  journal={Science},
  volume={327},
  number={5968},
  pages={1018--1021},
  year={2010},
  publisher={American Association for the Advancement of Science}
}

@article{barnett2020inferring,
  title={Inferring mobility measures from GPS traces with missing data},
  author={Barnett, Ian and Onnela, Jukka-Pekka},
  journal={Biostatistics},
  volume={21},
  number={2},
  pages={e98--e112},
  year={2020},
  publisher={Oxford University Press}
}

@inproceedings{rankin1985effects,
  title={Effects of missing data on the statistical analysis of clinical time series},
  author={Rankin, Eric D and Marsh, Jeanne C},
  booktitle={Social Work Research and Abstracts},
  volume={21},
  number={2},
  pages={13--16},
  year={1985},
  organization={Oxford University Press}
}

@article{tawn2020missing,
  title={Missing data in wind farm time series: Properties and effect on forecasts},
  author={Tawn, Rosemary and Browell, Jethro and Dinwoodie, Iain},
  journal={Electric Power Systems Research},
  volume={189},
  pages={106640},
  year={2020},
  publisher={Elsevier}
}

@article{chen2019complete,
  title={Complete trajectory reconstruction from sparse mobile phone data},
  author={Chen, Guangshuo and Viana, Aline Carneiro and Fiore, Marco and Sarraute, Carlos},
  journal={EPJ Data Science},
  volume={8},
  number={1},
  pages={1--24},
  year={2019},
  publisher={Springer}
}

@article{karrer2014percolation,
  title={Percolation on sparse networks},
  author={Karrer, Brian and Newman, Mark EJ and Zdeborov{\'a}, Lenka},
  journal={Physical review letters},
  volume={113},
  number={20},
  pages={208702},
  year={2014},
  publisher={APS}
}

@article{Nowzari2016AnalysisNetworks,
    title = {{Analysis and control of epidemics: A survey of spreading processes on complex networks}},
    year = {2016},
    journal = {IEEE Control Systems Magazine},
    author = {Nowzari, Cameron and Preciado, Victor M and Pappas, George J},
    number = {1},
    pages = {26--46},
    volume = {36},
    publisher = {IEEE}
}

@article{hariharan2004project,
  title={Project Lachesis: Parsing and Modeling Location Histories},
  author={Hariharan, Ramaswamy and Toyama, Kentaro},
  booktitle={Geographic Information Science},
  pages={106--124},
  year={2004},
  publisher={Springer}
}

@inproceedings{akiba2019optuna,
  title={Optuna: A next-generation hyperparameter optimization framework},
  author={Akiba, Takuya and Sano, Shotaro and Yanase, Toshihiko and Ohta, Takeru and Koyama, Masanori},
  booktitle={Proceedings of the 25th ACM SIGKDD international conference on knowledge discovery \& data mining},
  pages={2623--2631},
  year={2019}
}

@inproceedings{cuttone2014inferring,
  title={Inferring human mobility from sparse low accuracy mobile sensing data},
  author={Cuttone, Andrea and Lehmann, Sune and Larsen, Jakob Eg},
  booktitle={Proceedings of the 2014 ACM International Joint Conference on Pervasive and Ubiquitous Computing: Adjunct Publication},
  pages={995--1004},
  year={2014}
}

@article{stopczynski2014measuring,
  title={Measuring large-scale social networks with high resolution},
  author={Stopczynski, Arkadiusz and Sekara, Vedran and Sapiezynski, Piotr and Cuttone, Andrea and Madsen, Mette My and Larsen, Jakob Eg and Lehmann, Sune},
  journal={PloS one},
  volume={9},
  number={4},
  pages={e95978},
  year={2014},
  publisher={Public Library of Science San Francisco, USA}
}

@article{aslak2020infostop,
  title={Infostop: scalable stop-location detection in multi-user mobility data},
  author={Aslak, Ulf and Alessandretti, Laura},
  journal={arXiv preprint arXiv:2003.14370},
  year={2020}}

@article{cencetti2021digital,
  title={Digital proximity tracing on empirical contact networks for pandemic control},
  author={Cencetti, Giulia and Santin, Gabriele and Longa, Antonio and Pigani, Emanuele and Barrat, Alain and Cattuto, Ciro and Lehmann, Sune and Salathe, Marcel and Lepri, Bruno},
  journal={Nature communications},
  volume={12},
  number={1},
  pages={1655},
  year={2021},
  publisher={Nature Publishing Group UK London}}

@article{sapiezynski2019interaction,
  title={Interaction data from the copenhagen networks study},
  author={Sapiezynski, Piotr and Stopczynski, Arkadiusz and Lassen, David Dreyer and Lehmann, Sune},
  journal={Scientific Data},
  volume={6},
  number={1},
  pages={315},
  year={2019},
  publisher={Nature Publishing Group UK London}
}

@article{wu2008community,
  title={How community structure influences epidemic spread in social networks},
  author={Wu, Xiaoyan and Liu, Zonghua},
  journal={Physica A: Statistical Mechanics and its Applications},
  volume={387},
  number={2-3},
  pages={623--630},
  year={2008},
  publisher={Elsevier}
}

@article{watts2005multiscale,
  title={Multiscale, resurgent epidemics in a hierarchical metapopulation model},
  author={Watts, Duncan J and Muhamad, Roby and Medina, Daniel C and Dodds, Peter S},
  journal={Proceedings of the National Academy of Sciences},
  volume={102},
  number={32},
  pages={11157--11162},
  year={2005},
  publisher={National Academy of Sciences}
}

@article{Budak2015DissectingMovement,
    title = {{Dissecting the spirit of Gezi: Influence vs. selection in the Occupy Gezi movement}},
    year = {2015},
    journal = {Sociological Science},
    author = {Budak, Ceren and Watts, Duncan J},
    pages = {370--397},
    volume = {2}
}

@article{Wesolowski2013TheMobility,
    title = {{The impact of biases in mobile phone ownership on estimates of human mobility}},
    year = {2013},
    journal = {Journal of the Royal Society Interface},
    author = {Wesolowski, Amy and Eagle, Nathan and Noor, Abdisalan M. and Snow, Robert W. and Buckee, Caroline O.},
    number = {81},
    month = {4},
    volume = {10},
    publisher = {The Royal Society},
    url = {/pmc/articles/PMC3627108/ /pmc/articles/PMC3627108/?report=abstract https://www.ncbi.nlm.nih.gov/pmc/articles/PMC3627108/},
    doi = {10.1098/RSIF.2012.0986},
    issn = {17425662},
    pmid = {23389897}
}

@article{luca2021survey,
  title={A survey on deep learning for human mobility},
  author={Luca, Massimiliano and Barlacchi, Gianni and Lepri, Bruno and Pappalardo, Luca},
  journal={ACM Computing Surveys (CSUR)},
  volume={55},
  number={1},
  pages={1--44},
  year={2021},
  publisher={ACM New York, NY}
}

@book{Anderson1992InfectiousControl,
    title = {{Infectious diseases of humans: dynamics and control}},
    year = {1992},
    author = {Anderson, Roy M and May, Robert M},
    publisher = {Oxford university press}
}

@article{birge_controlling_2022,
	title = {Controlling {Epidemic} {Spread}: {Reducing} {Economic} {Losses} with {Targeted} {Closures}},
	volume = {68},
	issn = {0025-1909},
	shorttitle = {Controlling {Epidemic} {Spread}},
	url = {https://pubsonline.informs.org/doi/abs/10.1287/mnsc.2022.4318},
	doi = {10.1287/mnsc.2022.4318},
	number = {5},
	urldate = {2025-11-19},
	journal = {Management Science},
	author = {Birge, John R. and Candogan, Ozan and Feng, Yiding},
	month = may,
	year = {2022},
	pages = {3175--3195},
}

@misc{safegraph_what_2019,
	title = {What {About} {Bias} in the {SafeGraph} {Dataset}?},
	howpublished = {https://www.safegraph.com/blog/what-about-bias-in-the-safegraph-dataset},
	urldate = {2022-05-05},
	author = {SafeGraph},
	year = {2019},
}

@article{ross2021household,
  title={Household visitation during the COVID-19 pandemic},
  author={Ross, Stuart and Breckenridge, George and Zhuang, Mengdie and Manley, Ed},
  journal={Scientific Reports},
  volume={11},
  number={1},
  pages={22871},
  year={2021},
  publisher={Nature Publishing Group UK London}
}

@article{pei2020differential,
  title={Differential effects of intervention timing on COVID-19 spread in the United States},
  author={Pei, Sen and Kandula, Sasikiran and Shaman, Jeffrey},
  journal={Science Advances},
  volume={6},
  number={49},
  pages={eabd6370},
  year={2020},
  publisher={American Association for the Advancement of Science},
  doi={10.1126/sciadv.abd6370}
}

@article{athey_estimating_2021,
  title={Estimating experienced racial segregation in {US} cities using large-scale {GPS} data},
  author={Athey, Susan and Ferguson, Billy and Gentzkow, Matthew and Schmidt, Tobias},
  journal={Proceedings of the National Academy of Sciences},
  volume={118},
  number={46},
  pages={e2026160118},
  year={2021},
  doi={10.1073/pnas.2026160118},
  url={https://www.pnas.org/doi/10.1073/pnas.2026160118}
}

@article{yabe_behavioral_2023,
  title={Behavioral changes during the {COVID}-19 pandemic decreased income diversity of urban encounters},
  author={Yabe, Takahiro and Bueno, Bernardo Garcia Bulle and Dong, Xiaowen and Pentland, Alex and Moro, Esteban},
  journal={Nature Communications},
  volume={14},
  number={1},
  pages={2310},
  year={2023},
  doi={10.1038/s41467-023-37913-y},
  url={https://www.nature.com/articles/s41467-023-37913-y}
}

@article{li_understanding_2024,
  title={Understanding the bias of mobile location data across spatial scales and over time: {A} comprehensive analysis of {SafeGraph} data in the {United States}},
  author={Li, Zhenlong and Ning, Huan and Jing, Fengrui and Lessani, M. Naser},
  journal={PLOS ONE},
  volume={19},
  number={1},
  pages={e0294430},
  year={2024},
  doi={10.1371/journal.pone.0294430},
  url={https://journals.plos.org/plosone/article?id=10.1371/journal.pone.0294430}
}

@inproceedings{coston2021leveraging,
	location = {Virtual Event Canada},
	title = {Leveraging Administrative Data for Bias Audits: Assessing Disparate Coverage with Mobility Data for {COVID}-19 Policy},
	isbn = {978-1-4503-8309-7},
	url = {https://dl.acm.org/doi/10.1145/3442188.3445881},
	doi = {10.1145/3442188.3445881},
	shorttitle = {Leveraging Administrative Data for Bias Audits},
	eventtitle = {{FAccT} '21: 2021 {ACM} Conference on Fairness, Accountability, and Transparency},
	pages = {173--184},
	booktitle = {Proceedings of the 2021 {ACM} Conference on Fairness, Accountability, and Transparency},
	publisher = {{ACM}},
	author = {Coston, Amanda and Guha, Neel and Ouyang, Derek and Lu, Lisa and Chouldechova, Alexandra and Ho, Daniel E.},
	urldate = {2024-11-26},
	year = {2021},
	date = {2021-03-03},
	langid = {english},
}

@article{li_substantial_2020,
	title = {Substantial undocumented infection facilitates the rapid dissemination of novel coronavirus ({SARS}-{CoV}-2)},
	volume = {368},
	issn = {0036-8075},
	url = {https://pmc.ncbi.nlm.nih.gov/articles/PMC7164387/},
	doi = {10.1126/science.abb3221},
	number = {6490},
	urldate = {2025-11-19},
	journal = {Science (New York, N.y.)},
	author = {Li, Ruiyun and Pei, Sen and Chen, Bin and Song, Yimeng and Zhang, Tao and Yang, Wan and Shaman, Jeffrey},
	month = may,
	year = {2020},
	pmid = {32179701},
	pmcid = {PMC7164387},
	pages = {489--493},
}

@article{Chinazzi2020TheOutbreak,
    title = {{The effect of travel restrictions on the spread of the 2019 novel coronavirus (COVID-19) outbreak}},
    year = {2020},
    journal = {Science},
    author = {Chinazzi, Matteo and Davis, Jessica T and Ajelli, Marco and Gioannini, Corrado and Litvinova, Maria and Merler, Stefano and y Piontti, Ana Pastore and Mu, Kunpeng and Rossi, Luca and Sun, Kaiyuan and {others}},
    number = {6489},
    pages = {395--400},
    volume = {368},
    publisher = {American Association for the Advancement of Science}
}

@article{eagle_inferring_2009,
	title = {Inferring friendship network structure by using mobile phone data},
	volume = {106},
	url = {https://www.pnas.org/doi/abs/10.1073/pnas.0900282106},
	doi = {10.1073/pnas.0900282106},
	number = {36},
	urldate = {2025-11-29},
	journal = {Proceedings of the National Academy of Sciences},
	author = {Eagle, Nathan and Pentland, Alex (Sandy) and Lazer, David},
	month = sep,
	year = {2009},
	pages = {15274--15278},
}

@inproceedings{zheng2008understanding,
  title={Understanding mobility based on GPS data},
  author={Zheng, Yu and Li, Quannan and Chen, Yukun and Xie, Xing and Ma, Wei-Ying},
  booktitle={Proceedings of the 10th international conference on Ubiquitous computing},
  pages={312--321},
  year={2008}
}

@article{kraemer2020effect,
  title={The effect of human mobility and control measures on the COVID-19 epidemic in China},
  author={Kraemer, Moritz UG and Yang, Chia-Hung and Gutierrez, Bernardo and Wu, Chieh-Hsi and Klein, Brennan and Pigott, David M and Open COVID-19 Data Working Group† and Du Plessis, Louis and Faria, Nuno R and Li, Ruoran and others},
  journal={Science},
  volume={368},
  number={6490},
  pages={493--497},
  year={2020},
  publisher={American Association for the Advancement of Science}}

@article{vanni2021use,
  title={On the use of aggregated human mobility data to estimate the reproduction number},
  author={Vanni, Fabio and Lambert, David and Palatella, Luigi and Grigolini, Paolo},
  journal={Scientific reports},
  volume={11},
  number={1},
  pages={23286},
  year={2021},
  publisher={Nature Publishing Group UK London}
}

@article{klein_characterizing_2024,
    title = {Characterizing collective physical distancing in the {U}.{S}. during the first nine months of the {COVID}-19 pandemic},
    volume = {3},
    issn = {2767-3170},
    url = {https://journals.plos.org/digitalhealth/article?id=10.1371/journal.pdig.0000430},
    doi = {10.1371/journal.pdig.0000430},
    language = {en},
    number = {2},
    urldate = {2025-09-18},
    journal = {PLOS Digital Health},
    author = {Klein, Brennan and LaRock, Timothy and McCabe, Stefan and Torres, Leo and Friedland, Lisa and Kos, Maciej and Privitera, Filippo and Lake, Brennan and Kraemer, Moritz U. G. and Brownstein, John S. and Gonzalez, Richard and Lazer, David and Eliassi-Rad, Tina and Scarpino, Samuel V. and Vespignani, Alessandro and Chinazzi, Matteo},
    month = feb,
    year = {2024},
    note = {Publisher: Public Library of Science},
    pages = {e0000430},
}

@article{ozaki2022direct,
  title={Direct modelling from GPS data reveals daily-activity-dependency of effective reproduction number in COVID-19 pandemic},
  author={Ozaki, Jun’ichi and Shida, Yohei and Takayasu, Hideki and Takayasu, Misako},
  journal={Scientific Reports},
  volume={12},
  number={1},
  pages={17888},
  year={2022},
  publisher={Nature Publishing Group UK London}}

@article{pappalardo2022scikit,
  title={Scikit-mobility: A Python library for the analysis, generation, and risk assessment of mobility data},
  author={Pappalardo, Luca and Simini, Filippo and Barlacchi, Gianni and Pellungrini, Roberto},
  journal={Journal of Statistical Software},
  volume={103},
  pages={1--38},
  year={2022}
}

@article{yabe2020understanding,
  title={Understanding post-disaster population recovery patterns},
  author={Yabe, Takahiro and Tsubouchi, Kota and Fujiwara, Naoya and Sekimoto, Yoshihide and Ukkusuri, Satish V},
  journal={Journal of the Royal Society Interface},
  volume={17},
  number={163},
  year={2020},
  publisher={The Royal Society}
}

@article{ozella2021using,
  title={Using wearable proximity sensors to characterize social contact patterns in a village of rural Malawi},
  author={Ozella, Laura and Paolotti, Daniela and Lichand, Guilherme and Rodr{\'\i}guez, Jorge P and Haenni, Simon and Phuka, John and Leal-Neto, Onicio B and Cattuto, Ciro},
  journal={EPJ Data Science},
  volume={10},
  number={1},
  pages={46},
  year={2021},
  publisher={Springer Berlin Heidelberg}
}

@article{liu2013face,
  title={Face-to-face proximity estimationusing bluetooth on smartphones},
  author={Liu, Shu and Jiang, Yingxin and Striegel, Aaron},
  journal={IEEE Transactions on Mobile Computing},
  volume={13},
  number={4},
  pages={811--823},
  year={2013},
  publisher={IEEE}
}

@article{kheifetz2022parametrization,
  title={On the parametrization of epidemiologic models—lessons from modelling covid-19 epidemic},
  author={Kheifetz, Yuri and Kirsten, Holger and Scholz, Markus},
  journal={Viruses},
  volume={14},
  number={7},
  pages={1468},
  year={2022},
  publisher={MDPI}
}

@misc{nyt_covid19_data,
  author       = {{The New York Times}},
  title        = {Coronavirus (COVID-19) Data in the United States},
  year         = {2021},
  howpublished = {\url{https://github.com/nytimes/covid-19-data}},
  note         = {Data based on reports from state and local health agencies. Licensed under CC BY-NC 4.0},
}

@article{makinoshima2025large,
  title={Large-scale geolocation data reveal evacuation behaviour during the 2024 Noto Peninsula earthquake and tsunami},
  author={Makinoshima, Fumiyasu and Yotsui, Saki and Sato, Shosuke and Imamura, Fumihiko},
  journal={Communications Earth \& Environment},
  volume={6},
  number={1},
  pages={891},
  year={2025},
  publisher={Nature Publishing Group UK London}
}

@article{giardini2023using,
  title={Using mobile phone data to map evacuation and displacement: a case study of the central Italy earthquake},
  author={Giardini, Francesca and Hadjidimitriou, Natalia Selini and Mamei, Marco and Bastardi, Giordano and Codeluppi, Nico and Pancotto, Francesca},
  journal={Scientific reports},
  volume={13},
  number={1},
  pages={22228},
  year={2023},
  publisher={Nature Publishing Group UK London}
}

@article{kieu2025modelling,
  title={Modelling social mobility disruptions and recovery during disasters: A mobile phone data approach},
  author={Kieu, Minh and Comber, Alexis and Quang, Thanh Bui and Malleson, Nick},
  journal={International Journal of Disaster Risk Reduction},
  pages={105812},
  year={2025},
  publisher={Elsevier}
}

@article{lu2026human,
  title={Human mobility in epidemic modeling},
  author={Lu, Xin and Feng, Jiawei and Lai, Shengjie and Holme, Petter and Liu, Shuo and Du, Zhanwei and Yuan, Xiaoqian and Wang, Siqing and Li, Yunxuan and Zhang, Xiaoyu and others},
  journal={Physics Reports},
  volume={1157},
  pages={1--45},
  year={2026},
  publisher={Elsevier}
}

@article{abbiasov202415,
  title={The 15-minute city quantified using human mobility data},
  author={Abbiasov, Timur and Heine, Cate and Sabouri, Sadegh and Salazar-Miranda, Arianna and Santi, Paolo and Glaeser, Edward and Ratti, Carlo},
  journal={Nature Human Behaviour},
  volume={8},
  number={3},
  pages={445--455},
  year={2024},
  publisher={Nature Publishing Group UK London}
}

@article{magyar2025utilizing,
  title={Utilizing mobile phone tracking data to estimate Intra-City modal mobility: A study on active mobility in two Finnish City regions},
  author={Magyar, Marton and Ala-Hulkko, Terhi and Antikainen, Harri and Lankila, Tiina and Kotavaara, Ossi},
  journal={Journal of Transport Geography},
  volume={128},
  pages={104326},
  year={2025},
  publisher={Elsevier}
}

@article{chafetz2022data4covid19,
  title        = {The\# Data4COVID19 Review: Assessing the Use of Non-Traditional Data During A Pandemic Crisis},
  author       = {Chafetz, Hannah and Zahuranec, Andrew J and Marcucci, Sara and Davletov, Behruz and Verhulst, Stefaan},
  year         = {2022},
  journal      = {Available at SSRN 4273229}
}

@article{aktay2020google,
  title        = {Google COVID-19 Community Mobility Reports: anonymization process description (version 1.1)},
  author       = {Aktay, Ahmet and Bavadekar, Shailesh and Cossoul, Gwen and Davis, John and Desfontaines, Damien and Fabrikant, Alex and Gabrilovich, Evgeniy and Gadepalli, Krishna and Gipson, Bryant and Guevara, Miguel and others},
  year         = {2020},
  journal      = {arXiv preprint arXiv:2004.04145}
}

@misc{apple2020covid,
  title        = {COVID-19 Mobility Trends Reports.},
  author       = {Apple},
  year         = {2020},
  url          = {https://covid19.apple.com/mobility}
}

@misc{descartes2020aggregated,
  title        = {Aggregated Mobility Index.},
  author       = {Descartes labs},
  year         = {2021},
  url          = {https://mktg.descarteslabs.com/mobility-tracking}
}

@article{bahr_missing_2022,
	title = {Missing {Data} and {Other} {Measurement} {Quality} {Issues} in {Mobile} {Geolocation} {Sensor} {Data}},
	volume = {40},
	issn = {0894-4393},
	url = {https://doi.org/10.1177/0894439320944118},
	doi = {10.1177/0894439320944118},
	language = {en},
	number = {1},
	urldate = {2024-01-06},
	journal = {Social Science Computer Review},
	author = {Bähr, Sebastian and Haas, Georg-Christoph and Keusch, Florian and Kreuter, Frauke and Trappmann, Mark},
	month = feb,
	year = {2022},
	note = {Publisher: SAGE Publications Inc},
	pages = {212--235},
}

@article{mccool_maximum_2024,
	title = {Maximum interpolable gap length in missing smartphone-based {GPS} mobility data},
	volume = {51},
	issn = {1572-9435},
	url = {https://doi.org/10.1007/s11116-022-10328-2},
	doi = {10.1007/s11116-022-10328-2},
	language = {en},
	number = {1},
	urldate = {2025-09-29},
	journal = {Transportation},
	author = {McCool, Danielle and Lugtig, Peter and Schouten, Barry},
	month = feb,
	year = {2024},
	pages = {297--327},
}

@article{sanchez2026correcting,
  title={Correcting temporal bias in mobility data using time-use surveys},
  author={Sanchez, Sarah A and Gibbs, Hamish and Yabe, Takahiro and O'Brien, Daniel T and Moro, Esteban},
  journal={arXiv preprint arXiv:2601.22330},
  year={2026}
}

@article{liu2024nonrepresentativeness,
  title={Nonrepresentativeness of human mobility data and its impact on modeling dynamics of the COVID-19 pandemic: Systematic evaluation},
  author={Liu, Chuchu and Holme, Petter and Lehmann, Sune and Yang, Wenchuan and Lu, Xin},
  journal={JMIR Formative Research},
  volume={8},
  pages={e55013},
  year={2024},
  publisher={JMIR Publications Toronto, Canada}
}

@article{chin2025bias,
  title={Bias in mobility datasets drives divergence in modeled outbreak dynamics},
  author={Chin, Taylor and Johansson, Michael A and Chowdhury, Anir and Chowdhury, Shayan and Hosan, Kawsar and Quader, Md Tanvir and Buckee, Caroline O and Mahmud, Ayesha S},
  journal={Communications Medicine},
  volume={5},
  number={1},
  pages={8},
  year={2025},
  publisher={Nature Publishing Group UK London}
}

@article{pullano2024mobility,
  title={Mobility resolution needed to inform predictive epidemic models for spatial transmission from mobile phone data},
  author={Pullano, Giulia and Bansal, Shweta and Rubrichi, Stefania and Colizza, Vittoria},
  journal={medRxiv},
  pages={2024--10},
  year={2024},
  publisher={Cold Spring Harbor Laboratory Press}
}

@article{hwang2022comparison,
  title={Comparison of GPS imputation methods in environmental health research},
  author={Hwang, Sungsoon and Webber-Ritchey, Kashica and Moxley, Elizabeth A},
  journal={Geospatial Health},
  volume={17},
  number={2},
  pages={1081},
  year={2022}
}

@misc{wu_location-based_2024,
    title = {Location-{Based} {Service} ({LBS}) {Data} {Quality} {Metrics} and {Effects} on {Mobility} {Inference}},
    url = {http://arxiv.org/abs/2411.16595},
    doi = {10.48550/arXiv.2411.16595},
    urldate = {2025-11-07},
    publisher = {arXiv},
    author = {Wu, Xinhua and Wang, Yanchao and Ugurel, Ekin and Chen, Cynthia and Huang, Shuai and Wang, Qi R.},
    month = nov,
    year = {2024},
    note = {arXiv:2411.16595 [cs]},
}

@article{chen_lbs_biases_2026,
  title = {From biases to opportunities: Leveraging {Location-Based-Service} ({LBS}) data for next-generation transportation planning},
  author = {Chen, Cynthia and Wang, Ryan and Bansal, Prateek and Chen, Lyra and Ugurel, Ekin and Zhang, Yuteng and Wu, Xinhua},
  journal = {Transportation Research Part C: Emerging Technologies},
  volume = {182},
  pages = {105416},
  year = {2026},
  doi = {10.1016/j.trc.2025.105416}
}

@article{ugurel_correcting_2024,
  title = {Correcting missingness in passively-generated mobile data with {Multi-Task Gaussian Processes}},
  author = {Ugurel, Ekin and Guan, Xiangyang and Wang, Yanchao and Huang, Shuai and Wang, Qi and Chen, Cynthia},
  journal = {Transportation Research Part C: Emerging Technologies},
  volume = {161},
  pages = {104523},
  year = {2024},
  doi = {10.1016/j.trc.2024.104523}
}

@article{wang_exploring_2025,
  title = {Exploring biases in travel behavior patterns in big passively generated mobile data from 11 {U.S.} cities},
  author = {Wang, Yanchao and Guan, Xiangyang and Ugurel, Ekin and Chen, Cynthia and Huang, Shuai and Wang, Qi R.},
  journal = {Journal of Transport Geography},
  volume = {123},
  pages = {104108},
  year = {2025},
  doi = {10.1016/j.jtrangeo.2024.104108}
}

@article{yoo2020quality,
  title={Quality of hybrid location data drawn from GPS-enabled mobile phones: Does it matter?},
  author={Yoo, Eun-Hye and Roberts, John E and Eum, Youngseob and Shi, Youdi},
  journal={Transactions in GIS},
  volume={24},
  number={2},
  pages={462--482},
  year={2020},
  publisher={Wiley Online Library}
}

@article{nadini2018epidemic,
  title={Epidemic spreading in modular time-varying networks},
  author={Nadini, Matthieu and Sun, Kaiyuan and Ubaldi, Enrico and Starnini, Michele and Rizzo, Alessandro and Perra, Nicola},
  journal={Scientific reports},
  volume={8},
  number={1},
  pages={2352},
  year={2018},
  publisher={Nature Publishing Group UK London}
}
\end{refsection}
\end{document}